\documentclass[lettersize,journal]{IEEEtran}
\usepackage{array}
\usepackage[caption=false,font=normalsize,labelfont=sf,textfont=sf]{subfig}
\usepackage{textcomp}
\usepackage{stfloats}
\usepackage{url}
\usepackage{verbatim}
\usepackage{graphicx}
\usepackage{cite}
\usepackage{longtable}   
\usepackage{booktabs}    
\usepackage{pdflscape} 
\usepackage{amsmath,amssymb,amsfonts}
\usepackage[caption=false,font=normalsize,labelfont=sf,textfont=sf]{subfig}
\usepackage[export]{adjustbox}
\usepackage{multirow,multicol}
\usepackage{enumitem}
\usepackage{colortbl}
\IEEEoverridecommandlockouts
\usepackage{changepage}
\usepackage{calc}
\usepackage{hyperref}
\hypersetup{
    colorlinks,
    citecolor=blue,
    filecolor=black,
    linkcolor=blue,
    urlcolor=blue
}
\usepackage{algorithm}
\usepackage{algpseudocode}
\algblock{Input}{EndInput}
\algnotext{EndInput}
\algblock{Output}{EndOutput}
\algnotext{EndOutput}

\usepackage{pifont}
\newcommand{\xmark}{\ding{53}}%
\usepackage[table]{xcolor}
\usepackage{circledsteps}

\usepackage[para,online,flushleft]{threeparttable}
\usepackage{bm}
\usepackage{rotating}
\usepackage{listings}
\usepackage{parcolumns}
\usepackage{float} 

\newcommand{\shamim}[1]{{\color{blue} #1}}

\begin{document}

\title{A systematic review of metaheuristics-based and machine learning-driven intrusion detection systems in IoT}

\author{Mohammad Shamim Ahsan, Salekul Islam, and Swakkhar Shatabda



\thanks{Mohammad Shamim Ahsan is with the Department of Computer Science and Engineering, Bangladesh University of Engineering and Technology (email: shamim19119@gmail.com)}

\thanks{Salekul Islam is with the Department of Electrical and Computer Engineering, North South University}

\thanks{Swakkhar Shatabda with the Department of Computer Science and Engineering, BRAC University (email: swakkhar.shatabda@bracu.ac.bd)}

}



\maketitle

\shamim{This paper has been published in \textit{Swarm and Evolutionary Computation}, available at \url{https://doi.org/10.1016/j.swevo.2025.101984}} 

\begin{abstract}
The widespread adoption of the Internet of Things (IoT) has raised a new challenge for developers since it is prone to known and unknown cyberattacks due to its heterogeneity, flexibility, and close connectivity. To defend against such security breaches, researchers have focused on building sophisticated intrusion detection systems (IDSs) using machine learning (ML) techniques. Although these algorithms notably improve detection performance, they require excessive computing power and resources, which are crucial issues in IoT networks considering the recent trends of decentralized data processing and computing systems. Consequently, many optimization techniques have been incorporated with these ML models. Specifically, a special category of optimizer adopted from the behavior of living creatures and different aspects of natural phenomena, known as metaheuristic algorithms, has been a central focus in recent years and brought about remarkable results. Considering this vital significance, we present a comprehensive and systematic review of various applications of metaheuristics algorithms in developing a machine learning-based IDS, especially for IoT. A significant contribution of this study is the discovery of hidden correlations between these optimization techniques and machine learning models integrated with state-of-the-art IoT-IDSs. In addition, the effectiveness of these metaheuristic algorithms in different applications, such as feature selection, parameter or hyperparameter tuning, and hybrid usages are separately analyzed. Moreover, a taxonomy of existing IoT-IDSs is proposed. Furthermore, we investigate several critical issues related to such integration. Our extensive exploration ends with a discussion of promising optimization algorithms and technologies that can enhance the efficiency of IoT-IDSs.            
\end{abstract}

\begin{IEEEkeywords}
Internet of Things (IoT), Intrusion Detection Systems (IDS), Machine Learning (ML), Deep Learning (DL), Metaheuristic Algorithms, Cybersecurity, Optimization Techniques.
\end{IEEEkeywords}

\section{Introduction}\label{sec:intro}
The popularity of the Internet of Things devices has spread surprisingly in the last few years. Nowadays, it offers a scalable platform not only for industry, healthcare, and home applications but also for agriculture, vehicular settings, and ultra-sophisticated systems like drone technologies \cite{intro_2_abualigah2021applications}. Alarmingly, this widespread adoption leads to unavoidable security issues as a negative side effect of close connectivity. Transferring susceptible information, such as personal data, patient data, and private business analysis makes such effects more severe and unimaginable. For example, millions of IoT and embedded devices (e.g., DVDs, printers, and IP cameras) were infected by a botnet in 2016, widely known as \textit{Mirai botnet}\cite{antonakakis2017understanding}. Notably, the attack was spread to several nations and manufacturing organizations and affected around 65 thousand IoT devices within the first 20 hours. Another concerning incident occurred in 2020, where an ADT employee pled guilty to accessing the security cameras of 220 women over 9,600 times during four years \cite{erin}. The defendant routinely added his email address to customers’ ``ADT Pulse'' accounts and got real-time access to the video feeds from their homes for sexual gratification. In such cases, an IoT-IDS can be used to detect and report on unique visitors, unauthorized access, and malicious activities.

Intrusion detection is one of the most crucial aspects of IoT security. An IDS typically identifies diverse attacks based on predefined rules or specific deviations from normal behavioral patterns. It can identify external and internal attacks on networks or computer systems, surpassing the performance of the traditional firewall. Typically, a firewall works on a set of specific rules, often based on IP addresses, port numbers, and protocols to check which packets are allowed to enter the network. Since firewalls filter packets relying on simple policies, they cannot detect internal or external attacks which require analyzing complex behavioral patterns. On the contrary, an IDS can identify malicious activities by observing the deviations from normal behaviors, which include both simple and complicated patterns. The entire IDS can be divided into two modules: feature engineering (mostly \textit{feature selection}) and \textit{classification} or detection. Feature selection aims to select a set of optimal features, discarding the least significant ones to efficiently and faster the classification process and reduce the computational overhead of the system. Feature selection methods can be categorized into three methods: \textit{filter}, \textit{wrapper}, and \textit{hybrid} methods. In the filter method, all features are statistically examined and rated with the help of data (input and target variables). Then based on the rating, less ranked features or features rated below a specific threshold are eliminated before classification. Information gain and linear correlation coefficient \cite{eid2013linear} are well-known metrics used in filter methods. The wrapper techniques outperform filter methods by training and testing a machine-learning model using each subset of features, iteratively. Specifically, these methods rank subsets of features based on their prediction accuracy generated from the machine learning algorithms. However, they are more expensive and time-consuming than the filter approaches. Sequential forward selection, sequential backward selection, stepwise selection \cite{hastie2017extended}, hill climbing, etc. are popular wrapper methods. Though filter methods are proven simple, fast, and scalable, they consider feature dependencies to a certain extent, resulting in an inappropriate feature set. Besides, wrapper methods offer better feature sets than filters. But, they become much slower and computationally expensive for a large number of features \cite{wah2018feature}. 

A group of wrapper methods fall into the category of optimization-based techniques, known as \textit{metaheuristics algorithms} that overcome the drawbacks of the prior methods. A metaheuristic is a general exploration method that applies to optimize an underlying heuristic. In the case of feature selection in ML, the optimal subset of features is searched in the feature space based on some heuristic or performance measure. Generally, a metaheuristic algorithm consists of two phases: exploitation and exploration. In the exploitation or intensification phase, the algorithm explores the neighborhood of an already promising solution in the search space. However, during exploration (a.k.a., diversification), the algorithm tries to traverse the unvisited regions of the search space. Although they do not make any hypothesis on the mathematical properties of the objective function, they gradually develop it through a continuous learning process. Among the major advantages, the utilization of parameters and comparatively faster convergence to the solutions are crucial. In addition, metaheuristics are efficient and effective in obtaining global optimal values, resulting in global optimal features. Moreover, even with large datasets, they perform significantly well~\cite{dokeroglu2022comprehensive}. However, metaheuristics are approximate and usually non-deterministic and do not guarantee the optimal (or, best) solution \cite{desale2015heuristic} like the exact algorithms (e.g., dynamic programming, branch and bound, branch and cut, linear programming, etc.). Still, they can provide near-optimal solutions in acceptable computing time (but the exact algorithms can not), which is highly essential for complex problems like detecting intrusions in dynamic environments. Interestingly, these techniques are mostly inspired by natural phenomena, including the instincts of living creatures. In addition to filter and wrapper techniques, a hybrid approach is another one that focuses on combining different aspects of existing feature selection methods \cite{intro_3_7745366}. 

Regarding attack or intrusion detection, traditional IDSs like statistical-driven (e.g., payload-based), rule-based, heuristics-based, etc. cannot detect the complex patterns of dynamic IoT systems. On the other hand, classical and deep machine learning models have been proven to generate notable results, even in heterogeneous environments such as IoT. The main purpose of using ML-based techniques is to handle large data sets and produce high accuracy, fast processing, and significant performance; thus enhancing security. However, they require high computational resources and a significant amount of time to achieve minor precision improvements \cite{bharati2022machine}. Despite these improvements offered by machine learning, the era of big data and the increasing use of IoT introduce new problems with traditional centralized cloud-based data storage and processing systems. In particular, low throughput, high latency, and data privacy are the most serious issues \cite{hua2023edge}. In addition, IoT devices contain sensitive and private data, such as financial or patient information. To address these issues, edge computing technology has become widely accepted, especially in the IoT context. In this technology, data are processed, stored, and computed closer to the location of devices. Consequently, not only the data transmission time, response time, and latency are reduced but also higher scalability and decentralization are achieved. Regarding IDSs, when machine learning models are trained in edge servers with large datasets, the computing power and the adequacy of energy support become crucial challenges since edge servers can hardly meet these requirements \cite{yu2017survey}.  

Although researchers always rely on utilizing optimization techniques to mitigate such problems and improve the effectiveness of ML-oriented IoT-IDSs, metaheuristics-based optimization has been a notable focus in recent years. Considering the outstanding facilities offered by these optimizers, they can play a pivotal role in designing IoT-IDSs. Specifically, these algorithms can be utilized not only to select optimal feature sets --- before being trained by an ML-based classifier --- but also to optimize the parameters (e.g., weights and biases) and hyperparameters (e.g., learning rate, number of neurons, layers volume, and amount of epochs) of the models during training. For these reasons, numerous recent works \cite{1_kareem2022effective, 3_alkanhel2023network, 7_sanju2023enhancing, 97_10.1007/s11276-023-03435-0, 50_jovanovic2022feature} have employed them to select an optimal set of features; while others have used these techniques to tune parameters \cite{35_baniasadi2022novel, 75_jothi2023wils} and hyperparameters \cite{29_bahaa2022novel, 71_basheri2023quantum, 85_vijayan2024original} in ML-driven classifiers.
\\\\ 
\noindent \textbf{Scope of this Review and Contributions} 

Among existing related studies, almost all have focused on one aspect: metaheuristics or machine learning techniques; not both. Although very few surveys mention the integration of these two, their coverage and classifications are considerably inadequate. Moreover, in notable cases, the selected works are not IoT-specific. Most importantly, these studies do not analyze the connections of optimization techniques with machine learning algorithms while developing an IoT-IDS. Furthermore, no studies have analyzed the different applications of metaheuristics for such detection systems. To address these gaps in the literature, we analyze a diverse range of metaheuristics, from swarm-based, nature-inspired, and evolutionary algorithms to search-based, human-inspired, physics-based, mathematics-based, and hybrid ones. Regarding machine learning techniques, conventional methods such as classifications, artificial neural networks, and ensemble learning, along with advanced algorithms, for instance, autoencoder, deep belief networks, deep neural networks, recurrent neural networks, convolutional neural networks, etc. are explored. Another significant contribution of this work is the analysis of the various applications of metaheuristics in the IoT IDSs, such as feature selection, parameter optimization, hyperparameter tuning, and their hybrid usage. Moreover, the correlations among metaheuristics, machine learning, and datasets used for the IoT-IDSs are figured out that distinguish this work from others. In summary, the following contributions are made:
 \begin{itemize}
     \item We present an extensive review of the existing applications of metaheuristics algorithms to develop machine learning-based intrusion detection systems, especially for IoT. In addition, a large-scale taxonomy of metaheuristics and ML-integrated IoT-IDSs is introduced.
     \item Hidden correlations among top-notch metaheuristics, ML techniques, and the most commonly used datasets are analyzed, and some insightful findings are disclosed. Importantly, these results also reflect the effectiveness of such metaheuristic-ML integration considering different applications, especially feature selection, parameter or hyperparameter tuning, and hybrid cases.
     \item Several crucial challenges that may arise when integrating metaheuristic algorithms with machine learning techniques are outlined. Accordingly, the viability of a few emerging technologies is discussed. Finally, some possible integrations of metaheuristics and ML, and their feasibility in IoT-IDSs are explored.
 \end{itemize}
The remainder of this paper proceeds as follows. Section \ref{sec:background} introduces the background, including the classification of intrusion detection systems, metaheuristics, and machine learning techniques. Section \ref{sec:related} provides the related surveys with their limitations, research gaps analysis, and differentiating aspects of our work. Then, in Section \ref{sec:methodology}, we discuss our research objectives along with search strategy and data assessment. Next, Section \ref{sec:results} introduces the results of the systematic literature review, including the technical and extensive exploration of the existing relevant detection systems and many insightful findings. After that, in Section \ref{sec:discussion}, our overall investigation with possible future challenges are summarized. 
Finally, Section \ref{sec:conclusion} concludes our work. Along with them, an \textit{Appendix} section at the end of the paper presents a performance tabulation of existing metaheuristics-based and ML-driven IoT-IDSs in tabulation form.

\section{Background\label{sec:background}}
In this section, intrusion detection methods, techniques, meta-heuristics, and machine learning-based models are briefly discussed. 
\subsection{Intrusion Detection Methods and Techniques in IoT}
\textbf{Methods.} An intrusion detection system (IDS) is a software or hardware system that detects traces of malicious activities on a computer system or network. Primarily, IDS can be categorized into three types: host-based IDS (HIDS), protocol-based intrusion detection system (PIDS), and network-based IDS (NIDS). In HIDS, the system is dedicated to working for a specific host. As a result, any insider as well as outsider attack is seamlessly detected. The most crucial limitation here is the necessity of one IDS for each host. Regarding PIDS, the system concentrates on identifying malicious behaviors in a specific protocol. Usually, a PIDS is executed either within a single or among multiple hosts. For example, a PIDS may inspect TCP or HTTP traffic to trace malicious content. Similarly, it can monitor traffic between a web server and a database to detect any suspicious SQL queries. In NIDS, intrusions within the network are detected by monitoring the patterns and contents of the incoming and outgoing traffic. Consequently, the outside intrusions can be identified and all hosts are protected. However, it becomes expensive whenever there is too much traffic in the network.

\begin{figure}
    \centering
    \includegraphics[width=1\linewidth]{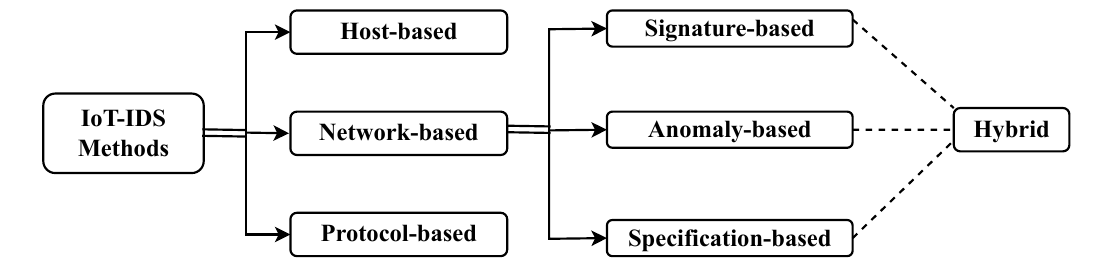}
    \caption{Categories of Intrusion Detection Methods in IoT.}
    \label{fig:IDS-cat}
\end{figure}

Regarding the IoT, most of the intrusion detection systems are network-based since attackers can seamlessly misuse the heterogeneous and dynamic characteristics of the IoT environment. In the literature, NIDS is classified into the following four methods: Signature-based Intrusion Detection System (SIDS), Anomaly-based Intrusion Detection System (AIDS), Specification-based Intrusion Detection System (SpIDS), and Hybrid Intrusion Detection System (HyIDS). In SIDS, an intrusion signature is checked with the previously known intrusion patterns, stored in the database, to find significant matching. It is also known as knowledge-based detection or misuse detection. AIDS is a dynamic intrusion detection approach that monitors the activity log of a system and reports anomalies whenever it observes any deviations from normal behaviors. This method provides the capabilities to detect not only known and unknown attacks but also any insider attacks. In SpIDS, a set of rules and thresholds are defined for network modules like nodes, protocols, firewalls, etc. Utilizing these specifications, the system detects intrusions while observing any discrepancies from the acceptable behaviors \cite{mitchell2014survey}. On the other hand, Hybrid IDS incorporates the advantages of SIDS, AIDS, and SpIDS to detect both familiar and novel intrusions utilizing limited computational resources. The classification of NIDS is showed in Figure~\ref{fig:IDS-cat}.  

\begin{figure*}
    \includegraphics[width=1\linewidth]{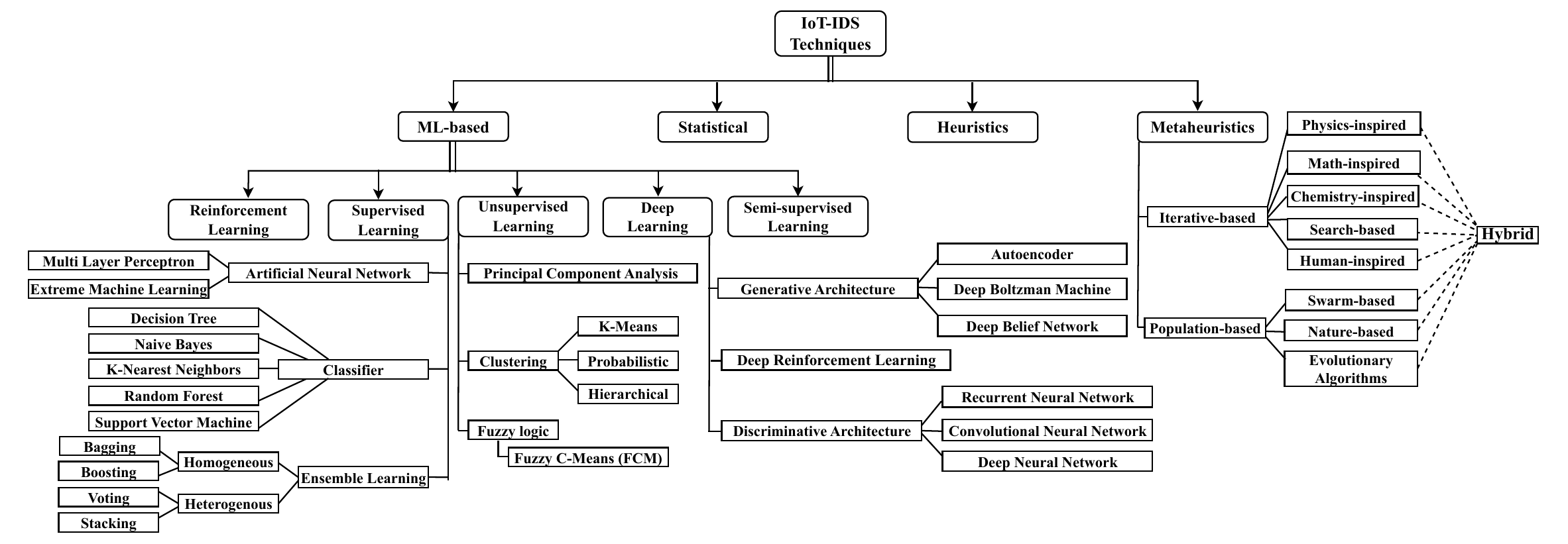}
    \caption{Classification of the existing IoT-IDS techniques.}
    \label{fig:iot_ids_classification}
\end{figure*}

\textbf{Techniques.} Considering the resource and energy-constrained characteristics, AIDS and HyIDS are the most appropriate and feasible methods in IoT \cite{fu2017automata, midi2017kalis}. Various machine Learning-based, statistical-driven (like payload-based), rule-based, and heuristics-based techniques are used to escalate the training process for AIDS. Recently, metaheuristics and hybrid approaches integrating different algorithms have been developed in this field, especially for IoT. In the next two Sections, some well-established techniques used in IoT-IDS are discussed. Figure~\ref{fig:iot_ids_classification} illustrates the classification of the existing IoT-IDS techniques as a whole.

\subsection{Metaheuristic Algorithms}
A metaheuristic is a general exploration (or diversification) method that can be applied to different problems in a similar way by visiting regions of the search space that are not already seen and evaluating candidate solutions. In general terms, metaheuristics are approximation algorithms that provide good or acceptable solutions within an acceptable computing time, that cannot be obtained with more specialized techniques, such as brute-force, linear programming, dynamic programming, randomization, quantum computation, exact algorithms, etc., but do not give formal guarantees about the quality of the solutions \cite{chopard2018introduction}. The main difference between heuristic and metaheuristic is that a heuristic algorithm utilizes some specially designed functions to explore the solution space intelligently; whereas, a metaheuristic is an iterative generation process that directs a supporting heuristic to explore and exploit the search space efficiently \cite{desale2015heuristic}. Moreover, heuristics can be applied to a specific problem; but, metaheuristics are more generalized and can be employed in the same way to many different problems. The metaheuristics algorithms can be classified into three groups as discussed below.
\begin{enumerate}
\item \textit{Population-based metaheuristics:} This type of metaheuristic utilizes global exploration and local exploitation ability for searching in global search space to discover new promising solutions and to refine the already discovered solutions. In this study, these algorithms are divided into three classes: (i) Swarm-based, (ii) Nature-based, and (iii) Evolutionary Algorithms (EAs). Swarm-based metaheuristics (a.k.a. swarm intelligence) are inspired by the social collective behavior of the birds, ants, bees, etc., where each animal (artificial agents) interacts with each other to achieve a particular goal in the environment \cite{sherinov2018multi}. Ant colony optimization (ACO), particle swarm optimization (PSO), Artificial bee colony (ABC), etc. are some of the most popular swarm-based metaheuristics. The second category is nature-inspired optimization, such as gorilla troops optimizer (GTO), crow search algorithm (CrSA), reptile search algorithm (RSA), butterfly optimization algorithm (BOA), moth–flame optimization (MFO), biogeography-based optimization (BBO), intelligent water drop (IWD), etc. Particularly, these algorithms mimic the successful characteristics of complex natural processes, including distinct animal behaviors, biological systems, natural calamities, etc. Though swarm intelligence also relies on nature, it specifically focuses on decentralized systems and collaborative behaviors; whereas, nature-inspired metaheuristics encompass diverse elements of nature, including the behavior of an individual animal. Another well-established population-based optimizations are evolutionary algorithms, which are based on the process of natural evolution like survival, reproduction, and mutation. Specifically, there are three important components of an EA: parent selection, variation operators (recombination/crossover and mutation), and replacement (evolution). Genetic algorithms (GAs), evolutionary programming (EP), and differential evolution (DE) are the most popular EAs in the literature. 

\item  \textit{Iterative-based metaheuristics:} The second major group of metaheuristics is iterative-based. These algorithms are inspired by the laws of physics, mathematics, chemistry, or social human behavior. Particularly, physics-based metaheuristics are based on the concepts of physical laws and principles, for example, classical mechanics, thermodynamics, optics, etc. Gravitational search algorithm (GSA), simulated annealing (SA), multi-verse optimizer (MVO), etc. are the most popular physics-based optimization techniques. Similarly, math-based metaheuristics adopt mathematical concepts like number theory, geometry, and algebra, along with modern mathematics. Arithmetic optimization algorithm (AOA) \cite{abualigah2021arithmetic} and sine cosine algorithm (SCA) \cite{mirjalili2016sca} are prominent techniques in literature that are based on the arithmetic operators and sine/cosine mathematical functions, respectively. Other interesting iterative-based metaheuristics are inspired by human interaction, intelligence, learning processes, and experiences. Some of these algorithms are teaching-learning-based optimization (TLBO), human-guided search (HGS) \cite{klau2010human}, and harmony search (HS). The rest algorithms in this category are search-based, for instance, local search (LS), tabu search (TS), neighborhood search (NS), etc. 

\item \textit{Hybrid metaheuristics:} Hybrid algorithms are integrated with different metaheuristics to utilize the advantages of distinct techniques for solving optimization problems. Algorithms in hybrid metaheuristics can focus on solving different problems simultaneously. For example, in hybridization with LS, global search is utilized to explore the search space, whereas LS is used to refine the areas of possible global optimum. On the other hand, the sub-metaheuristics within a hybrid approach can concentrate on optimizing different parts of the same problem, like a combination of PSO and GA, where PSO finds the optimal parameters used in GA.     

\end{enumerate}

 \subsection{Machine learning techniques}
 The relevant machine learning models are classified here. \textit{Supervised Learning (SL)} is the model that is trained with labeled data (a set of inputs and correct outputs) to learn the corresponding features, followed by an execution engine to predict using the test data. Supervised learning is used when the target has a similar pattern to the trained data. Different classification techniques like decision tree (DT), random forest (RF), k-nearest neighbor (KNN), etc., collaborating of multiple classifiers a.k.a ensemble learning (EL), along with artificial neural networks (ANN) fall into this category. In \textit{Unsupervised Learning (USL)} the desired outputs are not provided in the training phase. The main aim is to learn the similarity of the unlabeled data and further classify them into multiple groups. Some of the eminent algorithms are principal component analysis (PCA), and clustering techniques, such as k-means, probabilistic, and hierarchical clustering. \textit{Semi-supervised Learning (SSL)} encompasses the mechanism of both supervised and unsupervised learning. Specifically, it utilizes a combination of unlabeled and labeled inputs. The purpose of this is to make better predictions in discovered patterns. \textit{Reinforcement Learning (RL)} adopts the human learning process, especially learning from experiences. Particularly, it continuously optimizes feedback through actions after interaction with the environment. \textit{Deep Learning (DL)} models originate from the concept of information processing and distribution in the human brain. In brief, these types of architectures can be categorized into generative (unsupervised), discriminative (supervised), and deep RL architectures for IoT-IDSs.

 \section{Related Work\label{sec:related}}
  The systematic literature review (SLR) in \cite{hu2024deep} covers only a few population-based metaheuristics for intrusion detection in the IoT environment. Additionally, physical law-based, human-inspired, and hybrid optimization techniques are not discussed. Moreover, the authors do not concentrate solely on IoT. Rather, wireless, public networks, computer networks, Hadoop and MapReduce, and edge networks are also explored significantly. Furthermore, though they analyze the datasets used in developing the IDSs, no correlation is discovered among the metaheuristics, mostly used datasets and ML methods. Saadouni et al. \cite{saadouni2024intrusion} presents an SLR for IoT-IDS based on bio-inspired and ML-driven techniques. The authors vividly discuss the integration of ML methods with optimization algorithms. However, one of the most notable limitations of this work is the investigation of only 25 papers, whereas there are several well-established metaheuristics-assisted IDSs dedicated to IoT. Importantly, although the authors claim to study only IoT-based papers, we find that most of the articles are not focused on the IoT environment. In \cite{reddy2024systematic}, the authors study population-based optimizations, specifically swarm intelligence devised for detecting intrusions in IoT. Besides, they analyze the datasets used and the performances of the existing systems. The main drawback of the SLR is the lack of covering all categories of metaheuristics-driven IDSs. Moreover, only basic ML algorithms are discussed in the SLR; whereas many crucial deep learning-based IDSs are sorted out in our study. Sharma et al. \cite{sharma2024multi} aims to explore IoT-IDSs that rely on only multi-objective metaheuristics algorithms. Apart from this, they analyze different machine-learning models and popular datasets. However, the relation between these diverse techniques is still missing. Moreover, the study is not systematic. Heidari et al. \cite{heidari2023internet} introduces an SLR containing rigid comparison and exploration of different IDSs in the IoT environment. The first and foremost limitation is the missing metaheuristics and ML-based systems, which we aim to cover in our study. Verma et al. \cite{verma2020machine} propose a survey on ML-driven IDSs for IoT applications. Regrettably, the study does not include any optimization algorithms, rather it intends to analyze the machine learning classifiers commonly utilized in intrusion detection. Hajiheida et al. \cite{hajiheidari2019intrusion} also do not study the metaheuristics algorithms that are extensively employed in the IoT-IDS. Importantly, rather than focusing on the metaheuristics or ML-based systems, they categorize and discuss different systems, such as SIDS, AIDS, SpIDS, and HyIDS.

 The existing reviews either explore nature-inspired or ML-based IoT-IDS systems. Though there is only one survey \cite{sharma2024multi} that discusses the integration of metaheuristics and machine learning techniques for the IoT environment, the coverage and analysis are too inadequate regarding the volume, significance, and diversity of the related IDSs in literature. Most importantly, no reviews analyze the correlation between the optimization algorithms and machine learning models while experimenting on a specific dataset. Apart from these, the existing works do not categorize the applications of different metaheuristics regarding intrusion detection in the IoT environment. 

 To address all these issues, a systematic literature review is presented, which extensively explores the existing metaheuristics and ML-integrated IDSs, specific to the IoT environment. Additionally, we analyze these systems based on different applications of metaheuristics algorithms, such as optimal feature selection, parameter tuning, hyperparameter tuning, etc. Moreover, the discovery of the connections among these algorithms, their outstanding performances, and the used datasets significantly distinguish this review from others. Furthermore, a new large-scale visualized taxonomy is demonstrated to provide researchers with an overview of the existing IoT-IDSs. Table~\ref{tab:research_comp} provides an in-depth comparison with the state-of-the-art reviews, highlighting the contributions of our work over others. 
 
 \begin{table*}[!htb]
    \begin{adjustwidth}{-1cm}{}
    \caption{Comparison with existing state-of-the-art surveys.}
    \def\arraystretch{1.2}
    \resizebox{1.11\textwidth}{!}{
     \begin{threeparttable}
      \begin{tabular}{l|c|c|c|c|c|c|c|c|c|c|c|c|c|c|c}
        \hline
        \textbf{Work} & \textbf{Year} & \textbf{SLR} & \multirow{2}{1.5cm}{\textbf{\# of articles}} & \multirow{2}{1.25cm}{\textbf{IoT-specific}} & \multicolumn{4}{c|}{\textbf{Metaheuristics}} & \multicolumn{3}{c|}{\textbf{Machine Learning Models}} & \multirow{2}{1.65cm}{\textbf{Usage of Datasets}} & \multirow{2}{1.75cm}{\textbf{Perfor. Analysis}} & \multirow{2}{2.15cm}{\textbf{Correlation Analysis}} & \multirow{2}{2.15cm}{\textbf{Application Analysis}}\\\cline{6-12}
        & & & & & \textbf{EA} & \textbf{SI} & \textbf{PhA} & \textbf{Others} & \textbf{Basic ML} & \textbf{DL} & \textbf{Hybrid} & & & & \\\hline
        \cite{hu2024deep} & 2024 & \checkmark & 145 & \xmark & \xmark & \checkmark & \xmark & \xmark & \xmark & \xmark & \xmark & \xmark & \checkmark & \xmark & \xmark\\
        \cite{saadouni2024intrusion} & 2024 & \checkmark & 25 & \xmark & \checkmark & \checkmark & \xmark & \xmark & \checkmark & \checkmark & \xmark & \xmark & \checkmark & \xmark & \xmark\\
        \cite{reddy2024systematic} & 2024 & \checkmark & 101 & \xmark & \xmark & \checkmark & \xmark & \xmark & \checkmark & \xmark & \xmark & \checkmark & \checkmark & \xmark & \xmark\\
        \cite{sharma2024multi} & 2024 & \xmark & 37 & \checkmark & \checkmark & \checkmark & \xmark & \checkmark & \checkmark & \checkmark &  \xmark & \xmark & \xmark & \xmark & \xmark\\
        \cite{heidari2023internet} & 2023 & \checkmark & 24 & \checkmark & \xmark & \xmark & \xmark & \xmark & \xmark & \xmark & \xmark & \xmark & \checkmark & \xmark & \xmark\\
        \cite{verma2020machine} & 2020 & \xmark & 25* & \checkmark & \xmark & \xmark & \xmark & \xmark & \checkmark & \xmark & \xmark & \xmark & \checkmark & \xmark & \xmark\\
        \cite{hajiheidari2019intrusion} & 2019 & \checkmark & 43 & \checkmark & \xmark & \xmark & \xmark & \xmark & \checkmark & \checkmark & \xmark & \xmark & \checkmark & \xmark & \xmark\\
        \textbf{Ours} & 2025 & \checkmark & 111 & \checkmark & \checkmark & \checkmark & \checkmark & \checkmark & \checkmark & \checkmark &  \checkmark & \checkmark & \checkmark & \checkmark & \checkmark\\\hline
      \end{tabular}
      \begin{tablenotes}
            *Not explicitly mentioned in the paper.
        \end{tablenotes}
        \end{threeparttable}
        }
      \label{tab:research_comp}
      \end{adjustwidth}
    \end{table*}

\section{Review methodology\label{sec:methodology}}
This section presents the review methodology used in this paper. 
\subsection{ Objectives and Research Questions}
This research aims to investigate the existing integration of metaheuristics and ML algorithms to detect intrusions in the IoT ecosystem. Besides, uncovering the hidden correlations among the top-performing optimization techniques and ML models, considering specific datasets is another goal of similar importance. To achieve these objectives, firstly, it requires understanding the necessity and exploring different applications of the metaheuristics-assisted ML architectures for developing IoT-IDSs. Secondly, an investigation is needed on the most popular datasets and evaluation metrics utilized to measure these detection systems. The next plan is to discover which optimization techniques and ML models come up with excellent performances, considering the tested datasets. After that, we plan to analyze and sort out the challenges and issues that arise because of the collaboration of these different conceptual techniques. Finally, the unexplored areas of metaheuristics and machine learning techniques need to be studied to facilitate future research. Table~\ref{tab:rqs} outlines the specific research questions identified to achieve the objectives mentioned above.
\\
\begin{table*}[!htb]
    \centering
    \caption{Research questions and objectives.}
    \def\arraystretch{1.15}
    \resizebox{1\textwidth}{!}{
    \begin{tabular}{p{0.5cm}cc}
        \hline \textbf{RQ\#} & \textbf{Research Questions}  & \textbf{Objectives}\\\hline
        \multirow{3}{0.5cm}{RQ1} & \multirow{3}{8.1cm}{What are the need and existing applications of metaheuristics optimizations in developing ML-based IDS, especially for IoT?} 
		& \multirow{3}{8.1cm}{To understand the necessity and explore the existing metaheuristics and ML algorithms, incorporated for developing IoT-IDSs.}\\
        & & \\
        & & \\
        \multirow{3}{0.5cm}{RQ2} & \multirow{3}{8.1cm}{What are the most commonly used datasets and evaluation metrics for integrated IoT-IDS assessment?} 
		& \multirow{3}{8.1cm}{To identify the well-known datasets and their coverages, and define the popular performance metrics.}\\
         & & \\
        & & \\
        \multirow{3}{0.5cm}{RQ3} & \multirow{3}{8.1cm}{What are the relations between the optimization algorithms and classification methods with the datasets?} 
		& \multirow{3}{8.1cm}{To find out the best-performing IoT-IDSs, analyze the metaheuristics and ML algorithms they used, and synthesize the outcomes.}\\
         & & \\
        & & \\
        \multirow{3}{0.5cm}{RQ4} & \multirow{2}{8.1cm}{What are open issues raised by the integration of metaheuristics algorithms with ML?} 
		& \multirow{3}{8.1cm}{To discuss the unavoidable challenges as well as possible solutions while combining machine learning with metaheuristics.}\\
         & & \\
          & & \\
        \multirow{3}{0.5cm}{RQ5} & \multirow{2}{8.1cm}{{What are the unexplored metaheuristics optimization algorithms for IDS in IoT?}} 
		& \multirow{2}{8.1cm}{To facilitate future research in this field, the undiscovered techniques need to be mentioned.}\\
        & & \\\hline
    \end{tabular}
    }
    \label{tab:rqs}
\end{table*}

\subsection{Search strategy}
In this review, we search relevant works in established and well-known online sources, such as IEEE Xplore Digital Library, ACM Digital Library, Elsevier, Springer, Nature, Wiley, Taylor \& Francis, MDPI, and World Scientific. Initially, 765 articles are selected in total. For searching,  specific keywords related to IoT-IDS which incorporate metaheuristics and ML have been used (See Table \ref{tab:keywords}). Since we aim to cover the most recent techniques, we search and select only the relevant, effective, and scholarly papers that were published between 2020 and 2024. Importantly, this focused and extensive investigation differentiates our work from previous SLRs or surveys in the same field. Next, the papers that are deemed irrelevant and found to be duplicates are removed. Particularly, several research has been conducted on the metaheuristics and ML-driven detection systems that incorporate blockchain, fog computing, and supercomputing technologies, which are not relevant to this review. Moreover, several works focus on specific attack detection, for example, botnet, ransomware, malware, denial of service (DoS), etc. These articles are also removed from our dataset. Furthermore, intrusion detection systems that are not IoT-specific, are also eliminated since our concentration is solely on the IoT-IDSs. All of these keywords for exclusion are listed in Table \ref{tab:excluded}. Apart from this, the related surveys are not included in the dataset. Additionally, the papers published in foreign languages are excluded from this study. Finally, 111 articles are chosen for this literature review. \\
\begin{table*}[htbp]
\centering
\def\arraystretch{1}
\resizebox{0.75\textwidth}{!}{
\begin{tabular}{cc}
\hline
{\textbf{Key}} & \textbf{Criteria}  \\
\hline
Search string &   \multirow{3}{11cm}{(Metaheuristics-based) AND (Machine learning OR ML-based) AND (Intrusion Detection System OR IDS) AND (Internet of Things OR IoT)} \\
 &  \\
  &  \\
Limiters &     Article date between 2020 and 2024\\        
Search modes &  Search words occur either in the title, abstract, or in the introduction of the article\\\hline
\end{tabular}
}
\caption{Search criteria.}
\label{tab:keywords}
\end{table*}
\\
\begin{table*}[htbp]
\centering
\def\arraystretch{1}
\resizebox{0.65\textwidth}{!}{
\begin{tabular}{ccc}
\hline
\multicolumn{3}{c}{\textbf{Excluded keywords}} \\\hline
{Botnet detection} &   {Malware detection}   &   {Ransomware detection}    \\
 Data mining  & Blockchain       &    Remote Sensing Images      \\
Wireless Sensor Networks (in general) &   {DoS/DDoS detection}  &         Fog Computing        \\
Supercomputing     & Selective forwarding attack & \\
\hline               
\end{tabular}
}
\caption{Exclusion keywords.}
\label{tab:excluded}
\end{table*}

\subsection{Data analysis}
\begin{figure*}[!htb]
  \center
\includegraphics[width=5in, height=3.7in] {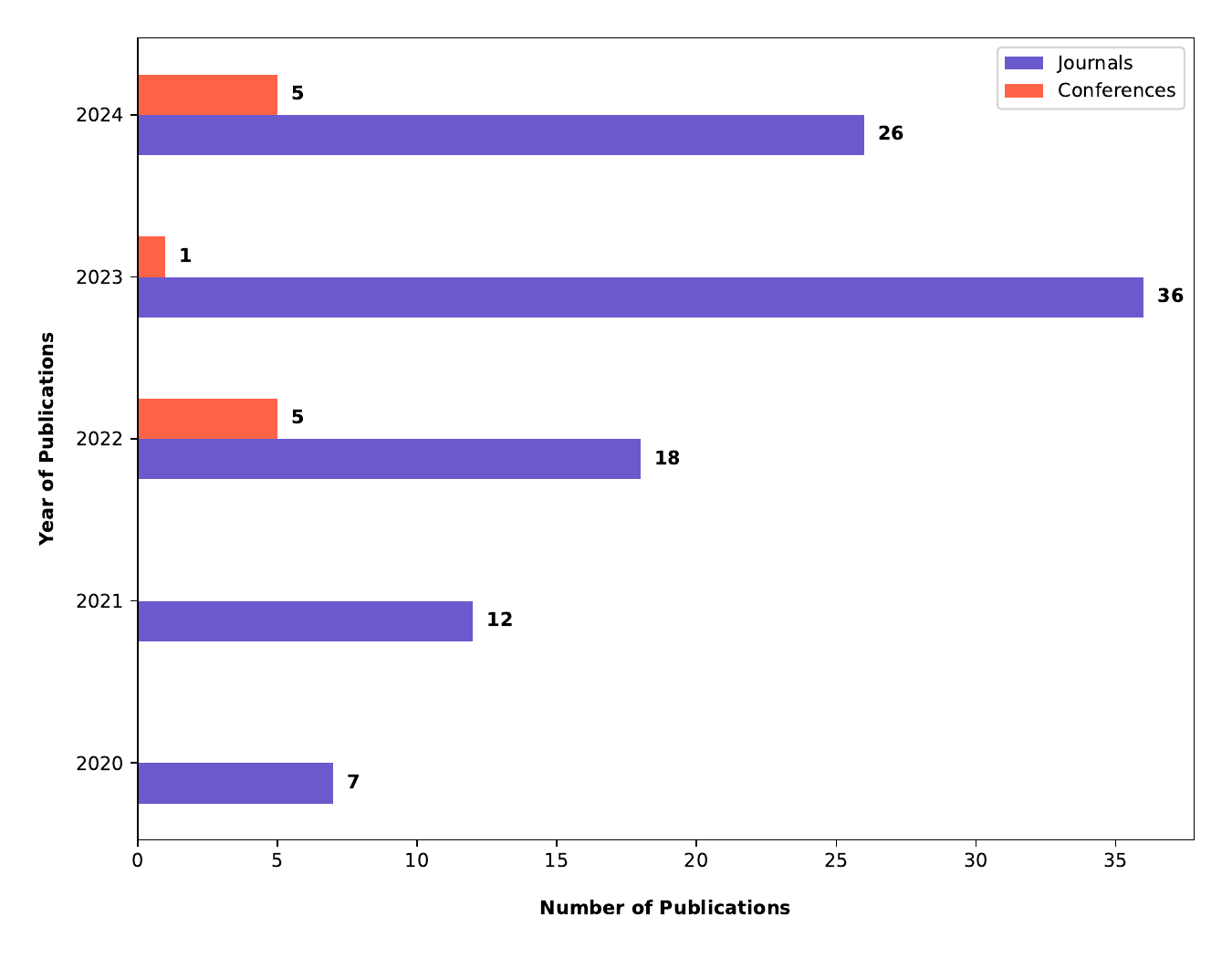}
   \caption{Year-wise publications for related IoT-IDSs, are included in this literature review.}
   \label{fig:year_wise_pub}  
\end{figure*}

After applying the search strategy, a dataset consisting of 111 papers are collected, containing both conference and journal works on metaheuristics and ML-based intrusion detection systems for IoT. Among these, 90\% are journal papers, and the rest are conference papers. According to Figure~\ref{fig:year_wise_pub}, in recent times, researchers have tended to give more attention to metaheuristics-assisted machine learning while developing IoT-IDS.  It is also discovered that the highest number of quality articles was published in 2023 with 37 papers, followed by 31 papers in 2024, and 24 papers in 2022. Before 2021, metaheuristics were not studied in such a focused way as depicted in the Figure.

\subsection{Investigation of Journal Papers}
\begin{figure}[!htb]
  \center
\includegraphics[scale=0.6] {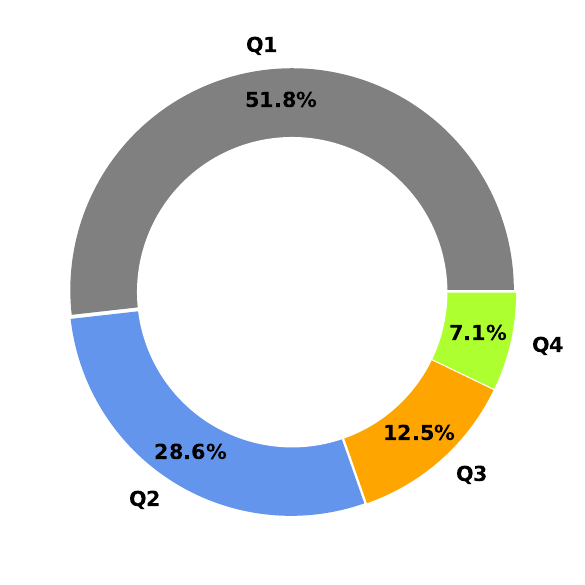}
   \caption{Percentile of different quartile journals included in this literature review [Q1=29, Q2=16, Q3=7, Q4=4].}
   \label{fig:journal_quartile_donut}  
\end{figure}

We further analyze the 100 journal papers. Specifically, they are collected from 56 journals in total, where almost half of them are ranked as Q1 journals. Simultaneously, the amount of Q2 and Q3 journals is also significant, which indicates the inclusion of high-quality and well-established research in this literature review. A donut plot is drawn in Figure~\ref{fig:journal_quartile_donut} to illustrate the percentage of different quartile journals studied in this work.

\section{Results of the review}\label{sec:results}
This section provides the results of the review answering each research question. 
\subsection{\textit{\textbf{RQ1: What are the need and existing applications of metaheuristics optimization in developing ML and DL-based IDS, especially for IoT? }}}
Though machine learning algorithms offer high accuracy, enhanced security, and better performance, they require excessive computational power like high-performance GPUs (Graphics Processing Units), and a large volume of storage for generating, executing, and managing large datasets during training and testing phases. Undoubtedly, deep learning algorithms can seamlessly handle enormous datasets and offer fast processing; but, these require significant time to achieve minor precision improvements. Moreover, parameter-tuning is another unavoidable critical issue since adjusting the number of layers with the expected accuracy is entirely correlated \cite{bharati2022machine}. In this regard, metaheuristics can be utilized to obtain near-optimal solutions in a shorter time than the exact algorithms. Specifically, a metaheuristic algorithm is an iterative generation process that guides a subordinate heuristic to explore and exploit the search space efficiently \cite{desale2015heuristic}. Besides, these algorithms terminate when specific conditions (e.g., the number of iterations, elapsed time, etc.) are satisfied. As a result, there is no possibility of running the algorithm for a long period; even the likelihood of being stuck in the local minima is negligible.

\par
\begin{figure*}
    \centering
    \includegraphics[width=1\linewidth]{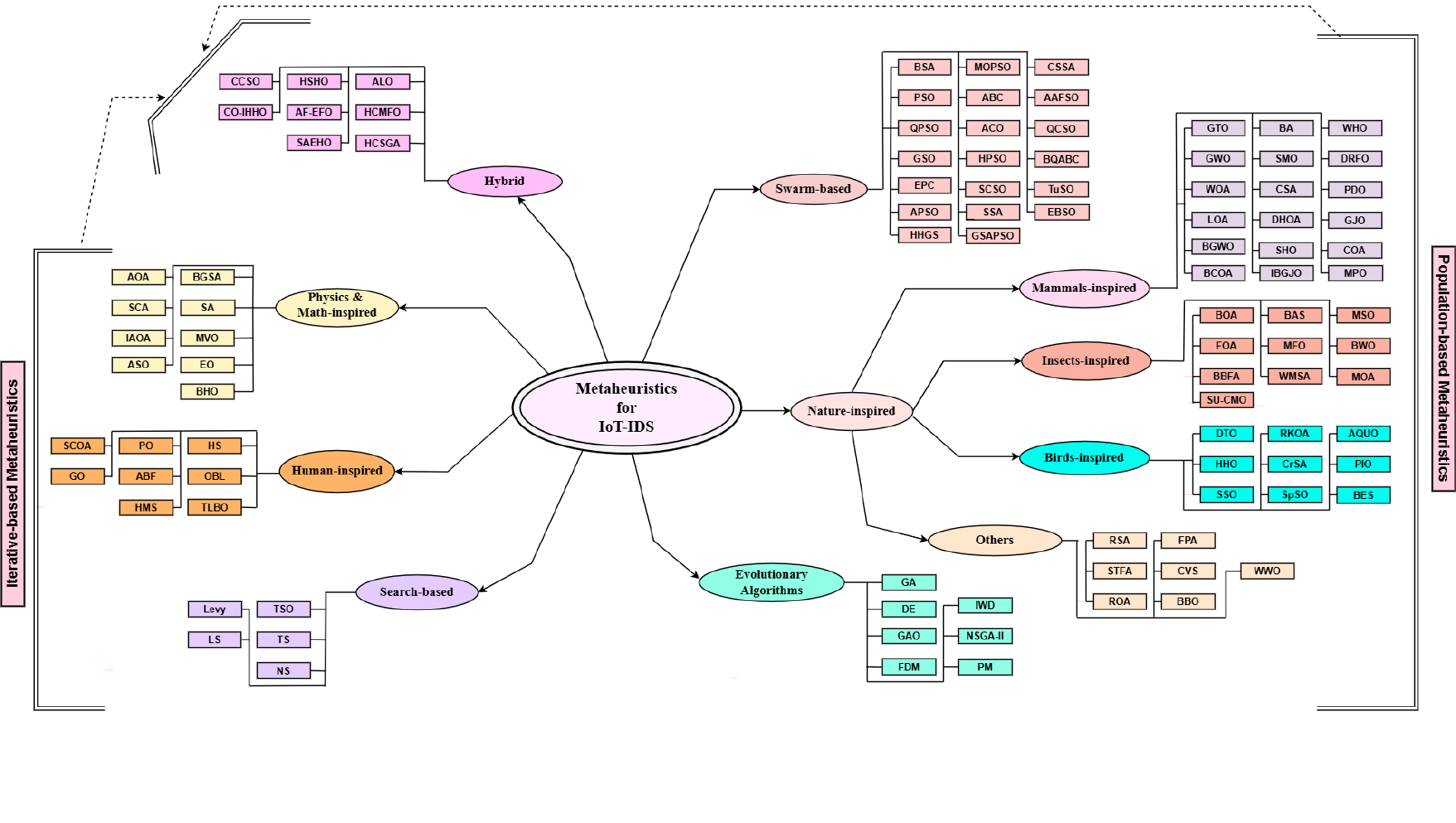}
    \caption{Taxonomy of the existing metaheuristics and ML-integrated IoT-IDSs.}
    \label{fig:taxonomy}
\end{figure*}

We discuss the existing integrated IoT-IDSs categorized into different metaheuristics algorithms. Besides, these optimization techniques are also analyzed by classifying them into diverse applications, such as future selection, parameter tuning, hyperparameter optimization, and hybrid applications. In this work, swarm intelligence techniques are differentiated from nature-based optimizations considering the vastness and significance of these two distinct categories of metaheuristics. Though swarm-based algorithms are also based on nature, they especially concentrate on decentralized systems and collaborative behaviors; contrarily, nature-inspired metaheuristics encompass diverse elements of nature, including the discrete behaviors of individual animals. The taxonomy of the existing metaheuristics and ML-integrated IoT-IDSs is illustrated in Figure~\ref{fig:taxonomy}.

\subsubsection{Population-based Metaheuristics and ML} .\\ \\
\textbf{Swarm-based Metaheuristics.} The well-known particle swarm optimization (PSO) technique is widely used in IoT-IDSs. Particularly, Elmasry et al. \cite{17_elmasry2020evolving} leverage a double PSO to optimize features and hyperparameters. The efficiency of this metaheuristic is examined by discriminative and generative ML methods, especially CNN, LSTM, and DBN. Among these, DBN excels over others by 2\% to 6\% while testing on the NSL-KDD and CICIDS2017 datasets. In another paper, Saheed et al. \cite{74_saheed2023novel} integrate an AE with PSO to optimally select features from the BoT-IoT and UNSW-NB15 datasets. After that, they also modify the inertia weight of PSO to optimize the hyperparameters of DNN, resulting in the efficient classification of attacks (an accuracy of 97.61\% and 94.62\% in BoT-IoT and UNSW-NB15, respectively). However, the combination of genetic self-adjusted PSO (GSAPSO) and EGB and KNN classifiers does not result in considerable performances~\cite{111_dakic2024intrusion}.

Apart from these, bird swarm optimization (BSA), salp swarm algorithm (SSA), and golden jackal optimization algorithm (GJOA) are found to be reliable for detecting intrusions in IoT environments. In particular, a DBN-driven IoT-IDS is presented in \cite{90_biju2024evaluated}, which offers 98.96\% accuracy, 99.4\% precision, and 98.87\% recall on the NSL-KDD dataset when parameters are tuned by evaluated BSA (EBSA). Aljehane et al. \cite{94_aljehane2024golden} leverage GJOA to find the most significant features from the CICIDS-2017 dataset and SSA to optimize hyperparameters of the attention-driven bi-directional LSTM (A-BiLSTM). This dual usage of population-based optimizers secures 99.69\% accuracy, 98.92\% f1-score, and 98.74\% MCC. Another variant of SSA, referred to as chaotic salp swarm optimization (CSSA) performs well with the incorporation of LightGBM, having an accuracy of 98.35\%$\sim$99.38\% on the MC-IoT, MQTT-IoT-IDS2020, and MQTTset datasets~\cite{57_prajisha2022efficient}. However, adopting BSA and social group optimization algorithm (SCOA) for selecting features and optimizing parameters in kernel extreme machine learning model (KELM) fails to deliver decent results across all metrics (99.45\% accuracy, 80.26\% precision, 82.67\% recall, and 80.95\% f1-score).  

Interestingly, some detection systems demonstrate different performances in various datasets. For example, IDS with artificial bee colony (ABC) and extreme machine learning classifier performs well on CICIDS-2017; whereas significantly fails on UNSW-NB15 (accuracy 98.71\% vs. 71.54\% )~\cite{34_stankovic2022feature}. Similar results appear in swarm-inspired sand cat swarm optimizer (SCSO) and ELM-based system~\cite{50_jovanovic2022feature}. However, integration of glow-swarm optimization (GSO) with PCA~\cite{22_anusha2022intrusion}, chimp chicken swarm-based optimization (CCSO) with deep LSTM~\cite{101_deore2022hybrid}, and improved ACO with ensemble classifier (DDT, ANFIS and MDSVM)~\cite{39_vanitha2023improved} does not result in considerable performance. 
\\ \\
\textbf{Nature-inspired Metaheuristics.} These optimization techniques are extensively integrated with machine learning-based IoT-IDSs, especially Grey-Wolf Optimization (GWO)~\cite{56_xu2023iot, 89_elsedimy2024novel, 91_saheed2024voting, 98_moizuddin2022bio, 2_rm2020effective, 61_davahli2020lightweight}, firefly optimization algorithm (FOA)~\cite{8_savanovic2023intrusion, 60_gangula2023intrusion, 87_harbi2024bio}, Capuchin Search Algorithm (CSA)~\cite{64_om2022effective, 73_abd2023intrusion}, and Whale Optimization Algorithm (WOA)~\cite{72_ramana2022wogru, 75_jothi2023wils, 79_mohy2023whale, 80_bajpai2024hybrid}. Regarding the GWO-based detection systems,~\cite{56_xu2023iot, 98_moizuddin2022bio, 2_rm2020effective, 61_davahli2020lightweight} utilize GWO for feature selection; whereas~\cite{89_elsedimy2024novel, 91_saheed2024voting} use for optimizing hyperparameters of quantum-based SVM classifier and EL method (comprised of DT, RF, KNN, and MLP), respectively. The former systems identify intrusions relying on XGBoost, elastic regularization-assisted contractive autoencoder (CAE), deep neural network (DNN), and SVM, respectively. Notably, all these IDSs come up with remarkable efficiency (e.g., 99\%$\sim$100\% accuracy). However, different datasets, preprocessing and other related techniques are employed within these IoT-IDSs. In~\cite{60_gangula2023intrusion, 87_harbi2024bio}, FOA is integrated for selecting near-optimal features, and detection is conducted by classifiers, especially ensemble, and DT, respectively. Though~\cite{60_gangula2023intrusion} shows notable detection ability,~\cite{87_harbi2024bio} severely under-performs. In contrast, Savanovic et al. \cite{8_savanovic2023intrusion} improve the original FOA for tuning hyperparameters in classification techniques like KNN, and XGBoost. As a consequence, their IoT-IDS provides an accuracy of 99.98\% and 99.6997\%, respectively on UNSW-NB15 and IoT healthcare datasets. Turing to CSA-based detection systems, Kumar et al.~\cite{64_om2022effective} optimize parameters of a capsule autoencoder (HKCAE) and Elaziz et al.~\cite{73_abd2023intrusion} select near-optimal features for CNN model using this metaheuristic. Importantly, in this case, parameter tuning turns to be more effective than feature selection since the former IDS offers outstanding accuracy on BoT-IoT and UNSW-NB15 datasets (99.9\% and 99.7\%); whereas the later system behaves inconsistently across various datasets (considerable for BoT-IoT, KDDCup-99, and CICIDS-2017; severe for NSL-KDD). Among the four WOA-based IDSs, ~\cite{72_ramana2022wogru} is the best-performing system with approximately 99.8\% accuracy, precision, recall, f1-score, and specificity. The authors leverage this optimization technique to tune the hyperparameters of a gated recurrent unit (GRU). However, optimizing parameters of an LSTM using WOA also turns out to be effective (99.1\%$\sim$99.5\% accuracy)~\cite{75_jothi2023wils}.

Considering other IOT-IDSs of this category, most of the nature-inspired metaheuristics are leveraged for feature selection. Further examining of these systems reveal that various types of optimizers, such as Moth–Flame Optimization (MFO)~\cite{66_gadekallu2023moth}, crow search algorithm (CrSA)~\cite{46_jayalatchumy2024improved}, chaotic vortex search (CVS)~\cite{51_geetha2024cvs}, decisive red fox optimization (DRFO)~\cite{88_rabie2024novel}, multi-objective prairie dog optimization (PDO~\cite{92_sharmamulti}, reptile search algorithm (RSA)~\cite{16_dahou2022intrusion}, binary multi-objective Capuchin search algorithm (BMECapSA)~\cite{42_asgharzadeh2023anomaly}, Aquila optimizer (AQUO)~\cite{49_fatani2021advanced}, BA~\cite{95_bella2024intrusion}, and Mayfly Optimization Algorithm (MOA)~\cite{96_vadakkethil2024mayfly} are prominent. Among them,~\cite{66_gadekallu2023moth, 46_jayalatchumy2024improved, 51_geetha2024cvs, 88_rabie2024novel, 92_sharmamulti, 42_asgharzadeh2023anomaly} demonstrate higher efficiency while experimenting on diverse datasets. However, no common trend is found in terms of machine learning-based classifiers.

Other than these systems, a few IDSs utilize nature-centric optimization techniques for tuning parameters~\cite{93_cherian2024iot, 53_qaddoura2024evolving, 82_kethineni2024intrusion} or hyperparameters~\cite{85_vijayan2024original, 114_alamro2023modeling} of the detection models; sometimes for hybrid applications~\cite{78_gopalakrishnan2022new, 100_pingale2023remora, 106_escorcia2023sea}. Although in discrete cases, these systems show considerable results, their performances are not generally satisfactory.
\\ \\
\textbf{Evolutionary Algorithms.} Only a small number of IoT-IDSs have utilized evolutionary algorithms (EAs) and they do not significantly surpass other detection systems. The best-performing IDS of this type is proposed by Latif et al.~\cite{112_latif2024dtl}, where hyperparameters of a CNN-based ensemble classifier are optimized by a genetic algorithm (GA). Ultimately, this system achieves 100\% accuracy, precision, recall, f1-score, and Cohen's kappa score in the Edge\_IIoTset dataset. Second best-performing IDSs presented in \cite{12_dey2023hybrid}, where feature selection is accomplished by a non-dominated sorting genetic algorithm (NSGA) and classification is done using a support vector machine (SVM). Importantly, the system achieves a remarkable accuracy of 99.48\% on the TON\_IoT dataset. Gupta et al. \cite{104_10317882} integrate an evolutionary algorithm intelligent water drop (IWD) and a nature-inspired biogeography-based optimization (BBO) technique with a feed-forward neural network (FNN). The IDS detects attacks more correctly when tested on CICIDS-2017 compared to the IoTID20 dataset (accuracy 98.2339\% vs. 96.7414\% and f1-score 99.0865\% vs. 95.4901\%). However, integration of an assimilated artificial fish swarm optimization (AAFSO) with genetic algorithm (GA)-tuned faster recurrent CNN (FRCNN) does not perform well across diverse datasets~\cite{65_anushiya2023new}.
\\ \\
\textbf{Population-based Hybrid Metaheuristics.} Numerous works have employed more than one population-based metaheuristics to improve the performance of IoT-IDSs. In most cases, such hybridization is utilized to select an optimal set of features. Regarding traditional machine learning classifiers, KNN~\cite{55_abu2021iot, 84_mahanipour2024enhancing, 1_kareem2022effective, 3_alkanhel2023network} and RF~\cite{13_krishna2021attack, 38_gaber2023industrial, 41_ethala2022hybrid, 33_phalguna2021hybrid, 44_keserwani2021smart} are widely used in these systems. Specifically, SSA+ALO~\cite{55_abu2021iot}, quantum-driven binary ABC+GA~\cite{84_mahanipour2024enhancing}, gorilla troops optimizer (GTO)+birds swarm algorithm (BSA)~\cite{1_kareem2022effective}, and GWO+dipper throated optimization (DTO)~\cite{3_alkanhel2023network, 27_alkanhel2023hybrid} are integrated in these KNN-based detection systems. Notably, all these IDSs provide substantial performance in terms of accuracy, precision, recall, and other related metrics. On the other hand, RF-based IDSs utilize LOA+FOA~\cite{13_krishna2021attack}, PSO+bat algorithm (BA)~\cite{38_gaber2023industrial}, spider monkey algorithm+hierarchical PSO~\cite{41_ethala2022hybrid}, and PSO+GWO~\cite{33_phalguna2021hybrid, 44_keserwani2021smart} for feature selection purpose. These systems also demonstrate significant efficiency in classifying various attacks. In addition to these, hunger game search (HGS) and remora optimization algorithm (ROA) in \cite{31_kumar2022intellectual}, and GA and GWO are hybridized in~\cite{30_davahli2020hybridizing} for selecting near-best features from AWID dataset for SVM classifier. Both of these combinations show notable performances with 99.1\% accuracy and negligible FPR. Turning to deep learning-oriented IoT-IDSs, three works are found that applied a Look Ahead Artificial Neural Network (LAANN), recurrent neural networks (RNNs), and a deep learning-based hybrid neural network (DLHCNN) respectively. Particularly, sea turtle foraging algorithm (STFA)+explorated PSO (EXPSO)~\cite{54_jeyaselvi2023highly}, Harris hawk optimization (HHO)+fractional derivative mutation (FDM)~\cite{7_sanju2023enhancing}, and Chicken Swarm Optimization (ChSO)+GA~\cite{67_gupta2022hybrid} are utilized in these systems to select optimal feature set. However, they fail to provide remarkable performance with 95\%$\sim$98\% considering all related metrics.

Apart from feature selection-based IDSs, a few research studies focus on tuning the parameters or hyperparameters of classification models. Khafaga et al.~\cite{4_khafaga2023voting} propose an innovative whale optimization (WOA) regulated by DTO to optimize parameters of KNN, RF, and NN. Experimental evaluation using the RPL-NIDS17 dataset results in 99\% AUC and 95.1\% accuracy. In~\cite{26_sagu2023design}, SAEHO and SU-CMO are also proposed for adjusting the parameters of the two hybrid classifiers, particularly CNN+DBN and Bi-LSTM+GRU. The integration of SAEHO and hybrid classifier offers better accuracy than that of SU-CMO and hybrid classifier (91.6\% vs 84.8\% ). Bahaa et al.~\cite{29_bahaa2022novel} integrate adaptive PSO and WOA for hyperparameter optimization in their CNN-based detection system. However, the system achieves only 94.54\% accuracy and 0.9 JSC. However, a small number of papers focus on employing hybrid population-based optimizers for accomplishing multiple purposes at a time~\cite{45_karthikeyan2024firefly, 40_alruwaili2023red, 86_aburasain2024enhanced}. One of the notable systems is introduced by Karthikeyan et al.~\cite{45_karthikeyan2024firefly}, where GWO is leveraged to optimize parameters and FOA to choose the most suitable features for the SVM classifier in the IoT-WSN environment. The separate use of these two metaheuristics crucially influences the system's accuracy (99.29\%). However, other such existing IoT-IDSs drastically fails, especially red kite optimization algorithm (RKOA)+Levy flight chaotic WOA with EL (LSTM, BiLSTM, and Bi-GRU)~\cite{40_alruwaili2023red} and black widow optimization (BWO)+BES with hybrid deep learning (HDL)~\cite{86_aburasain2024enhanced}. 

\subsubsection{Iterative-based Metaheuristics and ML}
\textbf{Physics and Math-based} Regarding math-inspired optimization techniques, the arithmetic optimization algorithm (AOA) is widely employed. In \cite{19_fraihat2023intrusion}, AOA is utilized to select optimal feature sets for random forest and extra trees classifiers. Experimental evaluations on four public datasets reveal that the IDS produces a much less false positive rate (0.002\%) for the tests conducted using the NF-ToN-IoT-v2 dataset. Makhadmeh et al. \cite{102_makhadmeh2024intrusion} also apply AOA for executing the same purpose using different classifiers, KNN. Interestingly, the accuracy of these two systems is almost identical (around 99.9\%). Though AOA and quantum-driven PSO (QPSO) are utilized in~\cite{11_malibari2022novel} for different purposes, the deep wavelet neural network (DWNN)-based detection system does not secure substantial accuracy (98.21\%). Turning to the physics-based IoT-IDSs, Atom Search Optimization (ASO) and Equilibrium Optimization (EO) techniques are utilized in \cite{110_maazalahi2024k} to select optimal features prior to applying K-means clustering. Importantly, this hybridization demonstrates remarkable performance on NSL-KDD, UNSW-NB15, and KDD-CUP99 datasets. In another paper~\cite{113_li2024cooperative}, a black hole optimization technique is employed to select near-optimal features from the UNSW-NB15 and NSL-KDD datasets. Two CNNs are utilized in parallel to identify intrusion and result in around 97.5\%$\sim$99.89\% efficiency in terms of common performance metrics.
\\ \\
\textbf{Human-inspired.} Considering the integration of metaheuristics adopted from human behaviors and their decision-making process,~\cite{20_alghamdi2022hybrid, 52_pan2021lightweight, 18_fatani2023enhancing, 115_otoum2021ids} IoT-IDSs are the most notable ones. Specifically, the political optimizer (PO) utilized for parameter tuning in a cascade forward neural network (CFNN) model, compact SCA (CSCA) for adjusting the parameters of the KNN classifier, and modified growth optimizer (GO) for selecting near-best features before training by CNN are significantly performed well with an accuracy of 99.86\%, 98.27\%$\sim$99.327\%, and 99.941\%), respectively. However, these groups of optimizers are not always effective, especially for intrusion detection in IoT environment~\cite{28_forestiero2021metaheuristic, 71_basheri2023quantum, 105_10541183}. 
\\ \\
\textbf{Search-based} Though IoT-IDSs of this category tend to bring significant performances, they are too rare in the literature. Only one of them is found~\cite{58_nazir2023novel}, the authors utilize tabu search with the idea of cellular automata to succeed in the features selection task. Consequently, it results in accuracy, precision, and FPR of 99.5\%, 97.92\%, and 0.004\%, respectively while classified using an ensemble learning method.

\subsubsection{Hybrid Metaheuristics and ML} In the context of IoT intrusion detection systems, existing hybrid optimization algorithms can be categorized into 6 small groups: swarm+physics-based~\cite{32_habib2020modified}, nature+physics-inspired~\cite{14_dey2023metaheuristic, 21_alweshah2022intrusion}, nature+math-inspired~\cite{24_rahmani2024improvement, 36_aljebreen2023binary}, nature+human-based~\cite{81_hanafi2024intrusion, 83_singh2024intrusion}, swarm+search-based~\cite{35_baniasadi2022novel}, and nature+search-inspired~\cite{62_alghanam2023improved, 63_anuradha2022intrusion, 77_om2021harmony}. Though utilizing physics-based techniques with nature-inspired ones (binary gravitational search (BGSA)+GWO~\cite{14_dey2023metaheuristic}, simulated annealing (SA)+shuffled shepherd optimization (SSO)~\cite{21_alweshah2022intrusion}) prove to be effective, integrating with swarm-based optimizers (multi-object PSO+Lévy flight~\cite{32_habib2020modified}) turns out to be inefficient. Regarding the third group, Rahmani et al.~\cite{24_rahmani2024improvement} employ grasshopper optimization (GAO) and AOA for parameters and hyperparameters tuning in a random neural network (RdNN), which generates an IDS having 99.56\% precision and 99.37\% detection rate. In another work~\cite{36_aljebreen2023binary}, a binary chimp optimization algorithm (BCOA) is integrated with the sine cosine algorithm (SCA) for securing the IoT-WSN network. Although the IDS generates high accuracy and specificity (99.63\% and 99.67\%, respectively), the f1-score is not convincing (94.52\%). The well-known human-based metaheuristic object-based learning (OBL) is hybridized with Harris Hawk Optimization (HHO) and Golden Jackal Optimization Algorithm (GJOA), respectively in~\cite{83_singh2024intrusion} and~\cite{81_hanafi2024intrusion} to select best possible features. These DT-based and LSTM-driven IoT-IDSs provide remarkable performances with 99.65\%$\sim$100\% and 98.93\% accuracy, respectively. Baniasadi et al.~\cite{35_baniasadi2022novel} adjust parameters of deep CNN (DCNN) utilizing neighborhood search (NS)-based PSO. They get a negligible amount of mean square error (0.00053\%) with 98.86\% accuracy and 95.32\% specificity on the UNSW-NB15 and Bot-IoT datasets. Turning to the last group of this category, it can be concluded that selecting features using a combination of nature-inspired and search-based metaheuristics is not effective according to the results provided in~\cite{62_alghanam2023improved, 63_anuradha2022intrusion, 77_om2021harmony}. It is worth mentioning that these IDSs use EL classifier, variational autoencoder (VAE), and deep RL, respectively.
 \par
Table~\ref{tab:meta_works} shows the list of works in the existing literature that leverage different metaheuristics-based techniques and ML algorithms. According to the table, most systems leverage nature-inspired optimization techniques to increase the efficiency of the classifiers. Besides, swarm-based, population-based hybrid, and hybrid metaheuristics are utilized significantly. Moreover, a thorough analysis of the IoT-IDSs based on different performance metrics, metaheuristics, their applications, ML algorithms, classification types, and datasets are presented in Appendix~A (Table~\ref{tab:performance_appendix}).
\begin{table*}[htbp]
\scriptsize
\caption{List of existing IoT-IDSs based on different metaheuristics and ML models.}
\resizebox{1\textwidth}{!}{
\begin{tabular}{p{2.5cm}p{1.95cm}p{1.75cm}p{1.75cm}p{0.75cm}p{1.75cm}p{0.75cm}p{2.5cm}}
\hline\multirow{3}{*}{\textbf{Metaheuristics}} & \multicolumn{7}{c}{\textbf{Machine Learning Models}}    \\\cline{2-8}
   & \multicolumn{3}{c}{\textbf{SL}}    & \multicolumn{1}{c}{\multirow{2}{*}{\textbf{USL}}} & \multicolumn{3}{c}{\textbf{DL}}\\\cline{2-4}\cline{6-8}
   & \multicolumn{1}{c}{\textbf{Classification}} & \multicolumn{1}{c}{\textbf{EL}}& \multicolumn{1}{c}{\textbf{ANN}} & \multicolumn{1}{c}{} & \multicolumn{1}{c}{\textbf{Generative}}   & \multicolumn{1}{c}{\textbf{DRL}} & \multicolumn{1}{c}{\textbf{Discriminative}}\\\hline
Swarm-based  & \cite{5_saif2022hiids, 23_alweshah2023intrusion, 48_sarwar2022design, 111_dakic2024intrusion} & \cite{39_vanitha2023improved, 57_prajisha2022efficient, 111_dakic2024intrusion} & \cite{9_vaiyapuri2023metaheuristics, 34_stankovic2022feature, 50_jovanovic2022feature, 53_qaddoura2024evolving}  & \cite{22_anusha2022intrusion, 69_kannan2020intrusion}   & \cite{17_elmasry2020evolving, 90_biju2024evaluated} & & \cite{17_elmasry2020evolving, 65_anushiya2023new, 71_basheri2023quantum, 74_saheed2023novel, 94_aljehane2024golden, 101_deore2022hybrid}  \\\hline
Nature-inspired& \cite{25_ghasemi2024new, 37_lv2023binary, 59_alamiedy2020anomaly, 68_sandhya2021enhancing, 87_harbi2024bio, 89_elsedimy2024novel, 92_sharmamulti} & \cite{8_savanovic2023intrusion, 46_jayalatchumy2024improved, 56_xu2023iot, 60_gangula2023intrusion, 66_gadekallu2023moth, 78_gopalakrishnan2022new, 85_vijayan2024original, 91_saheed2024voting} & \cite{6_li2022improving, 51_geetha2024cvs, 53_qaddoura2024evolving, 70_bathula2022designing, 79_mohy2023whale, 88_rabie2024novel, 104_10317882} & & \cite{63_anuradha2022intrusion, 64_om2022effective, 93_cherian2024iot, 98_moizuddin2022bio, 106_escorcia2023sea} & & \cite{2_rm2020effective, 16_dahou2022intrusion, 42_asgharzadeh2023anomaly, 43_jayasankar2024intrusion, 47_chander2023metaheuristic, 49_fatani2021advanced, 72_ramana2022wogru, 73_abd2023intrusion, 75_jothi2023wils, 80_bajpai2024hybrid, 82_kethineni2024intrusion, 94_aljehane2024golden, 95_bella2024intrusion, 96_vadakkethil2024mayfly, 99_kaviarasan2023network, 100_pingale2023remora, 107_10.1016/j.comcom.2023.03.005, 114_alamro2023modeling} \\\hline
EA-based& \cite{5_saif2022hiids, 12_dey2023hybrid}  && \cite{104_10317882, 112_latif2024dtl}  && && \cite{65_anushiya2023new, 113_li2024cooperative}    \\\hline
Population-based hybrid  & \cite{1_kareem2022effective, 3_alkanhel2023network, 4_khafaga2023voting, 13_krishna2021attack, 27_alkanhel2023hybrid, 30_davahli2020hybridizing, 31_kumar2022intellectual, 33_phalguna2021hybrid, 38_gaber2023industrial, 41_ethala2022hybrid, 44_keserwani2021smart, 45_karthikeyan2024firefly, 55_abu2021iot, 61_davahli2020lightweight, 84_mahanipour2024enhancing} && \cite{4_khafaga2023voting, 54_jeyaselvi2023highly} & & \cite{26_sagu2023design} & & \cite{7_sanju2023enhancing, 26_sagu2023design, 29_bahaa2022novel, 40_alruwaili2023red, 67_gupta2022hybrid, 76_shahapure2021water, 86_aburasain2024enhanced, 114_alamro2023modeling} \\\hline
Phy/Math-based & \cite{19_fraihat2023intrusion, 102_makhadmeh2024intrusion} &&    & \cite{110_maazalahi2024k}  & && \cite{11_malibari2022novel}  \\\hline
Human-inspired & \cite{52_pan2021lightweight, 97_10.1007/s11276-023-03435-0} && \cite{20_alghamdi2022hybrid, 53_qaddoura2024evolving}  && \cite{105_10541183}    && \cite{18_fatani2023enhancing, 28_forestiero2021metaheuristic, 71_basheri2023quantum, 105_10541183} \\\hline
Search-based &   & \cite{58_nazir2023novel} &    && &&  \\\hline
Hybrid  & \cite{21_alweshah2022intrusion, 32_habib2020modified, 83_singh2024intrusion} & \cite{14_dey2023metaheuristic, 36_aljebreen2023binary, 62_alghanam2023improved} & \cite{24_rahmani2024improvement}  && \cite{63_anuradha2022intrusion} & \cite{77_om2021harmony} & \cite{15_fatani2021iot, 35_baniasadi2022novel, 81_hanafi2024intrusion}  \\\hline
\end{tabular}
}
    \label{tab:meta_works}
\end{table*}

\subsection*{Analysis on Different Applications}
\begin{figure}[htbp]
  \center
\includegraphics[scale=0.6] {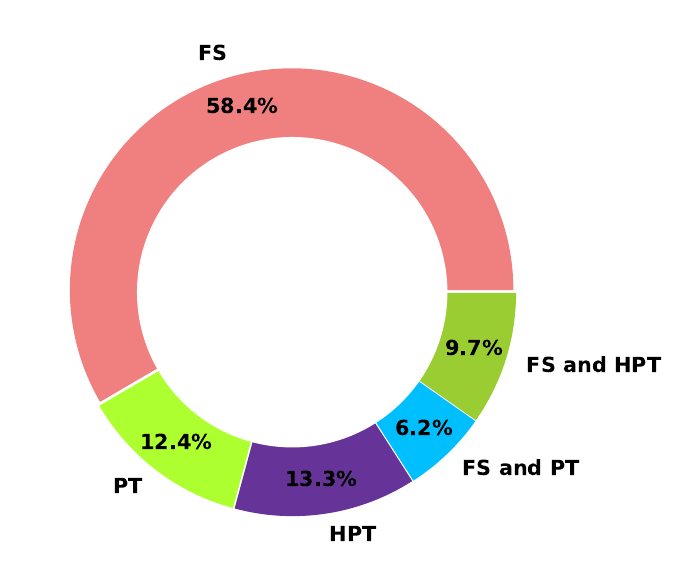}
   \caption{Percentile of different applications of metaheuristics algorithms.}
   \label{fig:meta_application}
\end{figure}

We find that most of the existing IoT-IDSs solely utilize different metaheuristics algorithms for selecting an optimal set of features. Besides, some of the systems employ them distinctly for optimizing the parameters and hyperparameters of the machine learning models. However, a few works focus on \textit{hybrid applications} by leveraging feature selection and parameter (FS-PT) or hyperparameter (FS-HPT) tuning in the same detection system. Figure \ref{fig:meta_application} illustrates the percentage of different applications leveraged by the IDSs, and Table~\ref{tab:meta_application} shows the list of corresponding existing works.
\begin{table*}[htbp]
    \centering
    \caption{List of existing IoT-IDSs based on different applications of metaheuristics.}
    \def\arraystretch{1.15}
    \resizebox{0.75\textwidth}{!}{
    \begin{tabular}{p{2.5cm}c}
        \hline
        \textbf{Application} & \textbf{Ref.}\\
        \hline
        {FS} & \multirow{5}{11cm}{\cite{1_kareem2022effective, 2_rm2020effective, 3_alkanhel2023network, 5_saif2022hiids, 6_li2022improving, 7_sanju2023enhancing, 12_dey2023hybrid, 13_krishna2021attack, 14_dey2023metaheuristic, 15_fatani2021iot, 16_dahou2022intrusion, 18_fatani2023enhancing, 19_fraihat2023intrusion, 21_alweshah2022intrusion, 22_anusha2022intrusion, 23_alweshah2023intrusion, 27_alkanhel2023hybrid, 30_davahli2020hybridizing, 31_kumar2022intellectual, 32_habib2020modified, 33_phalguna2021hybrid, 34_stankovic2022feature, 37_lv2023binary, 38_gaber2023industrial, 39_vanitha2023improved, 41_ethala2022hybrid, 42_asgharzadeh2023anomaly, 43_jayasankar2024intrusion, 44_keserwani2021smart, 46_jayalatchumy2024improved, 48_sarwar2022design, 49_fatani2021advanced, 50_jovanovic2022feature, 51_geetha2024cvs, 54_jeyaselvi2023highly, 55_abu2021iot, 56_xu2023iot, 57_prajisha2022efficient, 58_nazir2023novel, 59_alamiedy2020anomaly, 60_gangula2023intrusion, 61_davahli2020lightweight, 62_alghanam2023improved, 66_gadekallu2023moth, 67_gupta2022hybrid, 68_sandhya2021enhancing, 69_kannan2020intrusion, 73_abd2023intrusion, 76_shahapure2021water, 77_om2021harmony, 79_mohy2023whale, 81_hanafi2024intrusion, 83_singh2024intrusion, 84_mahanipour2024enhancing, 87_harbi2024bio, 88_rabie2024novel, 92_sharmamulti, 95_bella2024intrusion, 96_vadakkethil2024mayfly, 97_10.1007/s11276-023-03435-0, 98_moizuddin2022bio, 102_makhadmeh2024intrusion, 104_10317882, 105_10541183, 107_10.1016/j.comcom.2023.03.005, 110_maazalahi2024k, 113_li2024cooperative}} \\
        & \\
        & \\
        & \\
        & \\\hline
        {PT} & \multirow{1}{11cm}{\cite{4_khafaga2023voting, 20_alghamdi2022hybrid, 24_rahmani2024improvement, 26_sagu2023design, 35_baniasadi2022novel, 53_qaddoura2024evolving, 64_om2022effective, 75_jothi2023wils, 82_kethineni2024intrusion, 90_biju2024evaluated, 93_cherian2024iot, 99_kaviarasan2023network, 100_pingale2023remora, 101_deore2022hybrid}}
         \\\hline
        {HPT} & {\cite{8_savanovic2023intrusion, 24_rahmani2024improvement, 28_forestiero2021metaheuristic, 29_bahaa2022novel, 53_qaddoura2024evolving, 70_bathula2022designing, 71_basheri2023quantum, 72_ramana2022wogru, 80_bajpai2024hybrid, 85_vijayan2024original, 89_elsedimy2024novel, 91_saheed2024voting, 100_pingale2023remora, 111_dakic2024intrusion, 112_latif2024dtl}} 
         \\\hline
         {FS and PT} & {\cite{9_vaiyapuri2023metaheuristics, 11_malibari2022novel, 36_aljebreen2023binary, 45_karthikeyan2024firefly, 47_chander2023metaheuristic, 63_anuradha2022intrusion, 114_alamro2023modeling}} 
         \\\hline
         {FS and HPT} & {\cite{17_elmasry2020evolving, 25_ghasemi2024new, 36_aljebreen2023binary, 40_alruwaili2023red, 52_pan2021lightweight, 65_anushiya2023new, 74_saheed2023novel, 78_gopalakrishnan2022new, 86_aburasain2024enhanced, 94_aljehane2024golden, 106_escorcia2023sea}}
         \\\hline
    \end{tabular}
    }
    \label{tab:meta_application}
\end{table*}

\subsection{\textit{\textbf{RQ2: What are the most commonly used datasets and evaluation metrics for IoT-IDS assessment?}}}
In this study, we outline the most popular intrusion datasets used in the IoT context, specifically for testing metaheuristics and ML-driven systems. Interestingly, all works utilize public datasets, rather than creating on their own. 

\begin{figure}[htbp]
  \center
\includegraphics[scale=0.6] {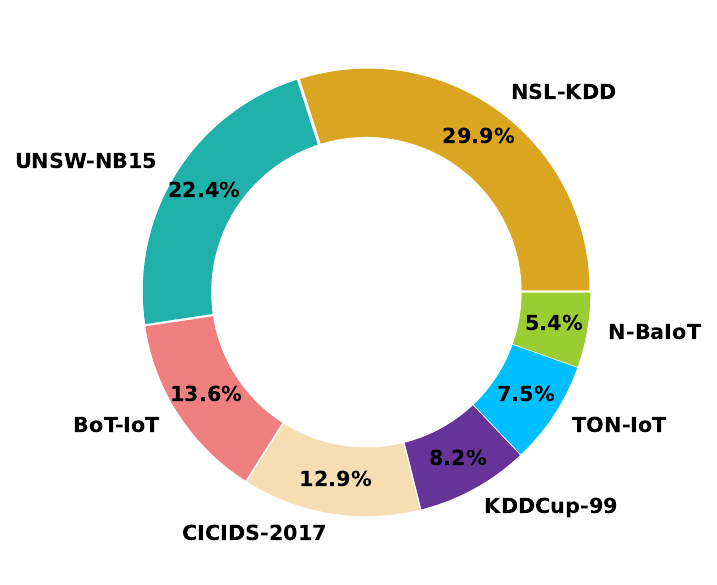}
   \caption{Percentile of the most used datasets for experimenting with the existing IoT-IDSs.}
   \label{fig:datasets}
\end{figure}

Our investigation reveals that the well-known NSL-KDD dataset \cite{nsl_tavallaee2009detailed} is extensively used by IoT-IDSs. It consists of 148,517 records extracted from the 5,209,458 samples of the oldest benchmark dataset, KDDCup-99 \cite{kdd_99} by removing redundant records. Both of these datasets contain 41 features and 5 target classes. In 2015, the UNSW-NB15 dataset \cite{unsw_moustafa2015unsw} was created having 49 features and 10 target classes. Importantly, since this dataset does not have any outdated features of attacks, researchers tend to test their works on this. Additionally, the BoT-IoT \cite{bot_koroniotis2019towards} and CICIDS-2017 \cite{cic_sharafaldin2018toward} datasets are widely used as well. However, Table~\ref{tab:datasets} states that CICIDS-2017 contains around 80 features and 8 classes; whereas BoT-IoT is a huge dataset with 73,360,900 records and 46 features and 7 classes. Several systems also utilize the TON-IoT \cite{ton_moustafa2021new} and N-BaIoT \cite{nbaiot_meidan2018n} datasets. TON-IoT is a large dataset with 22,339,021 records in total, of which 461,043 are dedicated to training and testing purposes. On the other hand, N-BaIoT consists of 23 features and 7,062,606 records. The proportion of each dataset used in the existing detection systems is shown in Figure \ref{fig:datasets}. Apart from these mostly used datasets, the IoTID20 dataset \cite{iotid20_ullah2020scheme} is used in 5 papers. Additionally, AWID \cite{awid_kolias2015intrusion}, WSN-DS \cite{wsnds_almomani2016wsn}, and RPL-NIDDS17 \cite{rpl_verma2019evaluation} datasets are employed by three systems each; whereas, the rest of the works utilize distinct datasets. \\
\begin{table*}[htbp]
    \centering
    \scriptsize
    \caption{The most used datasets for experimenting with the existing IoT-IDSs.}
    \begin{tabular}{p{2.5cm}|p{2.5cm}|p{2.1cm}|p{1.5cm}|p{1.25cm}|p{1.25cm}|p{1.25cm}|p{0.5cm}}
\hline
\textbf{Dataset (yr.)} & \textbf{Target classes} & \textbf{Distribution (\%)} & \textbf{\# of features} & \textbf{\# of training records}           & \textbf{\# of testing records}   & \textbf{Total \# of records}   & \textbf{Imbal-ance}          \\\hline
NSL-KDD (2009) \cite{nsl_tavallaee2009detailed}      & Benign, Probe, DoS, U2R, R2L & 51.89 + 35.94 + 9.48 + 2.52 + 0.17 & 41  & \multicolumn{1}{c|}{125,973}     & \multicolumn{1}{c|}{22,544}       & \multicolumn{1}{c|}{148,517} &  \checkmark     \\\hline
UNSW-NB15 (2015)  \cite{unsw_moustafa2015unsw}   & Benign, Fuzzers, Analysis, Backdoors, DoS, Exploits, Generic, Reconnaissance, Shellcode, Worms & 87.36 + 0.95 + 0.11 + 0.09 + 0.64 + 1.75 + 8.48 + 0.55 + 0.06 + 0.01 & 49  & \multicolumn{1}{c|}{175,341}     & \multicolumn{1}{c|}{82,332}       &     &  \checkmark  \\\hline
BoT-IoT (2019) \cite{bot_koroniotis2019towards}     & Benign, DoS, DDoS, Reconnaissance, Information Theft   & 0.01 + 44.96 + 52.54 + 2.48 + 0.002 & 46  & -   & -   & \multicolumn{1}{c|}{73,360,900}       & \checkmark  \\\hline
CICIDS-2017 (2017)  \cite{cic_sharafaldin2018toward}  & Benign, DoS Hulk, PortScan, DDoS, DoS GoldenEye, FTP-Patator, SSH-Patator, DoS slowloris, DoS Slowhttptest, Bot, Web Attack – Brute Force, Web Attack – XSS, Infiltration, Web Attack – SQL Injection, Heartbleed & 83.34 + 8.16 + 5.61 + 1.48 + 0.36 + 0.28 + 0.21 + 0.20 + 0.19 + 0.07 + 0.05 + 0.02 + 0.00 + 0.00 + 0.00 &  80  & -    & -     & \multicolumn{1}{c|}{2,830,743}  & \checkmark \\\hline
KDDCup-99 (1999)  \cite{kdd_99}  & Benign, Probe, DoS, U2R, R2L & 19.86 + 0.84 + 79.30 + 0.0 + 0.02 &  41  & \multicolumn{1}{c|}{4,898,431} & \multicolumn{1}{c}{311,027} & \multicolumn{1}{|c|}{5,209,458} &  \checkmark         \\\hline
TON-IoT  (2021) \cite{ton_moustafa2021new}   & Benign, Backdoor, DDoS, DoS, Injection, MITM, Password, Ransomware, Scanning, XSS & 3.56 + 2.27 + 27.60 + 15.11 + 2.03 + 0.00 + 7.69 + 0.33 + 31.96 + 9.44 & 46  & \multicolumn{2}{c}{461,043}   & \multicolumn{1}{|c|}{22,339,021}      & \checkmark  \\\hline
N-BaIoT (2018) \cite{nbaiot_meidan2018n}   &  Benign, mirai\_udp, gafgyt\_udp, gafgyt\_tcp, mirai\_syn, mirai\_ack, mirai\_scan, mirai\_udpplain, gafgyt\_combo, gafgyt\_junk, gafgyt\_scan & 7.87 + 17.41 + 13.40 + 12.17 + 10.38 + 9.11 + 7.62 + 7.41 + 7.29 + 3.71 + 3.61 & 23  & -       & -       & \multicolumn{1}{c|}{7,062,606}    & mode-rately \\\hline
\end{tabular}
      \label{tab:datasets}
\end{table*}
\textbf{Evaluation metrics}. The most prominent metrics widely used by the existing IoT-IDSs are enlisted here. These metrics include accuracy, precision, recall or sensitivity or true positive rate (TPR), f1-score, specificity or selectivity or true negative rate (TNR), false positive rate (FPR) or false alarm rate (FAR), area under curve (AUC), Matthew’s correlation coefficient (MCC), and G-mean. Equations [1-9] denote their standard mathematical representations, respectively.
\begin{equation}
    \text{Accuracy} = \frac{TP+TN}{TP+TN+FP+FN}
\end{equation}
\begin{equation}
    \text{Precision} = \frac{TP}{TP+FP}
\end{equation}
\begin{equation}
    \text{Recall} = \frac{TP}{TP+FN}
\end{equation}
\begin{equation}
    \text{F1-score} = \frac{2\times\text{Precision}\times\text{Recall}}{\text{Precision}+\text{Recall}}
\end{equation}
\begin{equation}
    \text{Specificity} = \frac{TN}{TN+FP}=1-\text{FPR}
\end{equation}
\begin{equation}
    \text{FPR} = \frac{FP}{TN+FP}
\end{equation}
\begin{equation}
    \text{AUC} = \int_{a}^{b} f(x) \,dx 
\end{equation}
\begin{equation}
    \text{MCC} = \frac{TP.TN - FP.FN}{\sqrt{(TN+FN)\times(TN+FP)\times(TP+FN)\times(TP+FP)}}
\end{equation}
\begin{equation}
    \text{G-Mean} = \sqrt{\text{Precision}\times\text{Recall}}
\end{equation}
Here, true positive (TP) is the number of detected attacks that occurred indeed, and false positive (FP) is the number of predicted intrusions which are not truly occurred. Similarly, true negative (TN) is the number of correctly classified benign or normal instances, and false negative (FN) is the volume of events that are wrongly categorized as Benign. AUC is the curve of TPR against FPR which indicates the quality of a classification model. Particularly, this indicator gives an idea of the general accuracy of the classifier for all false positive detection rates. MCC is used to evaluate the quality of both binary and multi-class classifications. Specifically, it balances the measurement considering TP, TN, FP, and FN equally. G-mean is a balanced geometric mean calculated as the square root of the product of precision and recall. It aims to correctly measure the classifier performance on any imbalanced dataset. Depending on various requirements, several other metrics such as mean square error (MSE), Jaccard similarity coefficient (JSC) or Jaccard index, negative predictive value (NPV), and many more are also measured.

\subsection{\textit\textbf{RQ3: What are the relations between the optimization algorithms and classification methods with the datasets?}}
To address this research question, we analyze the existing works and discover the best-performing IDSs tested using distinct highly-used datasets. All selected papers are published in Q1 or Q2 journals, indicating the correctness and effectiveness of our analysis.

\subsubsection{Connecting dots among Metaheuristics, Datasets, and Machine Learning}
\textit{NSL-KDD.} Turning to the NSL-KDD dataset, different classification techniques are used in the assessments of IDSs. Among these classifiers, random forest (RF) remarkably demonstrates high performance in the top-performing IDSs with 99.66\%$\sim$99.98\% accuracy and 99.85\%$\sim$99.96\% f1-score (see Table~\ref{tab:top_systems}). Notably, the RF algorithms are incorporated with population-based metaheuristics like GWO-PSO, LOA-FOA, and SMO-HPSO to select optimal features for the corresponding IDSs. Regarding deep learning algorithms, different discriminative methods like convolutional neural networks (CNNs), BiGRU, and LSTM are integrated to develop IoT intrusion detection systems. However, these models do not outperform RFs.
\\ \\
\textit{UNSW-NB15.} Interestingly, from Table~\ref{tab:top_systems}, it can be seen that the ensemble learning (EL) classification models are leveraged significantly in the tests conducted on the UNSW-NB15 dataset. The topmost systems generate an accuracy and f1-score of 99.41\%$\sim$100\% and 99.33\%$\sim$99.99\%, respectively. Importantly, these IDSs leverage metaheuristics for feature selection, parameter, and hyperparameter tuning in the machine learning models. In the majority of cases, EL techniques are integrated with either nature-based or population-based hybrid optimization algorithms, such as GWO, MFO, FOA, BGSA-BGWO, and LS-PIO. Besides, researchers also tend to explore diverse deep learning architectures though these systems are less effective than the previous ones.
\\ \\
\textit{BoT-IoT.} Regarding the experiments on the BoT-IoT dataset, Table~\ref{tab:top_systems} illustrates that most of the best-performing IDSs employ discriminative architectures, specifically CNNs. However, these IDSs do not outperform EL, SVM, AE, and ANN-based systems (accuracy of 98.86\%$\sim$99.15\% vs 99.68\%$\sim$99.98\%). Further investigation of the metaheuristics uncovers that CNNs are tested by combining with diverse categories of optimization techniques (hybrid, SI, and nature-based); whereas, other machine learning algorithms are integrated with nature-based metaheuristics and exhibit better performance. Among these optimization algorithms, roughly half of them are used for feature selection and the other half are utilized for parameter and hyperparameter optimization.
\\ \\
\textit{CICIDS-2017 and KDDCup-99.} Turning to the tests conducted on CICIDS-2017 and KDDCup-99 datasets, it can be observed that the same case as BoT-IoT concerning the utilization of deep learning techniques, that is, the abundant use of CNNs in the best-performing IDSs. On the CICIDS-2017 dataset, these algorithms work impressively since the accuracy and f1-score are between 99.77\%$\sim$99.93\%, and 99.72\%$\sim$99.93\%, respectively considering the topmost five IDSs. Table~\ref{tab:top_systems} depicts that these CNNs are consolidated with either hybrid (TSO-DE) or nature-based (RSA, AQUO, and CSA) metaheuristics. Notably, all of these algorithms are employed for selecting optimal features. On the contrary, when examined on the KDDCup-99 dataset, in most cases, CNNs do not surpass others in terms of f1-score. Other machine learning algorithms, specifically ANN and EL are synthesized with EXPSO-STFA and LS-PIO hybrid optimizers, respectively, which are also used for the optimal selection of the features. 
\\ \\
\textit{TON-IoT.} Investigation of the TON-IoT dataset reveals that both discriminative and generative-based models are leveraged to design IoT-IDSs. Distinctly, CNN and hybrid of AE-DNN models demonstrate higher performance with an accuracy of 99.99\% and 99.888\%, respectively. The corresponding f1-scores are almost identical. Concerning metaheuristics, these ML models incorporate binary multi-objective CSA (BMECapSA) and simulated annealing (SA), respectively, utilized for feature selection. 
\\ \\
\textit{N-BaIoT.} Similar to the experimental evaluations on the NSL-KDD dataset, plentiful use of classification algorithms, specifically RFs and KNN is noticed in the case of the N-BaIoT dataset. The accuracy and f1-score of these systems are satisfactory with 98.2\%$\sim$99.86\% and 99.4\%$\sim$99.86\%, respectively. However, the highest performance is achieved by the XGBoost technique. Importantly, KNNs tend to perform better when incorporated with a hybrid optimization algorithm, particularly SSO-SA. Regarding RF, it offers the same accuracy and f1-score when integrated with population-based hybrid metaheuristics (LOA-FOA and GWO-PSO). Notably, all these optimizers are employed for feature selection.
\onecolumn
\begin{center}
\scriptsize
\begin{longtable}{p{1.5cm}|p{0.5cm}|p{1.85cm}p{0.85cm}p{1.75cm}|p{1.25cm}|p{0.75cm}p{0.65cm}p{2.25cm}|p{0.5cm}}
\caption{The assessment of the best-performing IDSs in the IoT environment. Cells containing ``-" indicate that information is not explicitly mentioned, ``N/A" denotes ``not applicable''. FC = Features Count.}\\
\hline
\textbf{Dataset} & \textbf{Ref} & \textbf{Metaheuristics} & \textbf{Appli.} & \textbf{ML}  & \textbf{FC} & \textbf{Acc(\%)} & \textbf{F1(\%)} & \textbf{Others(\%)} & \textbf{Quar.} \\
\hline
\endfirsthead
\multicolumn{10}{c}%
{\tablename\ \thetable\ -- \textit{Continued from previous page}} \\
\hline
\textbf{Dataset} & \textbf{Ref} & \textbf{Metaheuristics} & \textbf{Appli.} & \textbf{ML}  & \textbf{FC} & \textbf{Acc(\%)} & \textbf{F1(\%)} & \textbf{Others(\%)} & \textbf{Quar.} \\
\hline
\endhead
\hline \multicolumn{10}{r}{\textit{Continued on next page}} \\
\endfoot
\hline
\endlastfoot

\multirow{14}{1.5cm}{NSL-KDD}  & \cite{33_phalguna2021hybrid} & GWO-PSO  & FS & RF & - & 99.97   & 99.96  & - & Q2 \\
& \cite{13_krishna2021attack} & LOA-FOA  & FS & RF & -& 99.98   & 99.73  & AUC=99.76 & Q2 \\
  & \cite{56_xu2023iot} & BGWO  & FS & EL (XGBoost) & - & 99.9427 & 99.9426  & - & Q2 \\
  & \cite{113_li2024cooperative} & BHO & FS & Parallel CNNs  & - & 99.8928   & 99.89  & & Q1 \\
  & \cite{42_asgharzadeh2023anomaly} & BMECapSA & FS & CNN  & 18 & 99.85   & 99.85  & FAR=0.0019, FNR=0.001 & Q1 \\
  & \cite{68_sandhya2021enhancing} & SMO & FS & RF & - & 99.675  & 99.9325  & AUC=99.3025 & Q2 \\
  & \cite{44_keserwani2021smart} & GWO-PSO  & FS & RF & - & 99.66   & - & - & Q1 \\
  & \cite{67_gupta2022hybrid} & HCSGA & FS & DLHNN & - & 99.52   & 97.16  & - & Q1 \\
  & \cite{75_jothi2023wils} & WOA & PT & LSTM & N/A & 99.5 & - & Specificity=98.45 & Q1 \\
  & \cite{52_pan2021lightweight} & Compact SCA   & HPT & KNN &  N/A & 99.327  & - & FAR=0.5848 &  Q2 \\
  & \cite{45_karthikeyan2024firefly} & GWO,FOA  & PT, FS  & SVM  & - & 99.29   & 96.23 & FAR=99.59, AUC=98.51 & Q1 \\
  & \cite{90_biju2024evaluated} & EBSA  & PT & DBN  & N/A & 98.96   & 99.13 & - & Q2 \\
  & \cite{41_ethala2022hybrid} & SMO-HPSO & FS & RF & 22 & 98.98   & 98.59 & AUC=99.81 & Q1 \\
  & \cite{81_hanafi2024intrusion} & IBGJO & FS & LSTM & - & 98.93   & 98.17 &  - & Q1 \\&  \cite{110_maazalahi2024k} & ASO-EO, FOA & FS  & k-means & - & 98.9   & 100 &  - & Q1\\
  & \cite{46_jayalatchumy2024improved} & enhanced CrSA & FS & EL & 11 & 99 & 98.14 &  - & Q1 \\\hline
\multirow{13}{1.5cm}{UNSW-NB15}   &  \cite{8_savanovic2023intrusion}  & modified FOA  & HPT & EL(XGBoost) & N/A & 99.98   & 99.99 &  AUC-ROC=1 & Q1 \\
&  \cite{91_saheed2024voting} & GWO & HPT  & EL & N/A & 100  & 99.745 & FAR=1.5, ROC=99.4 & Q1 \\
 &  \cite{66_gadekallu2023moth} & MFO & FS & EL & 14 & 100  & 99.75 &  - & Q1 \\
  &  \cite{64_om2022effective} & CSA & PT & HKCAE & N/A & 99.7 & 98.9  & Specificity=98.3  & Q2 \\
  &  \cite{20_alghamdi2022hybrid} & PO  & PT & CFNN & N/A & 99.46   & 99.76 & -  & Q2 \\
  &  \cite{14_dey2023metaheuristic} & BGSA-BGWO  & FS & EL & 4 & 99.41   & 99.33 &  FAR=0.03 & Q2 \\
  & \cite{110_maazalahi2024k} & ASO-EO, FOA & FS & k-means & - & 99.1  & 99.4 &  - & Q1\\
  &  \cite{75_jothi2023wils} & WOA & PT & LSTM & N/A & 99.1 & - & Specificity=98.99 & Q1 \\
  &  \cite{80_bajpai2024hybrid} & WOA & HPT & GRU  & N/A & 99 & - & - & Q2 \\
  &  \cite{60_gangula2023intrusion} & FOA & FS & EL & - & 98.89   & 98.91 & AUC=99.79  & Q2 \\
  &  \cite{88_rabie2024novel} & DRFO  & FS & DBRBF & - & 98.5 & 98.5  &  FAR=8.2 & Q1 \\
 & \cite{113_li2024cooperative} & BHO & FS & Parallel CNNs  & - & 97.7217   & 97.56  & & Q1 \\
  &  \cite{54_jeyaselvi2023highly} & EXPSO-STFA & FS & LAANN & - & 95.65   & 95.64 &  Specificity=92.74, FNR=10.23, MCC=92.56 & Q1 \\\hline
\multirow{14}{1.5cm}{BoT-IoT}  &  \cite{91_saheed2024voting} & GWO & HPT & EL & N/A & 99.98   & 99.955 &  FAR=1.3, ROC=99.99 & Q1 \\
  &  \cite{64_om2022effective} & CSA & PT & HKCAE & N/A & 99.9 & 98.2  &  Specificity=99.7 & Q2 \\
  &  \cite{51_geetha2024cvs} & CVS & FS & FLN-ANN & - & 99.68   & 99.21 &  Specificity=99.83 & Q1 \\
  &  \cite{73_abd2023intrusion} & CSA & FS & CNN  & - & 99.15   & 98.806 & -  & Q1 \\
  &  \cite{89_elsedimy2024novel} & IGWO  & HPT & QSVM & N/A & 99.11   & 97.48 & -  & Q1 \\
  &  \cite{15_fatani2021iot} & TSO-DE & FS & CNN  & - & 99.042  & 99.042 &  FAR=0.00301 & Q1 \\
  &  \cite{16_dahou2022intrusion} & RSA & FS & CNN & -  & 99.02   & 99.07 &  - & Q1 \\
  &  \cite{49_fatani2021advanced} & AQUO  & FS & CNN & 10  & 98.926  & 98.904 & - & Q1 
  \\
  &  \cite{35_baniasadi2022novel} & NS-BPSO  & PT & DCNN & N/A & 98.86   & - & Specificity=95.32, MSE=0.00053 & Q1 \\
  &  \cite{74_saheed2023novel} & HAEMPSO  & FS, {HPT}  & DNN &  - & 97.61   & - & - & Q1 \\
  &  \cite{89_elsedimy2024novel} & IGWO  & HPT & QSVM & N/A & 99.11   & 97.48 &  - & Q1 \\
  &  \cite{101_deore2022hybrid}  & CCSO  & PT & Deep LSTM & N/A & 96.71   & \multicolumn{1}{l}{-} & Specificity=91.985 & Q1 \\\hline
\multirow{12}{1.5cm}{CICIDS-2017} &  \cite{15_fatani2021iot} & TSODE & FS & CNN  & - & 99.93   & 99.93  & FAR=0.00009 & Q1 \\
   &  \cite{16_dahou2022intrusion} & RSA & FS & CNN  & - & 99.911  & 99.888 & - & Q1 \\
  & \cite{49_fatani2021advanced} & AQUO  & FS & CNN &  - & 99.911  & 99.888 & - & Q1 \\
  &  \cite{73_abd2023intrusion} & CSA & FS & CNN &  - & 99.911  & 99.888 &  - & Q1 \\
  &  \cite{51_geetha2024cvs} & CVS & FS & FLN-ANN & - & 99.77   & 99.72 &  Specificity=99.92 & Q1 \\
  &  \cite{81_hanafi2024intrusion} & IBGJO & FS & LSTM & - & 99.75   & 98.81 & -  & Q1 \\
  &  \cite{94_aljehane2024golden} & GJOA,SSA & FS, HPT & A-BiLSTM  & - & 99.69   & 98.92 &  MCC=98.74 & Q1 \\
  &  \cite{44_keserwani2021smart} & GWO-PSO  & FS & RF & - & 99.66   & - & - & Q1 \\

  & \cite{114_alamro2023modeling} & ALO, FPA  & FS, HPT & CNN+LSTM & - & 99.55   & 99.55 & AUC=99.55 & Q1 \\
  
  & \cite{105_10541183}  & IGC-SA   & FS & AE-DNN  & - & 99.4 & 99.4   & - & PAIS24 \\
  &  \cite{18_fatani2023enhancing} & MGO & FS & CNN  & - & 99.22   & 99.218 & G-Mean=99.218 & Q1 \\
  &  \cite{104_10317882}  & IWD-BBO  & FS & FNN  & - & 98.2339 & 99.0865 & -  & Q1 \\
  &  \cite{81_hanafi2024intrusion} & IBGJO & FS & LSTM & - & 99.75   & 98.81 &  - & Q1 \\\hline
\multirow{8}{1.5cm}{KDDCup-99} &  \cite{18_fatani2023enhancing} & MGO & FS & CNN &  - & 99.941  & 99.942 & G-Mean=99.942 & Q1 \\
  &  \cite{49_fatani2021advanced} & AQUO  & FS & CNN  & - & 99.919  & 89.987 & -  & Q1 \\
  &  \cite{73_abd2023intrusion} & CSA & FS & CNN  & - & 99.917  & 89.988 & - & Q1 \\
  &  \cite{62_alghanam2023improved} & LS-PIO   & FS & EL & 15 & 99.82   & 97.23 &  FAR=6.9, TPR=99.23, AUC=96.32 & Q1 \\
  &  \cite{44_keserwani2021smart} & GWO-PSO  & FS & RF & - & 99.66   & - & -  & Q1 \\
   & \cite{110_maazalahi2024k} & ASO-EO & FS & k-means & - & 96.1  & 100 &  - & Q1\\
  &  \cite{54_jeyaselvi2023highly} & EXPSO-STFA & FS & LAANN & - & 95.65   & 95.64 &  Specificity=92.74, FAR=14.52, FNR=10.23, MCC=92.56 & Q1 \\
  &  \cite{15_fatani2021iot} & TSODE & FS & CNN &  - & 92.064  & 90.007 &  FAR=0.01989 & Q1 \\
  &  \cite{16_dahou2022intrusion} & RSA & FS & CNN  & - & 92.04   & 89.985 & -  & Q1 \\\hline
\multirow{5}{1.5cm}{TON\_IoT}  &  \cite{42_asgharzadeh2023anomaly} & BMECapSA & FS & CNN &  12 & 99.99   & 99.99 &  FAR=0.0001, FNR=0.00002 & Q1 \\
  &  \cite{105_10541183}  & SA   & FS & AE-DNN & - & 99.888  & 99.875 &  - & PAIS24 \\
  &  \cite{82_kethineni2024intrusion} & WHO & PT & fused CNN-BiGRU & N/A & 99.71   & 99.05 &  - & Q1 \\
  &  \cite{106_escorcia2023sea}  & STFA,SpSO  & FS, HPT & DBN &  - & 99.51   & - & Specificty=99.36, MCC=60.36  & Q1 \\
  &  \cite{58_nazir2023novel} & TS  & FS & EL & 13 & 99.5 & - &  FAR=0.004 & Q1 \\\hline
\multirow{6}{1.5cm}{N-BaIoT} &  \cite{56_xu2023iot} & BGWO  & FS & EL (XGBoost) & - & 99.9941 & 99.9941 & -   & Q2 \\
  &  \cite{13_krishna2021attack} & LOA-FOA  & FS & RF & - & 99.86   & 99.86 &  - & Q2 \\
  &  \cite{33_phalguna2021hybrid} & GWO-PSO  & FS & RF & - & 99.86   & 99.86 &  - & Q2 \\
  &  \cite{21_alweshah2022intrusion} & SSO-SA   & FS & KNN &  - & 98.7 & 99.8 &  -  & Q2 \\
  &  \cite{23_alweshah2023intrusion} & EPC & FS & KNN &  - & 98.2 & 99.4  &  - & Q1 \\
  &  \cite{29_bahaa2022novel} & APSO-WOA & HPT   & CNN  & N/A & 94.54   & - & JCC= 0.9 & Q1  
  \label{tab:top_systems}
\end{longtable}
\end{center}
\twocolumn
\noindent\textbf{\underline{Takeaway.}} According to the correlation analysis, the existing best IoT-IDSs have achieved 99.97\%$\sim$99.99\% accuracy and 99.95\%$\sim$99.99\% f1-score with NSL-KDD, UNSW-NB15, BoT-IoT, TON-IoT, and N-BaIoT datasets. Whereas, the performance slightly reduces while experimenting on the CICIDS-2017 and KDDCup-99 datasets (99.93\%$\sim$99.94\% accuracy and f1-score). Overall, the metaheuristics and ML-integrated detection systems are effective in the IoT environment, irrespective of all widely used datasets. Another important observation is that existing intrusion detection datasets are extremely imbalanced since significant discrepancy is observed in data distribution between majority and minority classes (see Table~\ref{tab:datasets}). Consequently, it is not difficult to get a remarkable accuracy from any ML classifier~\cite{somasundaram2016data}, which can mislead researchers. Therefore, ``accuracy" should not be considered a trustworthy performance metric in such scenarios. In contrast, the F1-score is widely accepted as a reliable metric since it reflects the harmonic balance between precision and recall~\cite{bej2021loras}. For this reason, almost all top-notch IoT-IDSs measure F1-score along with accuracy and others.

\begin{figure}[htbp]
    \centering
    \includegraphics[scale=0.6]{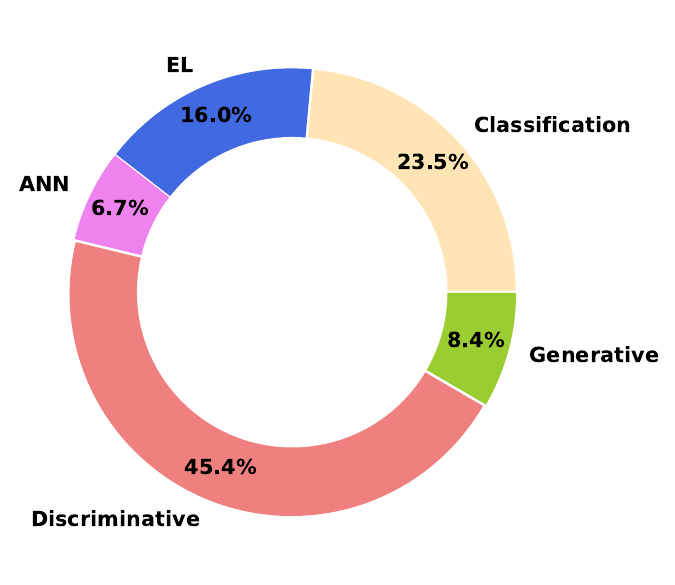}
    \caption{Usage of ML methods by the best performing IoT-IDSs.}
    \label{fig:ml_usage}
\end{figure}

\begin{figure}[htbp]
    \centering
    \includegraphics[scale=0.6]{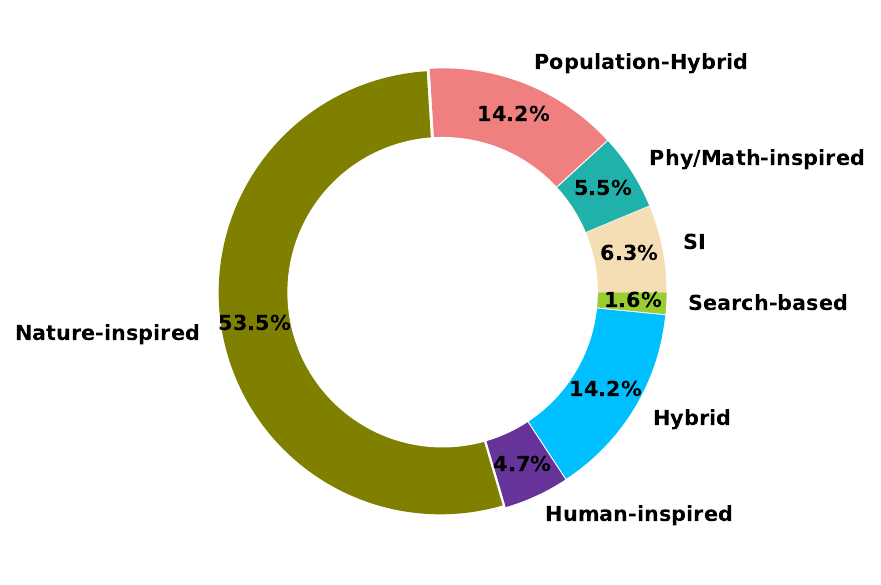}
    \caption{Usage of metaheuristics algorithms by the best performing IoT-IDSs.}
    \label{fig:meta_usage}
\end{figure}

Regarding the algorithms reveals that the classification techniques, especially RF and ensemble learning, and discriminative architectures, particularly CNN models are utilized in the top-performing IoT-IDSs, considering all the datasets as a whole. We find that most of the best-performing systems employ discriminative models to identify intrusions in IoT. Additionally, classification and EL strategies are also utilized on a notable scale. Turning to metaheuristics, nature-inspired techniques are likely to be the most suitable ones since they are leveraged in more than half of the top-notch systems. However, hybridization of metaheuristics also proves to be effective in detecting intrusions in IoT. Concerning the application of these optimization techniques, in most cases, they are integrated for selecting optimal features, except in some cases of parameter and hyperparameter tuning. Figure \ref{fig:ml_usage} and \ref{fig:meta_usage} demonstrate the usage of these techniques with corresponding percentages.

Further investigation in ensemble learning classifiers reveals the efficiency of using both traditional and deep machine learning algorithms. A significant portion of these classifiers are based on classic and deep ML, such as KNN+SVM+LSTM+MLP~\cite{60_gangula2023intrusion}, RF+DT+MLP+KNN~\cite{91_saheed2024voting}, and SVM+KNN+RF+LSTM~\cite{46_jayalatchumy2024improved}. In addition, a combination of classic ML classifiers is also observed, especially RFs~\cite{58_nazir2023novel}, LR+RF+XGBoost~\cite{66_gadekallu2023moth}, DT+AdaBoost+RF~\cite{14_dey2023metaheuristic}, and OC-SVM+IF+LOF~\cite{62_alghanam2023improved}. Interestingly, Latif et al.~\cite{112_latif2024dtl} leveraged a CNN-based bootstrap ensemble classifier (Generic CNN+Xception+Inception+InseptionResntV2+EffcientNetV2L). Regarding optimization techniques, there is a consistent trend of employing nature-inspired metaheuristics, specifically FOA~\cite{8_savanovic2023intrusion, 60_gangula2023intrusion}, MFO~\cite{66_gadekallu2023moth}, BGWO~\cite{56_xu2023iot}, CrSA~\cite{46_jayalatchumy2024improved}, GWO~\cite{91_saheed2024voting}, and GA~\cite{112_latif2024dtl}. In addition to these, search-based TS~\cite{58_nazir2023novel} along with hybrid metaheuristics BGSA-BGWO~\cite{14_dey2023metaheuristic} and LS-PIO~\cite{62_alghanam2023improved} are efficient in IoT-IDS. Moreover, almost all metaheuristics are used to select features; except in ~\cite{8_savanovic2023intrusion}, where modified FOA tunes hyperparameters of the XGBoost classifier.

\subsubsection{Delve Into the Parameter and Hyperparameter Tuning Application}
 An in-depth analysis is conducted on the parameter and hyperparameter tuning, which are the minority but crucial applications of the metaheuristics optimization algorithms. Simultaneously, we analyze the hybrid applications too. \\ \\
\noindent\textbf{\underline{Parameter Tuning Takeaway.}} Investigating the 20 articles that employ metaheuristics algorithms for optimizing parameters of the ML models (including hybrid applications), we find that most of the systems utilize nature-inspired and swarm intelligence techniques. Interestingly, several works leverage math-inspired optimizers, especially SCA, as well as a few systems rely on human-inspired techniques like PO and SCOA. Regarding ML architectures, deep learning-based models, such as DBN, LSTM, CNN, AE, GRU are mainly employed.  \par
However, the best performances are generated when nature-based optimizations are integrated with deep ML models (see Table~\ref{tab:par_hyp_diff}). Specifically, CSA+HKCAE \cite{64_om2022effective}, and WHO+fused CNN-BiGRU \cite{82_kethineni2024intrusion} come up with highest accuracies (99.7\%$\sim$99.9\%, and 99.71\%, respectively) as well as f1-scores (98.2\%$\sim$98.9\%, and 99.05, respectively). Apart from these IDSs, the human-inspired PO and CFNN-driven model \cite{20_alghamdi2022hybrid} produces 99.46\% accuracy and 99.76\% f1-score. Importantly, all of these detection systems utilize distinct metaheuristics only for tuning the parameters. No other metaheuristics-assisted optimization is performed for feature selection or hyperparameter tuning.
\\
\begin{table*}[htbp]

\scriptsize
\caption{The best IoT-IDSs that have used metaheuristics for tuning parameters and hyperparameters of the machine learning models.}
\begin{adjustwidth}{0cm}{0cm}
\resizebox{1.05\textwidth}{!}{
\begin{tabular}{p{0.1cm}p{1.1cm}p{0.79cm}p{1.1cm}p{0.35cm}p{0.35cm}|p{0.1cm}p{1.1cm}p{0.79cm}p{1.5cm}p{0.35cm}p{0.35cm}}
\hline
\multicolumn{6}{c}{\textbf{Parameter Optimization}}    & \multicolumn{6}{|c}{\textbf{Hyperparameter Optimization}}    \\\hline
\multicolumn{1}{l}{\textbf{Ref}} & \textbf{Dataset} & \textbf{Meta.} & \textbf{ML}    & \multicolumn{1}{l}{\textbf{Acc (\%)}} & \multicolumn{1}{l}{\textbf{F1 (\%)}} & \multicolumn{1}{|l}{\textbf{Ref}} & \textbf{Dataset} & \textbf{Meta.} & \textbf{ML} & \multicolumn{1}{l}{\textbf{Acc (\%)}} & \multicolumn{1}{l}{\textbf{F1 (\%)}} \\\hline
\cite{64_om2022effective}  & UNSW-NB15  & CSA & HKCAE & 99.7 & 98.9  & \cite{8_savanovic2023intrusion}  & UNSW-NB15  & modified FOA  & EL  & 99.98  & 99.99  \\\cline{2-12}
\multicolumn{1}{l}{}  & BoT-IoT  &    &     & 99.9 & 98.2  & \cite{91_saheed2024voting}  & UNSW-NB15  & GWO & EL  & 100  & 99.745  \\\cline{1-6} \cline{8-12}
\cite{82_kethineni2024intrusion}  & TON\_IoT  & WHO & fused CNN Bi-GRU & 99.71  & 99.05  & \multicolumn{1}{l}{}  & BoT-IoT  &    &  & 99.99  & 99.955  \\\hline
\cite{20_alghamdi2022hybrid}  & UNSW-NB15  & PO  & CFNN  & 99.46  & 99.76  & \cite{94_aljehane2024golden}  & CICIDS-2017    & SSA & A-BiLSTM   & 99.69  & 98.92  \\\hline
\end{tabular}
}
\end{adjustwidth}
\label{tab:par_hyp_diff}

\end{table*}
\\ \\
\noindent\textbf{\underline{Hyperparameter Tuning Takeaway.}} We examine the 27 works that leverage metaheuristics algorithms for optimizing the hyperparameters of the deep learning models, including hybrid applications. The hyperparameters optimized in the papers are shown in Table~\ref{tab:hyperparameters}, including the leveraged metaheuristics and ML models. From the table, it can be seen that four IDSs (\cite{24_rahmani2024improvement, 36_aljebreen2023binary, 53_qaddoura2024evolving, 100_pingale2023remora}) utilize metaheuristics for optimizing both parameters and hyperparameters in the ML models, at a time. \par
Interestingly,  Regarding best-performing systems, two nature-inspired and swarm-based metaheuristics are at the top. Particularly, modified FOA+EL(XGBoost) \cite{8_savanovic2023intrusion}, GWO+EL (DT,RF,KNN, and MLP) \cite{91_saheed2024voting}, and SSA+A-BiLSTM \cite{94_aljehane2024golden} demonstrate highest accuracies (99.98\%, 99.9\%$\sim$100\%, and 99.69\% respectively) and f1-scores (99.99\%, 99.745\%$\sim$99.955\%, and 98.92\% respectively). \par
Importantly, according to Table~\ref{tab:par_hyp_diff}, the first and second IDSs utilize the metaheuristics techniques only for hyperparameter tuning; whereas the last system simultaneously employs another nature-inspired optimizer (GJOA) to select an optimal feature set (hybrid application). Thus we observe hyperparameter tuning as a separate and hybrid application of the metaheuristics utilized in the top-performing IoT-IDSs. Another interesting finding is that the hybrid optimization technique is applied to a discriminative deep learning architecture; whereas the standalone hyperparameter tuning is employed to the ensemble classifiers.
\begin{table*}[t]
\centering
\scriptsize
\caption{List of hyperparameters optimized by the metaheuristics in the related works. “---” indicates that hyperparameters are not explicitly stated in the papers.}
\label{tab:hyperparameters}

\begin{tabular}{p{0.5cm}p{7.5cm}p{3.5cm}p{2.5cm}}
\hline
\textbf{Ref} & \textbf{Hyperparameters} & \textbf{ML} & \textbf{Metaheuristics} \\
\hline

\cite{8_savanovic2023intrusion} & learning rate, min\_child\_weight, subsample, colsample\_bytree, max\_depth, gamma & KNN, XGBoost & modified FOA \\
\hline

\cite{17_elmasry2020evolving} & learning rate, momentum, decay, dropout rate, number of hidden layers, neurons, epochs, batch size, optimizer, initialization, activation & DNN, LSTM-RNN, DBN & double PSO \\
\hline
\multicolumn{1}{r}{\cite{24_rahmani2024improvement}}   & \begin{tabular}[c]{@{}l@{}}(PT+HPT)\\ number of hidden layers, neurons of each layer, weights\end{tabular}   & RdNN   & GAO-AOA \\\hline
\multicolumn{1}{r}{\cite{25_ghasemi2024new}}   & Gamma (kernel coefficient parameter), C (the amount of regularization applied to the data)     & SVM & GWO \\\hline
\multicolumn{1}{r}{\cite{28_forestiero2021metaheuristic}} & ---  & IoT2Vec & ABF \\\hline
\cite{29_bahaa2022novel} & number of convolutional kernels, length of convolutional filter, activation functions in the convolutional layer, probability of nodes used between the convolutional and second layers, number of second-layer neurons, activation functions in the second layer, number of third-layer neurons, activation functions in the third layer, batch sample size, learning rate & CNN& APSO-WOA\\\hline
\cite{36_aljebreen2023binary} & \begin{tabular}[c]{@{}l@{}}(PT+HPT)\\number of hidden nodes, input weighted, biases, C$^+$ for \\ minority positive instances, C$^{-}$ for most negative instances \end{tabular}  & CCR-ELM            & SCA \\\hline
\cite{40_alruwaili2023red} & learning rate, number of epochs, batch size & EL  & LCWOA   \\\hline
\cite{52_pan2021lightweight} & k (Number of Neighbors), distance weight (w\_k) & kNN& Compact SCA \\\hline
\cite{53_qaddoura2024evolving} &   \begin{tabular}[c]{@{}l@{}}(PT+HPT)\\ weights, biases, regularization value, number of neurons, \\ type of activation function\end{tabular}    & RWNN   & GWO, PSO, MVO           \\\hline
\cite{65_anushiya2023new} & batch size, learning rate  & FR-CNN & GA\\
\cite{70_bathula2022designing} & number of hidden neurons   & ANN& SHO \\\hline
\cite{71_basheri2023quantum} & ---  & CRNN   & HS\\\hline
\cite{72_ramana2022wogru} &  earning rate, number of hidden layers, input weights, epochs  & GRU& WOA \\\hline
\cite{74_saheed2023novel} &   learning rate, input units, batch size, dropout, epochs, activation function, layers number, optimizer, units of hidden layer  & DNN& modified PSO            \\\hline
\cite{78_gopalakrishnan2022new} &  number of suitable hidden neurons of DNN, iterations of Adaboost, number of suitable bootstrap in the random forest  & HR-OELM        & AF-EFO  \\\hline
\cite{80_bajpai2024hybrid} & learning rate, sample sampling rate (subsample), maximum depth of the tree   & XGBoost            & WOA \\\hline
\cite{85_vijayan2024original} & ---  & EL & TuSO\\\hline
\cite{86_aburasain2024enhanced} & ---  & HDL& Enhanced BWO            \\\hline
\multicolumn{1}{r}{\cite{89_elsedimy2024novel}}   & num\_wolves, min range, max range, initial population, crossover rate, num\_qubits, depth, max fun, shots      & QSVM   & Improved GWO (IGWO)     \\\hline
\multicolumn{1}{r}{\cite{91_saheed2024voting}}   & ---  & EL & GWO \\\hline
\multicolumn{1}{r}{\cite{94_aljehane2024golden}}   &  learning rate, activation, epochs, dropout rate,  batch size & A-BiLSTM & SSA \\\hline
\multicolumn{1}{r}{\cite{100_pingale2023remora}}  & \begin{tabular}[c]{@{}l@{}}(PT+HPT)\\ weights and training parameters\end{tabular} & DMN& ROA \\\hline
\cite{106_escorcia2023sea}     &  learning rate, dropout, batch size, epoch count, activation function & DBN& SpSO\\\hline
\cite{111_dakic2024intrusion}     & \begin{tabular}[c]{@{}l@{}}(XGBoost): learning rate, min child weight, subsample, \\colsample by tree, max depth, gamma \\ (KNN): k, weights, distance \end{tabular} & XGBoost and KNN    & GSAPSO \\\hline
\cite{112_latif2024dtl}     &  optimizer, activation function, dense units, dropout, fine-tune layers, epochs & EL & GA \\\hline
\cite{114_alamro2023modeling}  & --- & CNN+LSTM & ALO, FPA
 \label{tab:hyperparameters}

\end{tabular}
\end{table*}

\subsection{\textit{\textbf{RQ4: What are open issues raised by the integration of metaheuristics with ML in IoT-IDS?}}}
Though metaheuristics and ML-integrated algorithms bring about significant evolution in the development of detection systems, there are still some issues and challenges that have to be addressed soon.
\begin{itemize}
    \item \textbf{Resource-constrained issue of IoT devices.} The most critical issue in the development of the IoT-IDS is the dynamic and heterogeneous characteristics of its ecosystem. IoT supports different large-scale networks with distinct communication protocols and applications, which have individual abilities and conditions. Moreover, the data has various degrees of complexities ranging from a simple sensor for observing blood pressure to a complex full-duplex video feed. Moreover, some devices build up with multiple sensors, for example, a smartphone has sensors like GPS, camera, fingerprinting, etc. Consequently, ensuring the security of this diverse IoT environment is an extremely challenging task. Even, there is no such evidence that the integration of metaheuristics and ML methods always guarantees to protect against all types of attacks. Besides, the hybridization of multiple methods imposes a severe effect on the computational power and energy resources. Additionally, though a few works are dedicated to healthcare and industrial IoT, the amount is too insufficient considering the importance of sensitive data protection and the new era of ``Industry 4.0". Moreover, it is indeed necessary to design unique IDSs for ITS, Internet of Medical Things (IoMT), Internet of Agriculture (IoA), Internet of Vehicle (IoV), and Internet of Done (IoD) by employing optimization-assisted ML techniques. 
    
    \item \textbf{Issues regarding datasets.}
    Most of the datasets, used in the existing papers, were created before 2020 (NSL-KDD: 2009, UNSW-NB15: 2015, BoT-IoT: 2019, CICIDS-2017: 2017, and KDDCup-99: 1999), which lack the features of the latest sophisticated attacks. Consequently, a question arises whether the existing techniques that integrate metaheuristics and ML, can detect these new intrusions or not. Alarmingly, there are not sufficient IoT-IDS datasets that contain the features or attributes of new intrusions, especially those generated after 2020.  Moreover, the well-known widely used intrusion detection datasets are imbalanced, which can severely affect machine learning models. Furthermore, in the papers, experiments are conducted in lab settings. As a result, the correctness and effectiveness of the existing algorithms in real-world scenarios are arguably a critical issue.
        
    \item \textbf{Amount and quality of the selected features.} Surprisingly, most of the IoT-IDS papers have not clearly stated the features that are selected by the metaheuristics. As a result, it becomes much more difficult to analyze the correctness and effectiveness of the optimization techniques. Additionally, introducing these features would give an intuition on which features are crucial for developing a generalized IDS in IoT. Moreover, though the ultimate classification results are supposed to indicate the validity of the algorithms, it does not necessitate proof of whether these optimization techniques have any impact or not. A possible solution could be the analysis of individual IDSs with and without applying those metaheuristics. Regrettably, these tactics are almost missing in the literature. Furthermore, maintaining a decent balance between the quality and quantity of the selected features is another vital issue in the IoT context.
    
    \item \textbf{Appropriate selection of parameters and hyperparameters.} The optimization of the parameters, in both classical ML and deep learning architectures, is another important thing of consideration while implementing an IoT-IDS. Especially, in neural networks, the optimal choice of parameters like weights and biases plays a pivotal role in enhancing the performance of the detection system. Additionally, a few works also utilized tuning the hyperparameters, such as learning rate, number of layers, neuron volume, number of epochs, regularization value, type of activation function, etc. of different deep neural networks, such as DBN, RNN, GRU, CNN, and LSTM. Therefore, working with inappropriate and less significant variables can result in utmost failure of classifying attacks. Although several works claimed to generate excellent accuracy through optimizing these parameters and hyperparameters, there is a crucial lack of dedicated analysis on them.
    
    \item \textbf{Issues regarding ML methods.} To improve classification performance, combining the advantages of multiple ML algorithms can be a promising approach. Though some research has already utilized this, more advanced techniques must be applied to keep the computational overhead in control, maintaining all security requirements since IoT devices can only operate with low power and limited resources. Another challenging thing is to make the proposed IDSs capable of analyzing real-time traffic. Alarmingly, most of the models in the literature are typically offline, that is, they are trained on different datasets and are tested on real-time data. Consequently, they need to retrain periodically, which is time-consuming and expensive. Considering the IoT environment, where the data are diverse and dynamically evolve over time, these traditional static IDSs are not sufficient in real-world big-data applications, especially at the enterprise level. In this case, incremental learning can be a viable solution \cite{liang2012group}, where the system dynamically learns continuously added features that were previously unknown. In \cite{hu2013online}, an online IDS is proposed for the dynamic distributed network. Specifically, at first, a local parameterized detection model is constructed in each node using the online Adaboost algorithm. Then, all of the local models are combined using PSO-based and SVM-driven algorithms to generate a global detection model, which achieves 99.99\% accuracy and 0.37\% FAR. Wahab et al. \cite{wahab2022intrusion} devise a technique to adjust the size of the DNN's hidden layers in an online manner so that the model can continuously learn and adapt new intrusions, and update predictions dynamically. The experimental evaluation states that their online DNN surpasses the static one in terms of false positives and false negatives by 6\% and 4.5\%, respectively.
\end{itemize}

\subsection{\textit{\textbf{RQ5: What are the unexplored metaheuristics optimization algorithms for IDS in IoT?}}}
In this section, some possible optimization techniques are discussed that can be utilized for either feature selection, parameter, or hyperparameter optimization. 
\begin{itemize}
    \item \textbf{Microbiology-inspired metaheuristics.} These types of optimization techniques rely on the life cycle, immune system, social behavior, and collective behavior of viruses, bacteria, and other microorganisms. For example, the bacterial foraging optimization algorithm (BFOA) \cite{guo2021survey}, bacterial swarming algorithm (BSA), bacterial-GA foraging (BF), and quantum-inspired bacterial swarming optimization (QBSO) are some popular techniques of this category. Interestingly, a few metaheuristics are based on the replication and herd immunity concept of the coronavirus, known as coronavirus herd immunity optimizer (CHIO) \cite{al2021coronavirus}, and coronavirus optimization algorithm (COVIDOA). Other well-known microbiology-driven techniques are sperm swarm optimization algorithm (SSO), swine influenza models-based optimization (SIMBO), and symbiosis organisms search (SOS). No IoT-IDS is found to utilize any of these algorithms. However, a chaotic bacterial colony optimization (CBCO) technique is utilized for tuning the weights, biases, and number of neurons of the Elman recurrent neural network (ERNN), hence generalizing the model's performance \cite{hussan2023ddos}. As a result, the proposed IoT DDoS attack detection system surpasses other related systems. In \cite{khayyat2023improved}, an improved bacterial foraging optimization is employed for feature selection in the smart city anomaly detection system. Specifically, the classic BFOA is enhanced using the simulated annealing technique by incorporating decisions based on probability to achieve better convergence to the global optima and to handle the local extrema. However, a Bayesian optimization algorithm is applied to tune the hyperparameters of the multiplicative long short-term memory (MLSTM) model. The experimental evaluations show that IBFOA provides better classification accuracy with less computational complexity. Since these works are related to the intrusion detection system, microbiology-inspired metaheuristics can be a good choice to test in the IoT-IDS environment with diverse ML models.

    \item \textbf{Chemistry-based metaheuristics.} Though many physics and math-based metaheuristics are utilized for developing robust IoT-IDSs, we find no chemistry-inspired algorithms, adopting the concepts of chemical reactions and laws, such as molecular reaction, motion, radiation, etc. Kinetic gas molecules optimization (KGMO) \cite{moein2014kgmo}, artificial chemical reaction optimization algorithm (ACROA), and ions motion optimization algorithm (IMOA) are well-established optimization algorithms of this category. Intuitively, these algorithms can produce excellent performances like physics and math-inspired ones. However, these metaheuristics have been applied to other domains, such as cyber-physical systems ~\cite{tavolato2020analytical}, clustering and routing algorithms in WSN ~\cite{asha2020efficient}, clustering in big data environment ~\cite{ding2020fuzzy}, etc. Asha and Gowrishankar \cite{asha2020efficient} utilize glowworm swarm optimization (GSO) and kinetic gas molecule optimization (KGMO) to increase network lifetime and number of transmissions in WSN. This hybrid energy-efficient algorithm outperforms the existing PSO-PSO-WSN and PSO-GSO-WSN. Moreover, KGMO generally offers fast convergence and is suitable for complex real-world optimization tasks. In \cite{ding2020fuzzy}, an extended CRO (real-coded CRO) is employed to find the optimal clusters for fuzzy clustering. Consequently, the false negative activities are reduced significantly compared to the earlier models (e.g., SVM, NB, DT, RF, and FCM). Besides, the accuracy and convergence speed are improved, especially in the big data environment. based on these remarkable outcomes in various domains, chemistry-based metaheuristics can be integrated into IoT-IDSs.
    
    \item \textbf{Miscellaneous metaheuristics.} Some other popular metaheuristics are inspired by the sunflower behavior to find the best orientation towards the sun (sunflower optimization (SFO) \cite{ehteram2023application}), players' intelligence to find the best position to score a goal (football game optimization (FGO)), optimization inspired from mother's care for her children~\cite{matouvsova2023mother} and the trading method of the stock exchanges (exchange market algorithm (EMA) \cite{ghorbani2014exchange}). Recently, Prashanth et al. \cite{prasanth2024effective} have devised an efficient routing technique for wireless sensor networks through load balancing by an SFO. While compared with the existing approaches (e.g., CRCGA, GECR, OMPFM, and GADA-LEACH), the optimization demonstrates better results considering packet delivery ratio, packet loss, throughput, average residual energy, and delay. Taking the similar characteristics (like resource-constrainedness) of both WSN and IoT, this variant of optimization algorithms can be a suitable solution for IoT-IDSs.

    \item \textbf{Metaheuristics and GAN-based Detection Systems.} Generative Adversarial Network (GAN) is one of the most popular detection methods regarding the development of an IDS \cite{ferdowsi2019generative}. Particularly, it facilitates synthetic data creation and better learning of the minority classes. Moreover, it can generate samples faster than DL methods, and is capable of identifying zero-day attacks in IoT since it learns from a wide range of attack scenarios \cite{salimans2016improved}. Ferdowsi and Saad \cite{ferdowsi2019generative} develop a GAN-based distributed IDS to identify malicious activities, independent of a centralized controlling process. They utilize different ANNs for both the generator and discriminator. Recently, Rahman et al. \cite{rahman2024syn} propose a GAN-based NIDS, called SYN-GAN, aiming to handle the disproportion in the existing imbalance datasets and mimic the real-world network intrusion data. Notably, the model demonstrates 91\%, 84\%, and 100\% accuracy in the  UNSW-NB15,  NSL-KDD, and BoT-IoT datasets, respectively. However, GAN faces difficulties while training with high dimensional data since the developed generator and discriminator are often complex and unstable \cite{lakshminarayanan2017simple}. Incorporating metaheuristics with GAN can be a promising solution to resolve these issues since these optimizers have already proved efficient in feature selection, parameter optimization, and hyperparameter tuning phases. Researchers, focusing on implementing IoT-IDSs, can get insight from the work \cite{anusha2024cyber, senthilkumar2024cloud}, where GAN is integrated with different optimization techniques like war strategy optimization (WSO), gazelle optimization algorithm (GOA), and archerfish hunting optimizer (AHOA), etc. for parameter tuning and feature selection, resulting in an improved performance in the attack classification process. Though these works integrate GAN and metaheuristics in WSN and cloud, they do not solely focus on the IoT environment. Considering the scale, uniqueness, and difficulties of the IoT ecosystem, as well as the necessity of the metaheuristics and ML-integrated IoT-IDSs (as described in Section~\ref{sec:intro}), we recommend applying GAN and these optimization techniques to develop IoT-IDSs.

\end{itemize}

\section{Discussions} \label{sec:discussion}
The Internet of Things (IoT) environment consists of a diverse range of sensitive and private data from various devices and sensors. Besides, the wide variety of protocols, technologies, and platforms make IoT more heterogeneous, complicated, and dynamic. Traditional intrusion detection systems, relying on specific rules, statistics, or heuristics often fail to identify these complex patterns of the IoT ecosystem. To overcome these problems, researchers have focused on machine learning-driven IDSs, especially for IoT. However, these methods come up with new problems, such as the requirement of high computational resources and significant time consideration for a small precision improvement. Moreover, the emerging trend of de-centralized edge computing technology-based IoT-IDSs faces challenges, for example, limited computing power and inadequate energy support, which are unavoidable requirements for the training and testing of ML models using large datasets. For these reasons, in recent years, a growing trend of utilizing optimization techniques, especially metaheuristics algorithms can be seen to select optimal features and tune parameters or hyperparameters in the ML-based IoT-IDSs. \par
Considering the recent trends and the significance of leveraging optimization and ML techniques, this study presents a systematic literature review on the metaheuristics and machine learning-integrated intrusion detection systems for IoT. The review includes 111 relevant papers, of which 96 are high-quality journal articles (51.8\% Q1 and 28.6\% Q2), and covers almost every recent optimization technique utilized for IoT-IDSs. The distinct analysis of different applications, such as feature selection, parameter, and hyperparameter tuning as well as hybrid applications has technically enriched this work. Our extensive investigation reveals that the majority of the systems (74.3\%) apply these optimizers for selecting an optimal set of features from the popular public datasets. Simultaneously, a notable amount of works focus on tuning different parameters (e.g., weights and biases) of machine learning models and hyperparameters like learning rate, batch size, number of hidden layers, number of neurons, number of epochs, dropout rate, activation function, etc. are also optimized by some systems. \par
Apart from these, we also discover that most of the high-performing relevant detection systems leverage hybrid metaheuristics. Moreover, well-established ML classifiers like RF, KNN, and EL are employed significantly to identify various types of attacks. Regarding deep learning models, different discriminative architectures, for instance, CNN, DBN, and LSTM have demonstrated remarkable results. One of the severe drawbacks of these systems is the use of comparatively older and imbalanced datasets. Consequently, they lack the features of the latest attacks. Furthermore, the majority of the articles in the literature lack a dedicated analysis of the selected features, parameters, and hyperparameters that influence IDSs' performance; even in most of the cases, the selected or optimized set are not mentioned explicitly. Moreover, the existing systems operate in offline mode with specific datasets, which may drastically fail in real-world scenarios where traffic behaviors are extremely varied. To address these challenges, the dynamically updated features need to be trained continuously as done in the incremental learning. Moreover, we suggest integrating metaheuristics with GAN to perform optimum feature selection, keeping the IoT constraints and security conditions consistent.\par
Multimodal data processing has gained significant momentum with the advent of alignment models, particularly vision-language models, which excel at creating efficient vector-based data structures for storage, retrieval, and utilization. This is especially beneficial in IoT applications where resources such as processing power and memory are limited. Real-time detection leveraging edge and fog computing has also advanced, providing scalable solutions while minimizing latency. Multimodal approaches are crucial for managing diverse datasets and reducing false positives \cite{li2023gasto,xu2024fast,li2024novel}. Emerging research directions include privacy-preserving federated learning, which enables decentralized model training while safeguarding data privacy, and explainable AI, which enhances trust by making IDS outputs interpretable. Additionally, quantum-inspired optimization is being explored for faster convergence, and multi-objective optimization aims to balance detection accuracy, latency, and energy efficiency in IoT environments. In addition, metaheuristic algorithms can provide an alternative to the neural architecture search algorithms for constructing hybrid novel deep learning models, handling multi-objective optimization problems, etc.

\section{Conclusion} \label{sec:conclusion}
Considering the importance and recent widespread use of metaheuristics algorithms in developing machine learning-based intrusion detection systems, we aim to technically analyze the existing integrated IoT-IDS models in this study. Specifically, we have investigated the metaheuristics-assisted and machine learning-driven systems, categorizing them into various applications like feature selection, parameter optimization, and hyperparameter tuning. One of the significant findings of this review is the establishment of hidden relations between top-notch optimization techniques and ML architectures concerning the most used datasets. Moreover, the introduction of a large-scale visualized taxonomy of these integrated IoT-IDSs, also adds value to the literature. In the end, several technical issues of metaheuristics and ML integration are discussed and some insightful directions are proposed to address these challenges in the coming days.

\section*{List of Acronyms}
\onecolumn
\footnotesize
\begin{longtable}{p{1cm}|p{6.5cm}|p{1.65cm}|p{7.1cm}}
\hline
    \textbf{Abbr.} & \textbf{Elaboration} & \textbf{Abbr.} & \textbf{Elaboration}\\
    \hline
\endfirsthead
\hline
    \textbf{Abbr.} & \textbf{Elaboration} & \textbf{Abbr.} & \textbf{Elaboration}\\
    \hline
\endhead
\hline\endfoot
\hline
\endlastfoot

AAFSO & Assimilated Artificial Fish Swarm Optimization  &
ABC & Artificial Bee Colony \\
ABF & Activity-based Footprinting &
A-BiLSTM & Attention-based Bidirectional LSTM  \\
ACO & Ant Colony Optimization &
AE & Auto Encoder   \\
ALO& Ant Lion Optimization &
ANN & Artificial Neural Network   \\
AOA & Arithmetic Optimization Algorithm &
APSO & Adaptive Particle Swarm Optimization  \\
AQUO& Aquila optimizer  &
ASO & Atom Search Optimization \\
BA & Bat algorithm &
BAS & Beetle Antenna Search \\
BBFA  & Binary Bee Foraging Algorithm  &
BBO &Biogeography-based Optimization \\
BCOA & Binary Chimp Optimization Algorithm &
BES& Bald Eagle Search  \\
BGSA & Binary Gravitational Search Algorithm &
BGWO & Binary Grey Wolf Optimization  \\
BHO & Black Hole Optimization &
Bi-GRU & Bidirectional Gated Recurrent Unit  \\
BiLSTM & Bidirectional Long-Short Term Memory &
BOA & Butterfly Optimization Algorithm \\
BQABC& Binary Quantum-inspired ABC &
BSA & Bird Swarms Algorithm \\
BWO& Black Widow Optimization &
CCSO& Chimp Chicken Swarm Optimization   \\
CD& Canberra Distance  &
CFNN& Cascade Forward Neural Network \\
ChSO & Chicken Swarm Optimization  &
CNN& Convolutional Neural Network \\
COA& Chimp Optimization Algorithm &
CRNN & Cascaded Recurrent Neural Network   \\
CrSA & Crow Search Algorithm &
CSA & Capuchin Search Algorithm \\
CSSA& Chaotic Salp Swarm Optimization &
CVS & Chaotic Vortex Search  \\
DBN & Deep Belief Network  &
DBRBF & Descriptive Back Propagated RBF \\
DCNN & Deep CNN &
DDT & Distance Decision Tree  \\
DE & Differential Evaluation &
DFWA & Dynamic Search Fireworks Optimization \\
DHOA & Deer Hunting Optimization Algorithm  &
DL & Deep Learning \\
DLHNN & Deep Learning-based Hybrid NN &
DMN & Deep Maxout Network \\
DRFO & Decisive Red Fox Optimization &
DT & Decision Tree \\
DTO & Dipper Throated Optimization &
DWNN & Deep Wavelet Neural Network \\
EA & Evolutionary Algorithm  &
EBSA& Evaluated Bird Swarm Optimization  \\
EFO & Electric Fish Optimization &
EGB& Extreme Gradient Boosting \\
EL& Ensemble Learning  &
ELM& Extreme Learning Machine  \\
EO & Equilibrium Optimization &
EPC & Emperor Penguin Colony \\
ET & Extremely Randomized Trees or Extra Trees  &
FDM & Fractional Derivative Mutation  \\
FL & Federated Learning  &
FLN & Fast-Learning Network \\
FNN & Feed-forward Neural Network &
FOA & Firefly Optimization Algorithm \\
GA & Genetic Algorithm &
GAO & Grasshopper Optimization  \\
GJOA & Golden Jackal Optimization Algorithm &
GO & Growth Optimizer  \\
GRU&Gate Recurrent Unit &
GSO & Glow-Worm Swarm Optimization  \\
GTO & Gorilla Troops Optimizer &
GWO & Grey-Wolf Optimization  \\
HCSGA& Hybrid Chicken Swarm Genetic Algorithm &
HDL & Hybrid Deep Learning \\
HHO & Harris Hawk Optimization &
HKCAE& Capsule AE with a Hybrid Kernel function  \\
HMS & Human Mental Search &
HNM & Hierarchical Network Model  \\
HPSO & Hierarchical Particle Swarm Optimization &
HR-OELM & High Ranking-based Optimized EL \\
HS& Harmony Search &
HSHO & Harmony Search Hawks Optimization  \\
IAOA & Improved Arithmetic Optimization Algorithm &
IBGJO & Improved Binary Golden Jackal Optimization  \\
IGC &Information Gain Calculation &
IoT & Internet of Things \\ 
IDS & Intrusion Detection System   &
ITS & Intelligent Transportation Systems \\
IWD& Intelligent Water Drop   &
KELM & Kernel Extreme Machine Learning Model  \\
KNN & K-Nearest Neighbor   &
LAANN& Look Ahead Artificial Neural Network  \\
LCWOA & Lévy-fight Chaotic Whale Optimization Algorithm   &
LOA & Lion Optimization Algorithm \\
LR & Linear Regression  &
LS& Local Search  \\
LSTM & Long Short-Term Memory   &
MDSVM & Mahalanobis Distance SVM \\
MFO & Moth–Flame Optimization   &
ML & Machine Learning  \\
MOA& Mayfly Optimization Algorithm  &
MOPSO & Multi-Objective Particle Swarm Optimization \\
MPO& Marine Predator Optimization   &
MSO & Moth Search Optimization \\
MVO & Multi-Verse Optimizer  &
NS & Neighborhood Search  \\
OBL& Opposition-based Learning  &
PCA & Principal Component Analysis \\
PDO& Prairie Dog Optimization  &
PIO& Pigeon-inspired Optimization  \\
PO & Political Optimizer   &
PM & Polymorphic Mutation  \\
PSO & Particle Swarm Optimization   &
QCSO& Quantum Cat Swarm Optimization  \\
QPSO & Quantum Behaved Particle Swarm Optimization   &
QSVM & Quantum Support Vector Machine  \\
RBFNN & Radial Basis Function Neural Network   &
RBM & Restricted Boltzmann machine \\
RdNN & Random Neural Network   &
RF& Random Forest \\
RKOA & Red Kite Optimization Algorithm   &
RL & Reinforcement Learning  \\
RNN & Recurrent Neural Network  &
ROA & Remora Optimization Algorithm  \\
RSA & Reptile Search Algorithm   &
RWNN & Random Weight Neural Network  \\
SA & Simulated Annealing   &
SAEHO & Seagull Adapted Elephant Herding Optimization \\
SCA & Sine Cosine Algorithm  &
SCOA & Social Group Optimization Algorithm  \\
SCSO & Sand Cat Swarm Optimizer  &
SHO & Spotted Hyena Optimization \\
SMO & Spider Monkey Optimization  &
SpSO & Sparrow Search Optimization \\
SSA& Salp Swarm Algorithm   &
SSO & Shuffled Shepherd Optimization \\
STFA& Sea Turtle Foraging Algorithm  &
SU-CMO & Self-Upgraded Cat and Mouse Optimizer  \\
SVM & Support Vector Machine   &
TLBO& Teaching-Learning-based Optimization  \\
TS & Tabu Search  &
TSO & Transient Search Optimization \\
TuSO& Tuna Swarm Optimization   &
VAE & Variational Autoencoder \\
WHO& Wild Horse Optimization   &
WMSA& Water Moth Search algorithm \\
WOA & Whale Optimization Algorithm   &
WWO & Water Wave optimization 
\end{longtable}
\twocolumn

\begingroup\scriptsize
\makeatletter
\renewcommand\@openbib@code{\itemsep\z@}
\makeatother      
\bibliographystyle{IEEEtran} 
\bibliography{doe}
\endgroup

\appendix

\begin{appendices}\label{sec:appendix}
\textbf{\textit{\appendixname{A}}. {Performances tabulation of the existing metaheuristics-based and ML-driven IoT-IDSs studied in this SLR.}}\par
Though recall and detection rate can be used interchangeably, some papers explicitly mention DR rather than recall. Therefore, we use these two terms separately in this study to avoid conflict and confusion.
\onecolumn
\scriptsize
    \begin{landscape}
    \setlength\LTleft{-0.77cm}
\setlength\LTright{\fill}
\begin{longtable}{p{0.25cm}p{3.7cm}p{0.95cm}p{1.5cm}p{3.7cm}p{3cm}p{0.15cm}p{0.75cm}p{0.75cm}p{0.75cm}p{0.75cm}p{0.75cm}p{0.75cm}p{0.75cm}p{1.5cm}}
\caption{Analysis of the IoT-IDSs based on different performance metrics, metaheuristics, applications, ML algorithms, datasets, and classification types. Here, ``FC" and ``CLT" represent selected feature count and classification types, respectively.}\\
\hline
\textbf{Ref.}     & \textbf{Meta.}    & \textbf{App$^n$}  & \textbf{FC} & \textbf{ML}    & \textbf{Dataset used}  & \multicolumn{1}{c}{\textbf{CLT}} & \multicolumn{8}{c}{\textbf{IoT-IDS Performance (\%)}}     \\
\multicolumn{1}{l}{}  & \multicolumn{1}{l}{} & \multicolumn{1}{l}{} & \multicolumn{1}{l}{} & \multicolumn{1}{l}{} & \multicolumn{1}{l}{}      &  & \textbf{Acc}   & \textbf{Prec}             & \textbf{\begin{tabular}[c]{@{}c@{}}Rec\end{tabular}} & \textbf{F1} & \multicolumn{1}{c}{\textbf{Speci}} & {\textbf{DR}} & {\textbf{FPR}} & {\textbf{Others}} \\
\hline
\endfirsthead
\multicolumn{15}{c}%
{\tablename\ \thetable\ -- \textit{Continued from previous page}} \\
\hline
\textbf{Ref.}     & \textbf{Meta.}    & \textbf{App$^n$}  & \textbf{FC} & \textbf{ML}    & \textbf{Dataset used}  & \multicolumn{1}{c}{\textbf{CLT}} & \multicolumn{8}{c}{\textbf{IoT-IDS Performance (\%)}}     \\
\multicolumn{1}{l}{}  & \multicolumn{1}{l}{} & \multicolumn{1}{l}{} & \multicolumn{1}{l}{} & \multicolumn{1}{l}{} & \multicolumn{1}{l}{}      &  & \textbf{Acc}   & \textbf{Prec}             & \textbf{\begin{tabular}[c]{@{}c@{}}Rec\end{tabular}} & \textbf{F1} & \multicolumn{1}{c}{\textbf{Speci}} & {\textbf{DR}} & {\textbf{FPR}} & {\textbf{Others}} \\
\hline
\endhead
\hline\endfoot
\hline
\endlastfoot

\cite{1_kareem2022effective}   & GTO-BSA    & FS & -  &  KNN & NSL-KDD  &  & 95.5  & \multicolumn{1}{l}{}  & 91.4 & \multicolumn{1}{l}{}  & \multicolumn{1}{c}{97.4}  & & & \\\hline
  & & & &  & CICIDS2017  &  & 98.7  & \multicolumn{1}{l}{}  & 97.3 & \multicolumn{1}{l}{}  & \multicolumn{1}{c}{99.7}  & & &  \\
   & & & & & UNSW-NB15  &  & 81.5  & \multicolumn{1}{l}{}  & 81.5 & \multicolumn{1}{l}{}  & \multicolumn{1}{c}{87.7}  & & &  \\
    & & & & & BoT-IoT  &  & 81.5  & \multicolumn{1}{l}{}  & 99.3 & \multicolumn{1}{l}{}  & \multicolumn{1}{c}{96.2}   & & & \\
\multirow{-4}{*}{\cite{2_rm2020effective}}  & \multirow{-4}{*}{hybrid PCA-GWO}     & \multirow{-4}{*}{FS} & \multirow{-4}{*}{-} &  \multirow{-4}{*}{DNN} & Kaggle dataset &  & 99.9  & \multicolumn{1}{l}{}  & 95.4 & \multicolumn{1}{l}{}  & \multicolumn{1}{c}{100}    & & & \\\hline
\cite{4_khafaga2023voting}   & new WOA, guided by DTO     & PT       & N/A&  KNN, RF, NN & RPL-NIDS17 &  & 95.1  & \multicolumn{1}{l}{}  & \multicolumn{1}{l}{}  & \multicolumn{1}{l}{}  &   & &  &  { AUC=99.0, MSE=2.50E-08}  \\\hline
\cite{5_saif2022hiids}  & PSO, GA, and DE & FS & 8-10 &  KNN,DT & NSL-KDD  &  & 95.71 & \multicolumn{1}{l}{}  & \multicolumn{1}{l}{}  & \multicolumn{1}{l}{}  &    & & & \\\hline
\cite{6_li2022improving} & BOA  & FS& - & ANN   & NSL-KDD  &  & 93.27 & \multicolumn{1}{l}{}  & 94.37    & \multicolumn{1}{l}{}  & \multicolumn{1}{c}{92.68}  & & & \\\hline
\cite{7_sanju2023enhancing}  & HHO-FDM         & FS & - &  LSTM +GRU = RNNs   & IoT-23   &  & 98.12 & 98.06    & 98.31    & 98.18    &   & &  &  {AUC-ROC=99.82 }   \\\hline
   & & & & & UNSW-NB15  &  & 99.98 & 99.99    & 99.98    & 99.99    &  & &  &  {AUC-ROC=100} \\
\multirow{-2}{*}{\cite{8_savanovic2023intrusion}} & \multirow{-2}{*}{modified FOA}      & \multirow{-2}{*}{HPT}  & \multirow{-2}{*}{N/A} &  \multirow{-2}{*}{KNN, XGBoost}   & IoT-healthcare-security-dataset   &  & 99.6997 & 99.6998  & 99.6997  & 99.6996  &  & & &   \\\hline
 
\cite{9_vaiyapuri2023metaheuristics} & BSA, SCOA        & FS, PT & - &  FL, KELM  & -    &  & 99.45 & 80.26    & 82.67    & 80.95    &    & & & \\\hline
 
\cite{11_malibari2022novel} & IAOA, QPSO          & FS+ PT  & - & DWNN  & CICIDS2017 &  & 98.21 & 96.53    & 98.22    & 97.92    &    & & & \\\hline
\cite{12_dey2023hybrid} & NSGA-II        & FS & 13 &  SVM & TON-IoT  &  & 99.48 & \multicolumn{1}{l}{}  & \multicolumn{1}{l}{}  & \multicolumn{1}{l}{}  &  & & &   \\\hline
   & & & & & NSL-KDD  &  & 99.98 & 99.87    & 100  & 99.73    &  & &  &  {AUC=99.76}  \\
\multirow{-2}{*}{\cite{13_krishna2021attack}} & \multirow{-2}{*}{LOA-FOA}       & \multirow{-2}{*}{FS} & \multirow{-2}{*}{-} &  \multirow{-2}{*}{RF} & NBaIoT   &  & 99.86 & 99.94    & 99.94    & 99.86    &  & &  &  { FN=7, FP=2 }  \\\hline
\cite{14_dey2023metaheuristic} & BGSA and BGWO & FS & 4 &  DT and EL (AdaBoost and RF)   & UNSW-NB15  &  & 99.41 & 99.92    & \multicolumn{1}{l}{}  & 99.33    &   &  {99.09}   &  {0.03} &   \\\hline
  & & & &  &  & \multicolumn{1}{c}{M}    & 92.064  & 89.943   & 92.064   & 90.007   &  & &  {0.01989}   &   \\
  & & & &  & \multirow{-2}{*}{KDDCup-99}    & \multicolumn{1}{c}{B}    & 92.451  & 94.32    & 92.451   & 92.851   &   & &  {0.07527}   &  \\ \cline{6-15}
  & & & &  &  & \multicolumn{1}{c}{M}    & 75.751  & 78.988   & 75.751   & 71.692   &  & &  {0.05868}   &  \\
  & & & &  & \multirow{-2}{*}{NSL-KDD} & \multicolumn{1}{c}{B}    & 77.381  & 83.637   & 77.381   & 77.08    &   & &  {0.19223}   &   \\ \cline{6-15}
& & & &    &  & \multicolumn{1}{c}{M}    & 99.042  & 99.042   & 99.042   & 99.042   & & &  {0.00301}   &   \\
& & & &    & \multirow{-2}{*}{BoT-IoT} & \multicolumn{1}{c}{B}    & 99.992  & 99.992   & 99.992   & 99.992   &  & &  {0.00007}  &   \\ \cline{6-15}
& & & &    &  & \multicolumn{1}{c}{M}    & 99.93 & 99.93    & 99.93    & 99.93    & & &  {0.00009}  &\\
\multirow{-8}{*}{\cite{15_fatani2021iot}} & \multirow{-8}{*}{TSODE}   & \multirow{-8}{*}{FS} & \multirow{-8}{*}{-} &  \multirow{-8}{*}{CNN} & \multirow{-2}{*}{CICIDS-2017}  & \multicolumn{1}{c}{B}    & 99.996  & 99.996   & 99.996   & 99.996   &   & &  {0.000029}  &  \\\hline
 & & & &   &  & \multicolumn{1}{c}{M}    & 92.04 & 89.684   & 92.04    & 89.985   &   & & &  \\
   & & & & & \multirow{-2}{*}{KDDCup-99}    & \multicolumn{1}{c}{B}    & 92.344  & 94.335   & 92.344   & 92.763   &  & & &   \\ \cline{6-15}
 & & & &   &  & \multicolumn{1}{c}{M}    & 76.107  & 82.171   & 76.107   & 71.731   &   & & &  \\
& & & &    & \multirow{-2}{*}{NSL-KDD} & \multicolumn{1}{c}{B}    & 77.814  & 83.83    & 77.814   & 77.545   &   & & &  \\ \cline{6-15}
 & & & &   &  & \multicolumn{1}{c}{M}    & 99.911  & 99.907   & 99.911   & 99.888   &   & & &  \\
& & & &    & \multirow{-2}{*}{CICIDS-2017}  & \multicolumn{1}{c}{B}    & 99.997  & 99.997   & 99.997   & 99.997   &   & & &  \\ \cline{6-15}
& & & &    &  & \multicolumn{1}{c}{M}    & 99.02 & 99.098   & 99.038   & 99.07    &   & & &  \\
\multirow{-8}{*}{\cite{16_dahou2022intrusion}} & \multirow{-8}{*}{RSA}       & \multirow{-8}{*}{FS} & \multirow{-8}{*}{-} &  \multirow{-8}{*}{CNN}  & \multirow{-2}{*}{BoT-IoT} & \multicolumn{1}{c}{B}    & 99.993  & 99.99    & 99.993   & 99.99    &   & & &  \\\hline
 & & & &   & NSL-KDD  &  & 98.77 &  98.1   &  92.29  &  95.11  &   & & &  \\
\multirow{-2}{*}{\cite{17_elmasry2020evolving}} & \multirow{-2}{*}{double PSO} & \multirow{-2}{*}{FS+HPT}  & \multirow{-2}{*}{10} &  \multirow{-2}{*}{DNN, LSTM-RNN, and DBN} & CICIDS2017 &  & 95.81 &  95.82  &  95.81  &  95.81  &   & & &  \\\hline
& & & &    & KDDCup-99  &  & 99.941  & 99.947   & 99.936   & 99.942   &  & &  &  {G-Mean=99.942}   \\
 & & & &   & NSL-KDD  &  & 92.04 & 90.841   & 91.04    & 90.941   &  & &  &  {G-Mean=90.941}   \\
 & & & &   & BoT-IoT  &  & 76.725  & 83.105   & 76.672   & 79.759   &  & &  &  {G-Mean=79.824}  \\
\multirow{-4}{*}{\cite{18_fatani2023enhancing}} & \multirow{-4}{*}{MGO, using WOA} & \multirow{-4}{*}{FS} & \multirow{-4}{*}{-} &  \multirow{-4}{*}{CNNs}  & CICIDS-2017    &  & 99.22 & 99.188   & 99.248   & 99.218   &   & &  &  {G-Mean=99.218}   \\\hline
& & & & &  & \multicolumn{1}{c}{M}    &  99.52   &  99.51   & \multicolumn{1}{l}{ } &  99.4 &  & & &  \\
& & & & & \multirow{-2}{*}{ NF-CSE-CIC-IDS2018-v2}   & \multicolumn{1}{c}{B}    &  99.54   &  99.54   & \multicolumn{1}{l}{ } &  99.54   &  & & &  \\ \cline{6-15}
 & & & & &  & \multicolumn{1}{c}{M}    &  97.21   &  97.15   & \multicolumn{1}{l}{ } &  97.16   &  &  { 99.52}   &  { 3.27} &   \\
& & & & & \multirow{-2}{*}{ NF-ToN-IoT-v2}  & \multicolumn{1}{c}{B}    &  99.99   &  99.99   & \multicolumn{1}{l}{ } &  99.99   &  &  { 99.54}   &  { 3.26} &  \\ \cline{6-15}
 & & & & &  & \multicolumn{1}{c}{M}    &  98.76   &  98.8 & \multicolumn{1}{l}{ } &  98.77   &  &  { 97.17}   &  { 0.32} &   \\
 & & & & & \multirow{-2}{*}{ NF-UNSW-NB15-v2}   & \multicolumn{1}{c}{B}    &  99.69   &  99.7 & \multicolumn{1}{l}{ } &  99.69   &  &  { 99.99}   &  { 0.02} &   \\ \cline{6-15}
& & & & &  & \multicolumn{1}{c}{M}    &  98.52   &  98.53   & \multicolumn{1}{l}{ } &  98.52   & &  { 98.76}   &  { 3.71} &  \\
\multirow{-8}{*}{\cite{19_fraihat2023intrusion}} & \multirow{-8}{*}{AOA}  & \multirow{-8}{*}{FS}  & \multirow{-8}{*}{7} &  \multirow{-8}{*}{RF and ET} & \multirow{-2}{*}{ NF-BoT-IoT-v2}  & \multicolumn{1}{c}{B}    &  99.98   &  99.98   & \multicolumn{1}{l}{ } &  99.98   &   &  { 99.69}   &  { 4.08} &  \\\hline
 & & & &   & NSL-KDD++  &  & 99.86 & 99.89    & 99.58    & 99.72    &   & & &  \\
    & & & & & UNSW-NB15  &  & 99.46 & 99.75    & 99.62    & 99.76    &   & & &  \\
\multirow{-3}{*}{\cite{20_alghamdi2022hybrid}} & \multirow{-3}{*}{PO}          & \multirow{-3}{*}{PT} & \multirow{-3}{*}{N/A} &  \multirow{-3}{*}{CFNN} & CIDCC-2017 &  & 99.38 & 99.69    & 99.66    & 99.69    &  & & &   \\\hline
 
\cite{21_alweshah2022intrusion} & SSO-SA1 and SSO-SA2 & FS & 39.11 &  KNN  & N-BaIoT  &  & 0.987 & 0.997    & 0.996    & 0.998    &   & & &  \\\hline
 
\cite{22_anusha2022intrusion} & GSO & FS & -&  PCA & NSL-KDD  &  & 93.35 & 91.9 & 94.02    & \multicolumn{1}{l}{ } &  &  { 95.12}   &  { 2.97} &  \\\hline
 
\cite{23_alweshah2023intrusion} & EPC & FS & 38.27 &  KNN  & N-BaIoT  &  & 98.2  & 99.7 & 99.2 & 99.4 &  & & &   \\\hline
\cite{24_rahmani2024improvement} & GAO-AOA              & PT+HPT  & N/A &  RdNN & TON\_IOT &  & \multicolumn{1}{l}{} & 99.56    & \multicolumn{1}{l}{}  & \multicolumn{1}{l}{}  & &  {99.37}   &  {4}   &   \\\hline
 
 & & & & & NSL-KDD  &  & 98 & 97 & 99 & 98 &  & & &   \\
 
\multirow{-2}{*}{\cite{25_ghasemi2024new}} & \multirow{-2}{*}{GWO} & \multirow{-2}{*}{FS+HPT} & \multirow{-2}{*}{-} &  \multirow{-2}{*}{SVM} & TON\_IOT &  & 81 & 82 & 84 & 83.57    &   & & &  \\\hline
\cite{26_sagu2023design}  & SAEHO, SU-CMO            & PT           & N/A &  CNN+DBN and Bi-LSTM+GRU &  UNSW-NB15&  & 92.8  & \multicolumn{1}{l}{}  & \multicolumn{1}{l}{}  & 81 &   & &  &  {MCC=0.786, Rand Index = 0.998 (dataset-2)}   \\\hline
\cite{27_alkanhel2023hybrid} & GWO-DTO & FS          & - &  KNN  & RPL-NIDS17 &  & 98.1  & 97.8 & 99.4 & 98.6 & \multicolumn{1}{c}{97.8}  & & &  \\\hline
 
\cite{28_forestiero2021metaheuristic} & ABF & HPT  & N/A &  IoT2Vec & CASAS dataset  &  & \multicolumn{1}{l}{ }    & \multicolumn{1}{l}{ } & \multicolumn{1}{l}{ } & 92.98    &  & & &  {avg entropy= 0.7478}   \\\hline
\cite{29_bahaa2022novel} & APSO-WOA & HPT     & N/A&  CNN & N-BaIoT  &  & 94.54 & 95.2 & \multicolumn{1}{l}{}  & \multicolumn{1}{l}{}  &   & &  & {kappa= 0.936, hamming loss= 0.944 , JSC= 0.9}  \\\hline
\cite{30_davahli2020hybridizing} & GA-GWO         & FS & 92 &  SVM  & AWID &  & 99.1  & 96.03    & \multicolumn{1}{l}{}  & 97.64    &  &  {99.32}   &  {0.69} &  \\\hline
\cite{31_kumar2022intellectual} & HHGS-ROA & FS & -&  SVM & AWID &  & 99.16 & 99.76    & 99.4 & 99.58    &  & &  {0.2} &  {MCC=99.97} \\\hline
\cite{32_habib2020modified} & MOPSO-Lévy      & FS & 44.33 &  KNN & N-BaIoT  &  & 97.06 & 88.69    & \multicolumn{1}{l}{}  & \multicolumn{1}{l}{}  &  & &  {1.66} & { TPR=0.7506, TNR=0.9834, G-mean=0.8317, AUC =0.867  }  \\\hline
  & & & &  &  & \multicolumn{1}{c}{M}    & 99.97 & 99.95    & 99.97    & 99.96    &  & & &   \\
  & & & &  & \multirow{-2}{*}{NSL-KDD} & \multicolumn{1}{c}{B}    & 99.98 & 99.87    & 100  & 99.73    &   & & &  \\ \cline{6-15}
\multirow{-3}{*}{\cite{33_phalguna2021hybrid}} & \multirow{-3}{*}{PSO-GWO} & \multirow{-3}{*}{FS}       & \multirow{-3}{*}{-} &  \multirow{-3}{*}{RF} & N-BaIoT  &  & 99.86 & 99.94    & 99.94    & 99.86    &  & & &   \\\hline
 
& & & & & CICIDS-2017    &  & 98.71 & \multicolumn{1}{l}{ } & 96.17    & \multicolumn{1}{l}{ } &   & & &  \\
 
\multirow{-2}{*}{\cite{34_stankovic2022feature}} & \multirow{-2}{*}{a hybrid ABC} & \multirow{-2}{*}{FS} & \multirow{-2}{*}{14.8, 11.9} &  \multirow{-2}{*}{ELM} & UNSW-NB15  &  & 71.54 & \multicolumn{1}{l}{ } & 80.58    & \multicolumn{1}{l}{ } &   & & &  \\\hline
\cite{35_baniasadi2022novel} & NSBPSO   & PT       & N/A &  DCNN & UNSW-NB15 and Bot-IoT &  & 98.86 & 99.03    & \multicolumn{1}{l}{}  & \multicolumn{1}{l}{}  & \multicolumn{1}{c}{95.32}  & &  &  {MSE=0.00053}  \\\hline
\cite{36_aljebreen2023binary} & BCOA, SCA  & FS, PT+HPT  & - &  CCR-ELM & WSN-DS   &  & 99.63 & \multicolumn{1}{l}{}  & 97.91    & 94.52    & \multicolumn{1}{c}{99.67} & & &  \\\hline 
\cite{37_lv2023binary} & BBFA   & FS& 23 & SVM & N-BaIoT  &  & 99.2   &  & & &   & 99 & 0.006 &  \\\hline
\cite{38_gaber2023industrial} & PSO-BA          & FS & 16 &  RF & WUSTL-IIOT-2021  &  & 99.99 & 99.6 & 99.6 & 99.6 &   & & &  \\\hline
\cite{39_vanitha2023improved} & IACO  & FS & -&  EL using DDT, ANFIS and MDSVM & UNSW-NB15  &  & 97.375  & \multicolumn{1}{l}{}  & \multicolumn{1}{l}{}  & \multicolumn{1}{l}{}  &   &  {92.365}  &  {6.67} &  \\\hline
\cite{40_alruwaili2023red} & RKOA, LCWOA & FS,HPT  & -&  EL using LSTM, BiLSTM, and BiGRU & WSN-DS   &  & 98.94 & \multicolumn{1}{l}{}  & 75.33    & 79.52    & \multicolumn{1}{c}{75.33} & &  &  {AUC=85.48}  \\\hline
 
& & & & & UNSW-NB  &  & 99.18 & 94.19    & 93.32    & 94.12    &  & &  {0.15} &  {AUC=99.78, MCC=0.19 } \\ \cline{6-15}
 
 & & & & &  & \multicolumn{1}{c}{ M} & 98.98 & 98.76    & 97.89    & 98.59    &    & & &  {AUC=99.81} \\
 
\multirow{-3}{*}{\cite{41_ethala2022hybrid}} & \multirow{-3}{*}{SMO-Hierarchical PSO(HPSO)}       & \multirow{-3}{*}{FS}       & \multirow{-3}{*}{22} &  \multirow{-3}{*}{RF}  & \multirow{-2}{*}{ NSL-KDD}  & \multicolumn{1}{c}{ B} & 98.31 & 98.61    & 98.41    & 98.56    &  & &  {0.21} &  {AUC=99.87, MCC=0.17 }   \\\hline
 
& & & & & NSL-KDD  &  & 99.85 & 99.85    & 99.85    & 99.85    &   & &  {0.0019}  &  {FNR=0.001}   \\
 
\multirow{-2}{*}{\cite{42_asgharzadeh2023anomaly}} & \multirow{-2}{*}{BMECapSA} & \multirow{-2}{*}{FS} & \multirow{-2}{*}{12,18} &  \multirow{-2}{*}{CNN}  & TON-IoT  &  & 99.99 & 99.99    & 99.99    & 99.99    &  & &  { 0.0001}  &  {FNR=0.00002}   \\\hline
 
\cite{43_jayasankar2024intrusion} & DFWA              & FS & -&  ODRNN & IDS dataset    &  & 96.11 & 96.11    & \multicolumn{1}{l}{ } & 97.21    &   &  {97.31}   &  {3.03} &  \\\hline
\cite{44_keserwani2021smart} & GWO-PSO  & FS & -&  RF &  KDDCup-99, NSL–KDD, CICIDS-2017   & \multicolumn{1}{c}{M} & 99.88, 99.25, 99.87 &  &  &  & & & &    \\\hline
\cite{45_karthikeyan2024firefly} & GWO, FOA   & PT, FS & -&  SVM & NSL-KDD  &  & 99.29 & \multicolumn{1}{l}{}  & 98.12    & 96.23    & \multicolumn{1}{c}{99.59} & &  &  {AUC=98.51}  \\\hline
 & & & &   & NSL-KDD  &  & 99 & 98.38    & 98.02    & 98.14    &   & & &  \\
\multirow{-2}{*}{\cite{46_jayalatchumy2024improved}} & \multirow{-2}{*}{enhanced CrSA}    & \multirow{-2}{*}{FS} & \multirow{-2}{*}{11,7} &  \multirow{-2}{*}{EL} & UNSW-NB15  &  & 97.75 & 83.57    & 83.39    & 81.66    &  & & &   \\\hline
 
 & & & & & UNSW-NB15  &  & 99.31 & 67.09    & 60.33    & 60.35    &   & & &  \\
 
\multirow{-2}{*}{\cite{47_chander2023metaheuristic}} & \multirow{-2}{*}{new DHOAF, SpSO}   & \multirow{-2}{*}{FS,PT}  & \multirow{-2}{*}{-} &  \multirow{-2}{*}{CRNN} & UCI SECOM  &  & 97.88 & 92.42    & 89.87    & 91.1 &  & & &   \\\hline
\cite{48_sarwar2022design} & PSO & FS & 17 &  RF & IoTID20  &  &  98 (B),     & \multicolumn{1}{l}{}  & \multicolumn{1}{l}{}  & \multicolumn{1}{l}{}  &   & & &  \\
 &   &  &   83 (M)     &  &   &  &   & & &  \\\hline
& & & &    &  & \multicolumn{1}{c}{M}    & 99.911  & 99.91    & 99.91    & 99.888   &  & & &   \\
    & & & & & \multirow{-2}{*}{CICIDS-2017} & \multicolumn{1}{c}{B}    & 99.997  & 99.997   & 99.997   & 99.997   &    & & & \\ \cline{6-15}
 & & & &   &  & \multicolumn{1}{c}{M}    & 76.002  & 81.719   & 76.002   & 71.602   &  & & &   \\
& & & &    & \multirow{-2}{*}{NSL-KDD} & \multicolumn{1}{c}{B}    & 77.382  & 83.692   & 77.382   & 77.077   &  & & &   \\ \cline{6-15}
 & & & &   &  & \multicolumn{1}{c}{M}    & 98.926  & 98.905   & 98.904   & 98.904   &  & & &   \\
 & & & &   & \multirow{-2}{*}{ BoT-IoT}  & \multicolumn{1}{c}{B}    & 99.994  & 99.992   & 99.993   & 99.992   &   & & &  \\ \cline{6-15}
 & & & &   &  & \multicolumn{1}{c}{M}    & 99.919  & 89.824   & 92.042   & 89.987   &  & & &   \\
\multirow{-8}{*}{\cite{49_fatani2021advanced}} & \multirow{-8}{*}{AQUO} & \multirow{-8}{*}{FS} & \multirow{-8}{*}{10} &  \multirow{-8}{*}{CNN} & \multirow{-2}{*}{ KDDCup-99}    & \multicolumn{1}{c}{B}    & 99.922  & 94.283   & 92.256   & 92.683   &  & & &   \\\hline
& & & & &  UNSW-NB15 & & 71.54 & \multicolumn{1}{l}{}  & 80.58    & \multicolumn{1}{l}{ } &  & & & \\
 
\multirow{-2}{*}{\cite{50_jovanovic2022feature}} & \multirow{-2}{*}{SCSO}          & \multirow{-2}{*}{FS} & \multirow{-2}{*}{14.5,11.7} &  \multirow{-2}{*}{ELM} & CICIDS-2017    &  & 98.7  & \multicolumn{1}{l}{ } & 96.17    & \multicolumn{1}{l}{ } &  & & &   \\\hline
 & & & &   & CIC IDS-2017   &  & 99.77 & 99.6 & \multicolumn{1}{l}{}  & 99.72    & \multicolumn{1}{c}{99.92} &  {99.81}   &  &  \\
\multirow{-2}{*}{\cite{51_geetha2024cvs}} & \multirow{-2}{*}{CVS} & \multirow{-2}{*}{FS} & \multirow{-2}{*}{-} &  \multirow{-2}{*}{FLN, an ANN} &  BoT-IoT   &  & 99.68 & 99.3 & \multicolumn{1}{l}{}  & 99.21    & \multicolumn{1}{c}{99.83} &  {99.11}   &  &   \\\hline
 & & & &   & NSL-KDD  &  & 99.327  & \multicolumn{1}{l}{}  & \multicolumn{1}{l}{}  & \multicolumn{1}{l}{}  &  &  {99.206}  &  {0.5848} &   \\
\multirow{-2}{*}{\cite{52_pan2021lightweight}} & \multirow{-2}{*}{Compact SCA}        & \multirow{-2}{*}{HPT}       & \multirow{-2}{*}{N/A} &  \multirow{-2}{*}{kNN} & UNSW-NB15  &  & 98.27 & \multicolumn{1}{l}{}  & \multicolumn{1}{l}{}  & \multicolumn{1}{l}{}  &    &  {97.94}   &  {5.82} &\\\hline
 
\cite{53_qaddoura2024evolving} & GWO, PSO, MVO         & PT+HPT & N/A&  RWNN & IoTID20  &  & \multicolumn{1}{l}{ }    & \multicolumn{1}{l}{ } & \multicolumn{1}{l}{ } & \multicolumn{1}{l}{ } &  & & &  {G-mean=0.7283 }   \\\hline
\cite{54_jeyaselvi2023highly} & EXPSO-STFA   & FS & -&  LAANN  &  KDDCup-99, NSL-KDD,   &  & 95.65 & 94.74    & 93.54    & 95.64    & \multicolumn{1}{c}{92.74}  & &  {14.52}  &  { FNR=10.2 }  \\
& CIDDS-001, and UNSW-NB15 & & & & & & & & & {MCC=92.56 }  \\\hline
\cite{55_abu2021iot} & SSA–ALO         & FS & -&  KNN  & N-BaIoT  &  & \multicolumn{1}{l}{} & \multicolumn{1}{l}{}  & \multicolumn{1}{l}{}  & \multicolumn{1}{l}{}  &  & &  {0.029}  &  {TPR=0.991, G-mean=0.984} \\\hline
& & & &    &  & \multicolumn{1}{c}{M}    & 99.9941 &  99.9941 &  99.9941 &  99.9941 &   & & &  \\
 & & & &   & \multirow{-2}{*}{N-BaIoT} & \multicolumn{1}{c}{B}    & 99.997  & 99.997   & 99.997   & 99.997   &   & & &  \\ \cline{6-15}
 & & & &   &  & \multicolumn{1}{c}{M}    & 99.9427 & 99.9426  & 99.9427  & 99.9426  &   & & &  \\
 & & & &   & \multirow{-2}{*}{NSL-KDD} & \multicolumn{1}{c}{B}    & 99.9427 &  99.9427 &  99.9427 &  99.9427 &   & & &  \\ \cline{6-15}
  & & & &  & WUSTL-IIOT-2021  & \multicolumn{1}{c}{M}    & 100  & 100  & 100  & 100  &   & & &  \\\cline{6-15}
\multirow{-6}{*}{\cite{56_xu2023iot}} & \multirow{-6}{*}{BGWO}         & \multirow{-6}{*}{FS}        & \multirow{-6}{*}{-} &  \multirow{-6}{*}{XGBoost} & WUSTL-EHMS-2020  & \multicolumn{1}{c}{M}    & 98.897  & 98.8923  & 98.897   & 98.8846  &   & & &  \\\hline
 
 & & & & & MC-IoT   &  & 99.38 & 99.25    & 98.8 & 98.76    & & & &    \\
 
 & & & & & MQTT-IoT-IDS2020 &  & 98.91 & 98.8 & 98.36    & 97.16    &   & & &  \\
 
\multirow{-3}{*}{\cite{57_prajisha2022efficient}} & \multirow{-3}{*}{ECSSA}        & \multirow{-3}{*}{FS} & \multirow{-3}{*}{-} &  \multirow{-3}{*}{LightGBM}  & MQTTset  &  & 98.35 & 97.38    & 97.68    & 98.56    &   & & &  \\\hline
\cite{58_nazir2023novel} & TS & FS & 13 &  EL using RF  & TON\_IoT &  & 99.5  & 97.92    & \multicolumn{1}{l}{}  & \multicolumn{1}{l}{}  &  & &  {0.004}  &  \\\hline
\cite{59_alamiedy2020anomaly} & multi-objective GWO               & FS & 4 &  SVM  & NSL–KDD  &  & 87.59 & \multicolumn{1}{l}{}  & \multicolumn{1}{l}{}  & \multicolumn{1}{l}{}  &  & & &   \\\hline
 
& & & & & UNSW-NB15  & \multicolumn{1}{c}{ B} & 98.89 & 99.68    & 99.32    & 98.91    &   & & &  {AUC=99.79}   \\ \cline{6-15}
 
& & & & &  & \multicolumn{1}{c}{ M} & 92.3  & 91.32    & 78.47    & 81.435   &    & &  {0.95} &  {AUC=90.2, MCC=0.45}  \\
 
\multirow{-3}{*}{\cite{60_gangula2023intrusion}} & \multirow{-3}{*}{FOA} & \multirow{-3}{*}{FS} & \multirow{-3}{*}{-} &  \multirow{-3}{*}{EL using KNN,SVM,LSTM,MLP} & \multirow{-2}{*}{ NSL-KDD}  & \multicolumn{1}{c}{ B} & 98.41 & 98.68    & 98.46    & 98.68    &    & &  {0.24} &  {AUC=99.79, MCC=0.26}\\\hline
 
\cite{61_davahli2020lightweight} & GABGWO & FS & 94 &  SVM & AWID and KDDCup-99 &  & 99.09 & 96.31    & 99.3 & 97.84    &   & &  {0.68} &  \\\hline
& & & &    & BoT-IoT  &  & 97.37 & \multicolumn{1}{l}{}  & \multicolumn{1}{l}{}  & 94.88    &   & &  {2.05} &  { TPR=98.78, AUC=95.68}   \\
    & & & & & UNSW-NB15  &  & 94.45 & \multicolumn{1}{l}{}  & \multicolumn{1}{l}{}  & 91.35    &   & &  {30.73}  &  {TPR=98, AUC=89.52}  \\
   & & & & & NLS-KDD  &  & 94.7  & \multicolumn{1}{l}{}  & \multicolumn{1}{l}{}  & 89.1 &  & &  {21.33}  &  {TPR=95.7, AUC=87.63  }   \\
\multirow{-4}{*}{\cite{62_alghanam2023improved}} &  \multirow{-4}{*}{LS-PIO} & \multirow{-4}{*}{FS}       & \multirow{-4}{*}{15,10,8,3} &  \multirow{-4}{*}{EL} & KDDCup-99  &  & 99.82 & \multicolumn{1}{l}{}  & \multicolumn{1}{l}{}  & 97.23    &  & &  {6.9} &  {TPR=99.23, AUC=96.32}  \\\hline
 & & & &   & UNSW-NB15  &  & 97.39 & 89.17    & 96.4 & 91.53    &  & & &   \\
\multirow{-2}{*}{\cite{63_anuradha2022intrusion}} & \multirow{-2}{*}{HCMFO,BAS}    & \multirow{-2}{*}{FS,PT} & \multirow{-2}{*}{24,15} &  \multirow{-2}{*}{VAE} & NSL-KDD  &  & 95.25 & 87.16    & 95.4 & 90.56    &   & & &  \\\hline
 
& & & & & BoT-IoT  &  & 99.9  & 98.7 & 99.7 & 98.2 & \multicolumn{1}{c}{ 99.7}  & & & \\
 
\multirow{-2}{*}{\cite{64_om2022effective}} & \multirow{-2}{*}{CSA}   & \multirow{-2}{*}{PT} & \multirow{-2}{*}{N/A} &  \multirow{-2}{*}{HKCAE} & UNSW-NB15  &  & 99.7  & 99.6 & 99.6 & 98.9 & \multicolumn{1}{c}{ 98.3} & & &  \\\hline
 & & & &   & UNSW-NB15  &  & 94.488  & 94.2942  & 94.5631  & 94.4284  &   & & &  \\
\multirow{-2}{*}{\cite{65_anushiya2023new}} & \multirow{-2}{*}{AAFSO,GA}       & \multirow{-2}{*}{FS,HPT}  & \multirow{-2}{*}{-} &  \multirow{-2}{*}{FR-CNN} & BoT-IoT  &  & 93.7756 & 86.6687  & 95.874   & 91.0393  &  & & &   \\\hline
\cite{66_gadekallu2023moth} & MFO & FS & 14 &  EL using LR, RF and XGBoost & UNSW-NB15  &  & 100   & 99.5 & 100  & \multicolumn{1}{l}{}  &  & & &   \\\hline
 
\cite{67_gupta2022hybrid} & HCSGA           & FS & -&  DLHNN & NSL-KDD  &  & 99.52 & 97.55    & \multicolumn{1}{l}{ } & 97.16    &    &  { 96.78}   & &   \\\hline
\cite{68_sandhya2021enhancing} & SMO      & FS &- &  RF & NSL-KDD  &  & 99.675  & 99.955   & 99.9425  & 99.9325  & & &  &  {AUC=99.3025}    \\\hline
\cite{69_kannan2020intrusion} & ACO     & FS       & -&  PCA  & KDDCup-99  &  & \multicolumn{1}{l}{} & \multicolumn{1}{l}{}  & \multicolumn{1}{l}{}  & \multicolumn{1}{l}{}  &  &  {91} &  {1.8} &  \\\hline
\cite{70_bathula2022designing} & SHO & HPT              & N/A&  ANN & KDDcup99  &  &  98.16   & 98.06 & 98.03 & 98.04 & 98.27   &  & 1.73 & FNR=1.97, NPV=98.27, MCC=96.30\\
& & & & &   IoTID20 &  &   98.83  & 99.79 & 98.96 & 99.38 & 96.67   & & 3.33 & FNR=1.04, NPV=96.67, MCC=90.20\\
& & & & &  IoT Botnet &  &  97.87   & 100 & 97.75 & 98.86 &  100  & & 0 & FNR=2.25, NPV=100, MCC=96.30\\\hline
\cite{71_basheri2023quantum} & QCSO, HS              & Clustering, HPT    & N/A&  CRNN  & KDDCup-99  &  & \multicolumn{1}{l}{} & \multicolumn{1}{l}{}  & \multicolumn{1}{l}{}  & \multicolumn{1}{l}{}  &   &  {92.04}   &  {6.86} &   \\\hline
 
\cite{72_ramana2022wogru} & WOA & HPT     & N/A&  GRU  & WSN-DS   &  & 99.804  & 99.868   & 99.83    & 99.866   & \multicolumn{1}{c}{ 99.826}  & & &  \\\hline
 
& & & &  &  & \multicolumn{1}{c}{ M} & 76.011  & 81.737   & 76.011   & 71.461   &   & & &  \\
 
 & & & & & \multirow{-2}{*}{ NSL-KDD}  & \multicolumn{1}{c}{ B} & 77.205  & 83.594   & 77.205   & 76.892   &    & & & \\ \cline{6-15}
 
 & & & & &  & \multicolumn{1}{c}{ M} & 99.15 & 98.807   & 98.806   & 98.806   &    & & & \\
 
& & & & & \multirow{-2}{*}{ BoT-IoT}  & \multicolumn{1}{c}{ B} & 99.994  & 99.993   & 99.993   & 99.992   &   & & &  \\ \cline{6-15}
 
& & & & &  & \multicolumn{1}{c}{ M} & 99.917  & 89.875   & 92.044   & 89.988   &   & & &  \\
 
 & & & & & \multirow{-2}{*}{ KDDCup-99}    & \multicolumn{1}{c}{ B} & 99.935  & 94.349   & 92.318   & 92.743   &   & & &  \\ \cline{6-15}
 
& & & & &  & \multicolumn{1}{c}{ M} & 99.911  & 99.91    & 99.91    & 99.888   &   & & &  \\
 
\multirow{-8}{*}{\cite{73_abd2023intrusion}} & \multirow{-8}{*}{CSA} & \multirow{-8}{*}{FS} &\multirow{-8}{*}{-} &  \multirow{-8}{*}{CNN} & \multirow{-2}{*}{ CICIDS-2017}  & \multicolumn{1}{c}{ B} & 99.997  & 99.997   & 99.997   & 99.997   &    & & & \\\hline
& & & &    & BoT-IoT  &  & 97.61 & \multicolumn{1}{l}{}  & \multicolumn{1}{l}{}  & \multicolumn{1}{l}{}  &   &  {97.81}   &  &   \\
\multirow{-2}{*}{\cite{74_saheed2023novel}} & \multirow{-2}{*}{HAEMPSO,modified PSO}      & \multirow{-2}{*}{FS,HPT} & -&  \multirow{-2}{*}{DNN}  & UNSW-NB15  &  & 94.62 & \multicolumn{1}{l}{}  & \multicolumn{1}{l}{}  & \multicolumn{1}{l}{}  &  &  {93.8} &  &  \\\hline
& & & & &  CIDDS-001 &  &  99.3 & \multicolumn{1}{l}{}  &  98.3 & \multicolumn{1}{l}{ } & \multicolumn{1}{c}{ 99}  & & & \\
 
& & & & & UNSW-NB15  &  & 99.1  & \multicolumn{1}{l}{ } & 98 & \multicolumn{1}{l}{ } & \multicolumn{1}{c}{ 98.99} & & & \\
\multirow{-3}{*}{\cite{75_jothi2023wils}} & \multirow{-3}{*}{WOA} & \multirow{-3}{*}{PT}  & \multirow{-3}{*}{N/A} &  \multirow{-3}{*}{LSTM} &  NSL-KDD   & &  99.5 & \multicolumn{1}{l}{ } &  98.7 & \multicolumn{1}{l}{}  & \multicolumn{1}{c}{ 98.45}& & &  \\\hline
  & & & &  & BoT-IoT  &  & 96 & \multicolumn{1}{l}{}  & 96 & \multicolumn{1}{l}{}  & \multicolumn{1}{c}{97.3} & & &   \\
\multirow{-2}{*}{\cite{76_shahapure2021water}} & \multirow{-2}{*}{WWO-MSO (WMSA)}         & \multirow{-2}{*}{FS} & \multirow{-2}{*}{-} &  \multirow{-2}{*}{DRNN} & KDDCup-99  &  & 94.5  & \multicolumn{1}{l}{}  & 92.9 & \multicolumn{1}{l}{}  & \multicolumn{1}{c}{96.4}  & & &  \\\hline
\cite{77_om2021harmony} & HSHO & FS& -&  DRL & \multicolumn{1}{l}{}  &  & 96.925  & \multicolumn{1}{l}{}  & \multicolumn{1}{l}{}  & \multicolumn{1}{l}{}  &    & &  &  {TPR=96.9, TNR=97.92}\\\hline
 
& & & & & NSL- KDD &  & 95.6  & 98.3 & 92.2 & 95.2 & \multicolumn{1}{c}{ 98.6} & &  { 0.4} &  {FNR=7.8,  NPV=98.6, MCC=91.3}   \\\cline{6-15}
 
 & & & & & Botnet   &  & 95 & 97.6 & 95.3 & 96.4 & \multicolumn{1}{c}{ 53.1} & &  {4.2} &  {FNR=4.7,  NPV=53.1, MCC=40.5 } \\\cline{6-15}
 
& & & & & CICIDS-2017    &  & 95.4  & 99.9 & 95.1 & 97.5 & \multicolumn{1}{c}{ 99.7} & &  {0}   &  {FNR=4.8, NPV=99.7, MCC=74.8 }  \\\cline{6-15}
\multirow{-4}{*}{\cite{78_gopalakrishnan2022new}} & \multirow{-4}{*}{AF-EFO}    & \multirow{-4}{*}{FS+PT}  & \multirow{-4}{*}{-} &   \multirow{-4}{*}{HR-OELM using DNN,RF,Adaboost} & CICIDS-2018    &  & 95.3  & 99.7 & 90.8 & 94.9 & \multicolumn{1}{c}{ 83.6} & &  {1.5} &  {FNR=05.7,  NPV=83.6, MCC=065.4}  \\\hline
   & & & & & NF-ToN-IoT &  & 96.83 & \multicolumn{1}{l}{}  & \multicolumn{1}{l}{}  & \multicolumn{1}{l}{}  &   & &  &  {MCC=89.74}  \\
 & & & &   & NF-Bot-IoT &  & 98.43 & \multicolumn{1}{l}{}  & \multicolumn{1}{l}{}  & \multicolumn{1}{l}{}  &   & &  &  {MCC=57.71}  \\
\multirow{-3}{*}{\cite{79_mohy2023whale}} & \multirow{-3}{*}{WOA}     & \multirow{-3}{*}{FS} & \multirow{-3}{*}{-} &  \multirow{-3}{*}{RBFNN} & Merged   &  & 95.93 & \multicolumn{1}{l}{}  & \multicolumn{1}{l}{}  & \multicolumn{1}{l}{}  &   & &  &  {MCC=82.68}  \\\hline
\cite{80_bajpai2024hybrid} & WOA   & HPT  & N/A &  XGBoost (EGB) & IoTID20 &  &   98.86 & 98.67 & 99.91 & 99.30 &   & & &   AUC=98.91\\
& & & & & UNSW-NB15 &  &   99.73 & 99.05 & 99.00 & 99.03 &   & & & AUC=99.01 \\\hline
 
& & & & & CICIDS2017 &  & 99.75 & 98.52    & 99.1 & 98.81    &  & & &   \\
 
\multirow{-2}{*}{\cite{81_hanafi2024intrusion}} &  \multirow{-2}{*}{GJO-OBL (IBGJO)} & \multirow{-2}{*}{FS} & \multirow{-2}{*}{32,20} &  \multirow{-2}{*}{LSTM} & NSL-KDD  &  & 98.93 & 97.98    & 98.37    & 98.17    &   & & &  \\\hline
& & & &    & APA-DDoS &  & 99.35 & 99.9 & \multicolumn{1}{l}{}  & 99.08    &  &  {98.99}   &  &   \\
\multirow{-2}{*}{\cite{82_kethineni2024intrusion}} & \multirow{-2}{*}{WHO}  & \multirow{-2}{*}{PT}       & \multirow{-2}{*}{N/A} &  \multirow{-2}{*}{a fused CNN model with Bi-GRU} & ToN-IoT  &  & 99.71 & 99.89    & \multicolumn{1}{l}{}  & 99.05    &  &  {99.02}   &  & \\\hline
\cite{83_singh2024intrusion} & CO-IHHO       & FS & -&  DT and KNN & BoT-IoT  &  &  100(B),  & \multicolumn{1}{l}{}  & \multicolumn{1}{l}{}  & \multicolumn{1}{l}{}  &  & & &   \\
  &  & & &  & &  &  99.65(M)  &   &   &  &   & & &  \\\hline
 & & & &   & NSL-KDD  &  & 98.89 & \multicolumn{1}{l}{}  & 97.03    & \multicolumn{1}{l}{}  & \multicolumn{1}{c}{98.76}  & &  {1.24} &  \\
& & & &    &  UNSW-NB15 &  & 90.22 & \multicolumn{1}{l}{}  & 94.83    & \multicolumn{1}{l}{}  & \multicolumn{1}{c}{88.06} & &  {11.94}  & \\
\multirow{-3}{*}{\cite{84_mahanipour2024enhancing}} & \multirow{-3}{*}{hybrid BQABC-GA}      & \multirow{-3}{*}{FS} & \multirow{-3}{*}{11,10.6,10.33} &  \multirow{-3}{*}{KNN} &  BoT-IoT   &  & 98.49 & \multicolumn{1}{l}{}  & 99.79    & \multicolumn{1}{l}{}  & \multicolumn{1}{c}{99.27} & &  {0.73} &  \\\hline
\cite{85_vijayan2024original} & TuSO     & HPT     & N/A&  EL combining RF, XGBoost, LightGBM (LGBM), and CatBoost & MQTT dataset   &  & 99.12 & 97.89    & 95.24    & 96.37    &    & & & \\\hline
  & & & &  & ToN-IoT  &  & 98.81 & 90.84    & 78.95    & 79.49    &  & & &   \\
\multirow{-2}{*}{\cite{86_aburasain2024enhanced}} & \multirow{-2}{*}{Enhanced BWO, BES}   & \multirow{-2}{*}{HPT,FS}  & \multirow{-2}{*}{-} &  \multirow{-2}{*}{HDL} &  Edge-IIoTset   &  & 98.35 & 84.85    & 80.95    & 82.79    &   & & &  \\\hline
\cite{87_harbi2024bio} & a modified FOA      & FS & 39 &  {DT}  &  Edge-IIoT &  & 79.64   &  &  &  &   & & &  \\\hline
 & & & &   & IoTID-20 &  & 99.8  & \multicolumn{1}{l}{}  & \multicolumn{1}{l}{}  & \multicolumn{1}{l}{}  &   & & &  \\
 & & & &   &  NetFlow-BoT-IoT-v2   &  & \multicolumn{1}{l}{} & 99.17    & 99 & 99.1 &   & & &  \\
 & & & &   &  NF-ToN-IoT-v2  &  & 99.9  & 99.9 & 99.8 & 99.8 &  & &  {0.001}  &   \\
 & & & &   &  NSL-KDD   &  & \multicolumn{1}{l}{} & \multicolumn{1}{l}{}  & \multicolumn{1}{l}{}  & \multicolumn{1}{l}{}  & &  {99.52}   &  &  \\
\multirow{-5}{*}{\cite{88_rabie2024novel}} & \multirow{-5}{*}{DRFO} & \multirow{-5}{*}{FS} & \multirow{-5}{*}{-} &  \multirow{-5}{*}{DBRF}  &  UNSW-NB 15  &  & 98.5  & 99 & 99 & 98.5 &   & &  {8.2} &   \\\hline
\cite{89_elsedimy2024novel} & Improved GWO (IGWO) & HPT & N/A&  QSVM & BoT-IoT  &  & 99.11 & 99.45    & 99.34    & 97.48    &   & & &  \\\hline
\cite{90_biju2024evaluated} & EBSA & PT       & N/A&  DBN (RBMs,MLPs)  & NSL-KDD  &  & 98.96 & 99.4 & 98.87    & \multicolumn{1}{l}{}  &   & & &  \\\hline
 & & & &   & BoT-IoT  &  & 99.98 & 99.94    & \multicolumn{1}{l}{}  & \multicolumn{1}{l}{}  &   &  {99.97}   &  {1.3} &  {ROC=99.99}   \\
\multirow{-2}{*}{\cite{91_saheed2024voting}}  & \multirow{-2}{*}{GWO}         & \multirow{-2}{*}{HPT}       & \multirow{-2}{*}{N/A} &  \multirow{-2}{*}{EL (DT,RF,KNN,MLP)} &  UNSW-NB15 &  & 100   & 99.59    & \multicolumn{1}{l}{}  & \multicolumn{1}{l}{}  &   &  {99.9} &  {1.5} &  {ROC=99.4}  \\\hline
 
& & & & &  & \multicolumn{1}{c}{SVM} & 97.842  & \multicolumn{1}{l}{ } & 97.921   & \multicolumn{1}{l}{ } &  & &  { 0.012}   &   \\
 
& & & & & \multirow{-2}{*}{NSL-KDD}  & \multicolumn{1}{c}{ KNN} & 98.975  & \multicolumn{1}{l}{ } & 99.959   & \multicolumn{1}{l}{ } &  & &  { 0.002}   &   \\ \cline{6-15}
 
& & & & &  & \multicolumn{1}{c}{SVM} & 71.673  & \multicolumn{1}{l}{ } & 75.992   & \multicolumn{1}{l}{ } &  & &  { 0.093}   &   \\
 
& & & & & \multirow{-2}{*}{CIC-IDS2017}  & \multicolumn{1}{c}{KNN} & 97.234  & \multicolumn{1}{l}{ } & 93.171   & \multicolumn{1}{l}{ } & & &  { 0.007}   & \\ \cline{6-15}
 
& & & & &  & \multicolumn{1}{c}{SVM} & 99.788  & \multicolumn{1}{l}{ } & 99.819   & \multicolumn{1}{l}{ } &   & &  { 0.027}   &  \\
 
\multirow{-6}{*}{\cite{92_sharmamulti}} & \multirow{-6}{*}{multi-objective PDO}         & \multirow{-6}{*}{FS} & \multirow{-6}{*}{10-9,11-15,12} &  \multirow{-6}{*}{SVM,KNN} & \multirow{-2}{*}{ IoTID20}  & \multicolumn{1}{c}{KNN} & 99.402  & \multicolumn{1}{l}{ } & 99.386   & \multicolumn{1}{l}{ } &   & &  {0.006}   &  \\\hline
 
\cite{93_cherian2024iot} & BOA   & PT  &N/A &  DBN  & UNSW-NB15  &  & 97.77 & 96.62    & 93.85    & 91.77    &  & & &  {MCC=91.23}    \\\hline
\cite{94_aljehane2024golden} & a new GJOA, SSA              & FS, HPT &- &  A-BiLSTM  & CICIDS-2017    &  & 99.69 & 98.92    & 98.92    & 98.92    &  & &  &  {MCC=98.74}     \\\hline
& & & & & UNSW-NB15  &  & 98.89 & &  &  &  & & &   \\
\multirow{-2}{*}{\cite{95_bella2024intrusion}} & \multirow{-2}{*}{BOA}       & \multirow{-2}{*}{FS} & \multirow{-2}{*}{-} &  \multirow{-2}{*}{DenseNet} & NSL-KDD  &  & 98.4  &  &  &  &   & & &  \\\hline
 
 & & & & & NSL-KDD  &  & 99.25 & 99.41    & 99.34    & 98.96    &  & & &   \\
 
& & & & & ToN-IoT  &  & 89.61 & 83.57    & 89.56    & 85.72    &    & & & \\
 
\multirow{-3}{*}{\cite{96_vadakkethil2024mayfly}} & \multirow{-3}{*}{MOA}    & \multirow{-3}{*}{FS} & \multirow{-3}{*}{-} &  \multirow{-3}{*}{BiLSTM} & UNSW-NB15  &  & 99.35 & 98.49    & 99.28    & 98.64    &  & & &   \\\hline
\cite{97_10.1007/s11276-023-03435-0} & TLBO  & FS       & -&  RF & UNSW-NB15  &  & 86.5  & \multicolumn{1}{l}{}  & \multicolumn{1}{l}{}  & \multicolumn{1}{l}{}  &   & & &  \\\hline
 
& & & & &  & \multicolumn{1}{c}{M} & 99.9  & 99.06    & 99.79    & 99.41    &   & &  {0.95} &  {MCC=99.39}   \\
& & & & & \multirow{-2}{*}{NSL-KDD}  & \multicolumn{1}{c}{ B} & 99.84 & 99.68    & 99.94    & 99.81    &   & & &  {MCC=99.67}   \\ \cline{6-15}
 
\multirow{-3}{*}{\cite{98_moizuddin2022bio}} & \multirow{-3}{*}{Generalized Mean GWO}  & \multirow{-3}{*}{FS} & \multirow{-3}{*}{15,5} &  \multirow{-3}{*}{ElasticNet Contractive AE} & BoT-IoT  & \multicolumn{1}{c}{ M} & 99.99 & \multicolumn{1}{l}{ } & \multicolumn{1}{l}{ } & 99.5 &  & & &  {AUC=100, MCC=99.51}   \\\hline
\cite{99_kaviarasan2023network} & COA  & PT  & N/A&  1D-CNN+COA (creating a HNM) & NSL-KDD  &  & 87.19 & 88.28    & 89.49    & 91.19    &  & & &   \\\hline
 
& & & & & NSL-KDD  &  & 92.1  & 92.3 & 92.9 & 93.4 &  & & &   \\
 
\multirow{-2}{*}{\cite{100_pingale2023remora}} & \multirow{-2}{*}{ROA}         & \multirow{-2}{*}{FS,PT+HPT} & \multirow{-2}{*}{-} &  \multirow{-2}{*}{DMN} & CICIDS-2018    &  & 94.5  & 93.1 & 93.9 & 93.2 &   & & &  \\\hline
 & & & &   & NSL-KDD  &  & 94.635  & \multicolumn{1}{l}{}  & 96.64    & \multicolumn{1}{l}{}  & \multicolumn{1}{c}{96.02}  & & & \\
\multirow{-2}{*}{\cite{101_deore2022hybrid}}  & \multirow{-2}{*}{CCSO} & \multirow{-2}{*}{PT}       & \multirow{-2}{*}{N/A} &  \multirow{-2}{*}{Deep LSTM}  &  BoT-IoT   &  & 96.71 & \multicolumn{1}{l}{}  & 96.35    & \multicolumn{1}{l}{}  & \multicolumn{1}{c}{91.985} & & & \\\hline
 
\cite{102_makhadmeh2024intrusion} & modified AOA         & FS & -&  KNN & BoT-IoT &  & 99.998  & \multicolumn{1}{l}{ } & \multicolumn{1}{l}{ } & \multicolumn{1}{l}{ } &   & & &  \\\hline
 
 & & & & & IOTID20  &  & 96.7414 & 91.3695  & 100  & 95.4901  &  & & &   \\
 
\multirow{-2}{*}{\cite{104_10317882}} & \multirow{-2}{*}{fuzzy and GA: IWD and BBO} & \multirow{-2}{*}{FS} & \multirow{-2}{*}{-} &  \multirow{-2}{*}{FNN} & CICIDS-2017    &  & 98.2339 & 99.4831  & 98.6334  & 99.0865  &   & & &  \\\hline
 
& & & & & UNSW-NB15  &  & 76.93 & 88.24    & 66.83    & 76.06    &  & & &   \\
 
& & & & & TON-IOTwin7    &  & 99.9  & 99.89    & 99.89    & 99.89    &   & & &  \\
 
& & & & & TON-IOTwin10   &  & 99.86 & 99.86    & 99.86    & 99.86    &   & & &  \\
 
\multirow{-4}{*}{\cite{105_10541183}}  & \multirow{-4}{*}{combination of IGC and SA}  & \multirow{-4}{*}{FS} & \multirow{-4}{*}{-} &  \multirow{-4}{*}{an AE with DNN} & CICIDS2017 &  & 99.4  & 99.47    & 99.34    & 99.4 &    & & & \\\hline
 & & & &   & UNSW-NB15  &  & 98.85 & \multicolumn{1}{l}{}  & 59.06    & 58.64    & \multicolumn{1}{c}{99.36} & &  &  {MCC=60.36}  \\
\multirow{-2}{*}{\cite{106_escorcia2023sea}}  & \multirow{-2}{*}{STFA,SpSO}           & \multirow{-2}{*}{FS,HPT}  & \multirow{-2}{*}{-} &  \multirow{-2}{*}{DBN}   & TON\_IoT &  & 99.51 & \multicolumn{1}{l}{}  & 99.73    & \multicolumn{1}{l}{}  &    & & & \\\hline
\cite{107_10.1016/j.comcom.2023.03.005}  & MPO    & FS &- &  RNN & NSL-KDD  &  & \multicolumn{1}{l}{} & \multicolumn{1}{l}{}  & \multicolumn{1}{l}{}  & \multicolumn{1}{l}{}  &   & &  &  {TNR=94, TPR=94}  \\\hline
& & & & &  KDDCup-99 &  &  96.1 &     &    &  100 &  99.3  & & 0.6 &  NPV=0.989, FNR=0.003, TNR=0.996, PPV=0.989, MCC=0.934, AUC=0.989 \\\cline{6-15}
 
 & & & & & UNSW-NB15    &  &  99.1  &     &    & 99.4 &  98.5  & & 0.9 & NPV=98.5, FNR=0.4, TNR=99.5, PPV=98.5, MCC=96.3, AUC=97.2  \\\cline{6-15}
 
\multirow{-3}{*}{\cite{110_maazalahi2024k}} & \multirow{-3}{*}{ASO-EO} & \multirow{-3}{*}{FS} & \multirow{-3}{*}{-} & \multirow{-3}{*}{k-means} & NSL-KDD & & 98.9 &   &     &  100  & 99.1 &   & 99.1 & NPV=98.9, FNR=0.3, TNR=99.7, PPV=99.1, MCC=97.1, AUC=98.8 \\\hline
 & & & & &  & \multicolumn{1}{l}{(EGB)} & 79.11 & 82.5267 & 79.1132  & 70.9564 &   & &   &  \\\cline{7-15}

\multirow{-2}{*}{\cite{111_dakic2024intrusion}}  & \multirow{-2}{*}{GSAPSO} & \multirow{-2}{*}{HPT} & \multirow{-2}{*}{N/A} & \multirow{-2}{*}{XGBoost and KNN} & \multirow{-2}{*}{CAN dataset}  & \multicolumn{1}{l}{(KNN)} & 79.10 & 82.9847 &  79.0996  & 70.8823 &   & &     &  \\\hline

\cite{112_latif2024dtl} & GA         & HPT & N/A &  EL (CNNs-based) & Edge\_IIoTset & M & 100  & \multicolumn{1}{l}{100} & \multicolumn{1}{l}{100} & \multicolumn{1}{l}{100} &   & & & Cohen’s Kappa score=100 \\\hline

& & & & &  \multicolumn{1}{l}{UNSW-NB15}  & & 97.7217 & 97.41 & 97.72  & 97.56 &   & &   &  \\\cline{6-15}

\multirow{-2}{*}{\cite{113_li2024cooperative}}  & \multirow{-2}{*}{BHO} & \multirow{-2}{*}{FS} & \multirow{-2}{*}{-} & \multirow{-2}{*}{Parallel CNNs} & \multicolumn{1}{l}{NSL-KDD}  &  & 99.8928 & 99.89 &  99.89  & 99.89 &   & &     &  \\\hline

& & & & &  \multicolumn{1}{l}{CICIDS-2017}  & & 99.55 & 99.55 & 99.55  & 99.55 &   & &   & AUC=99.55 \\\cline{6-15}

\multirow{-2}{*}{\cite{114_alamro2023modeling}}  & \multirow{-2}{*}{ALO, FPA} & \multirow{-2}{*}{FS, HPT} & \multirow{-2}{*}{-} & \multirow{-2}{*}{CNN+LSTM} & \multicolumn{1}{l}{ToN-IoT}  &  & 99.31 & 99.31 &  99.31  & 99.31 &   & &     & AUC=99.31 \\\hline

\cite{115_otoum2021ids} & HMS & clustering & N/A &  LightNet, Deep Q-learning & NSL-KDD &  &   & \multicolumn{1}{l}{} & \multicolumn{1}{l}{96.6} & \multicolumn{1}{l}{} & 96.8  &  96.9 & & 
\label{tab:performance_appendix}
\end{longtable}
    \end{landscape}
    \twocolumn

\end{appendices}

\end{document}